\documentclass[12pt,a4paper,final]{iopart}
\usepackage{iopams} 
\usepackage{graphicx}
\usepackage[breaklinks=true,colorlinks=true,linkcolor=blue,urlcolor=blue,citecolor=blue]{hyperref}
\usepackage{amssymb}
\usepackage{bm}
\usepackage{color}
\usepackage{relsize}
\usepackage{braket}
\usepackage[caption=false]{subfig}
\usepackage[all]{hypcap}
\usepackage[export]{adjustbox}

\renewcommand{\i}{\ensuremath{\mathrm{i}}}

\renewcommand{\d}{\ensuremath{\mathrm{d}}}

\begin{document}

\title[]{Dynamical screening in monolayer transition-metal dichalcogenides and its manifestations in the exciton spectrum}

\author{Benedikt Scharf$^{1}$}
\address{$^1$Institute for Theoretical Physics and Astrophysics, University of W\"{u}rzburg, Am Hubland, 97074 W\"{u}rzburg, Germany}

\author{Dinh Van Tuan$^{2}$}
\address{$^2$Department of Electrical and Computer Engineering, University of Rochester, Rochester, New York 14627, USA}

\author{Igor \v{Z}uti\'c$^{4}$}
\address{$^4$Department of Physics, University at Buffalo, State University of New York, Buffalo, NY 14260, USA}

\author{Hanan~Dery$^{2,5}$}
\address{$^2$Department of Electrical and Computer Engineering, University of Rochester, Rochester, New York 14627, USA}
\address{$^5$Department of Physics and Astronomy, University of Rochester, Rochester, New York 14627, USA}
\ead{hanan.dery@rochester.edu}

\begin{abstract}
Monolayer transition-metal dichalcogenides (ML-TMDs) offer exciting opportunities to test the manifestations of many-body interactions through changes in the charge density. The two-dimensional character and reduced screening in ML-TMDs lead to the formation of neutral and charged excitons with binding energies orders of magnitude larger than those in conventional bulk semiconductors. Tuning the charge density by a gate voltage leads to profound changes in the optical spectra of excitons in ML-TMDs. On the one hand, the increased screening at large charge densities should result in a blueshift of the exciton spectral lines due to reduction in the binding energy. On the other hand, exchange and correlation effects that shrink the band-gap energy at elevated charge densities (band-gap renormalization) should result in a redshift of the exciton spectral lines. While  these competing effects can be captured through various approximations that model long-wavelength charge excitations in the Bethe-Salpeter Equation, we show that a novel coupling between excitons and shortwave charge excitations is essential to resolve several experimental puzzles.

Unlike ubiquitous and well-studied plasmons, driven by collective oscillations of the background charge density in the long-wavelength limit, we discuss the emergence of shortwave plasmons that originate from the short-range Coulomb interaction through which electrons transition between the $\bm{K}$ and $-\bm{K}$ valleys. The shortwave plasmons have a finite energy-gap because of the removal of spin-degeneracy in both the valence- and conduction-band valleys (a consequence of breaking of inversion symmetry in combination with strong spin-orbit coupling in ML-TMDs). We study the coupling between the shortwave plasmons and the neutral exciton through the self-energy of the latter. We then elucidate how this coupling as well as the spin ordering in the conduction band give rise to an experimentally observed optical sideband in electron-doped W-based MLs, conspicuously absent in electron-doped Mo-based MLs or any hole-doped ML-TMDs. While the focus of this review is on the optical manifestations of many-body effects in ML-TMDs, a systematic description of the dynamical screening and its various approximations allow one to revisit other phenomena, such as nonequilibrium transport or superconducting pairing, where the use of the Bethe-Salpeter Equation or the emergence of shortwave plasmons can play an important role.

\end{abstract}

\pacs{00.00, 20.00, 42.10}
\vspace{2pc}
\noindent{\it Keywords}: Article preparation, IOP journals
\submitto{\JPA}
\maketitle

\section{Introduction}

\begin{figure}[htb!]
\centering
\includegraphics*[width=0.99\textwidth]{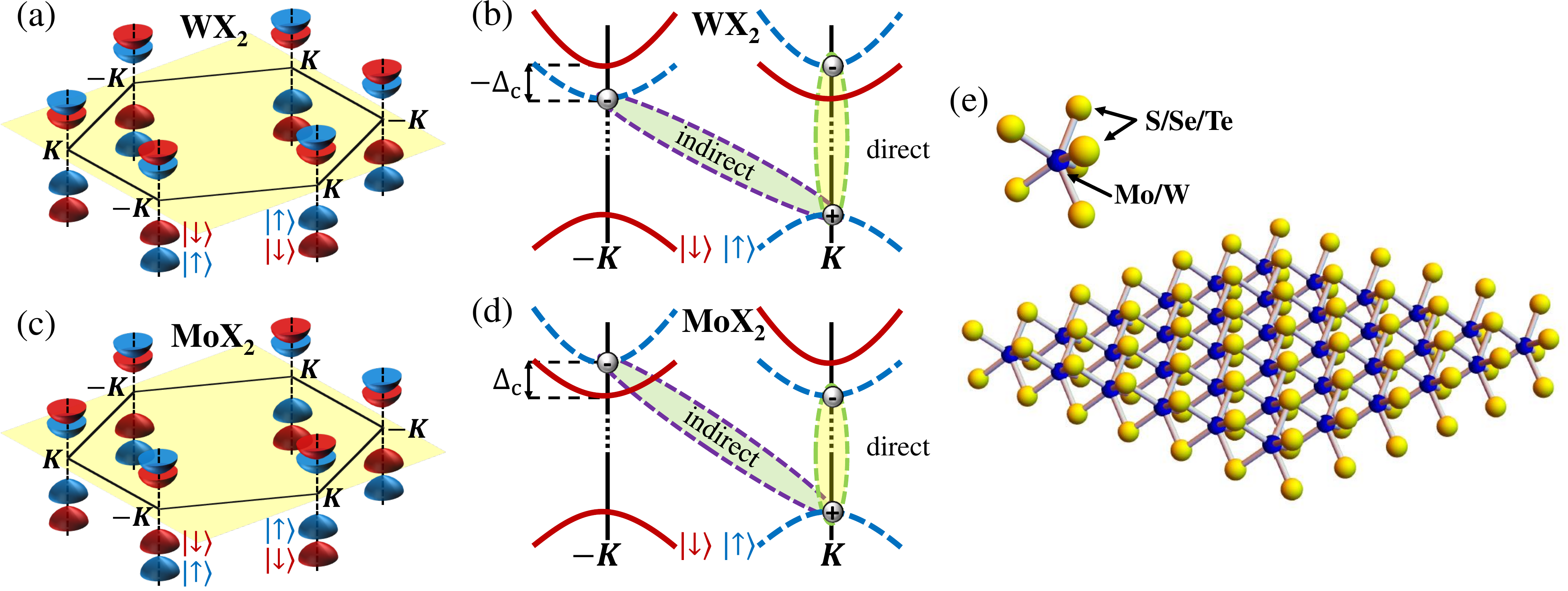}
\caption{Brillouin zone as well as low-energy conduction and valence bands around the time-reversed $\bm{K}$-point valleys for (a) W- and (c) Mo-based MLs, respectively. The two cases differ in the order of the spin-up and spin-down valleys in the conduction band.  (b),(d) The resulting direct and indirect excitons as well as the electronic states from which these excitons are primarily formed. The indirect exciton has lower energy in W-based MLs, while the direct exciton has lower energy in Mo-based MLs. The spin of the bands is color coded and $\Delta_c$ is the spin-splitting energy in the conduction band, which is typically much smaller than the one in the valence band. (e) Atomic structure of ML-TMDs.}\label{fig:Scheme}
\end{figure}

Among the rapidly expanding class of two-dimensional (2D) materials, monolayer transition-metal dichalcogenides (ML-TMDs), such as ML-WX$_2$ and ML-MoX$_2$ (X=S,Se,Te), have been the topic of particularly intense research interest \cite{Wang2012:NN,Geim2013:N,Xu2014:NP,Mak2016:NP,Wang_RMP18}. Their remarkable properties make them attractive candidates both for potential device applications as well as for exploring fundamental physical phenomena, not easily observable in other materials: TMDs become direct band-gap semiconductors in the limit of a single atomic ML~\cite{Mak2010:PRL,Splendiani2010:NL}, with possible applications in nanoscale electronics, optoelectronics, and energy harvesting~\cite{Lembke2012:ACSNano,Bao2013:APL,LopezSanchez2013:NN,Britnell2013:S,Pospischil2014:NN,Yin2014:S,Wang2015:NT,Cui2015:NN,Rathi2015:NL,Dumcenco2015:ACSNano,Yan_NatCommun16}. The hallmark of ML-TMDs is their spin-valley coupling, as shown in Fig.~\ref{fig:Scheme}. The coupling gives rise to several peculiar properties such as a valley-dependent helicity of the interband optical transitions~\cite{Xiao2012:PRL}, the valley Hall and valley Zeeman effects~\cite{Mak2014:S,Srivastava2015:NP,Stier2016:NC}, as well as strong magneto- and photo-luminescence~\cite{Mak2010:PRL,Splendiani2010:NL,Scrace2015:NN}; all of which have important implications for transport and qubits~\cite{Wang2012:NN,Xu2014:NP,Song2013:PRL,Dery_PRB16,Klinovaja2013:PRB,Rohling2014:PRL}. As 2D crystals, ML-TMDs offer several opportunities to tune their electronic or magnetic properties. For example, their 2D nature makes ML-TMDs susceptible to proximity effects that can fundamentally alter the properties of these layers~\cite{Zhao2017:NN,Zhong2017:SA,Scharf2017:PRL,Zutic_MT18}, while stacking them via van der Waals forces alleviates the need to fabricate logic and optoelectronic devices with lattice-matched crystals~\cite{Wang2012:NN,Geim2013:N,Mak2016:NP}.

\begin{figure}[htb!]
\centering
\includegraphics*[width=0.8\textwidth]{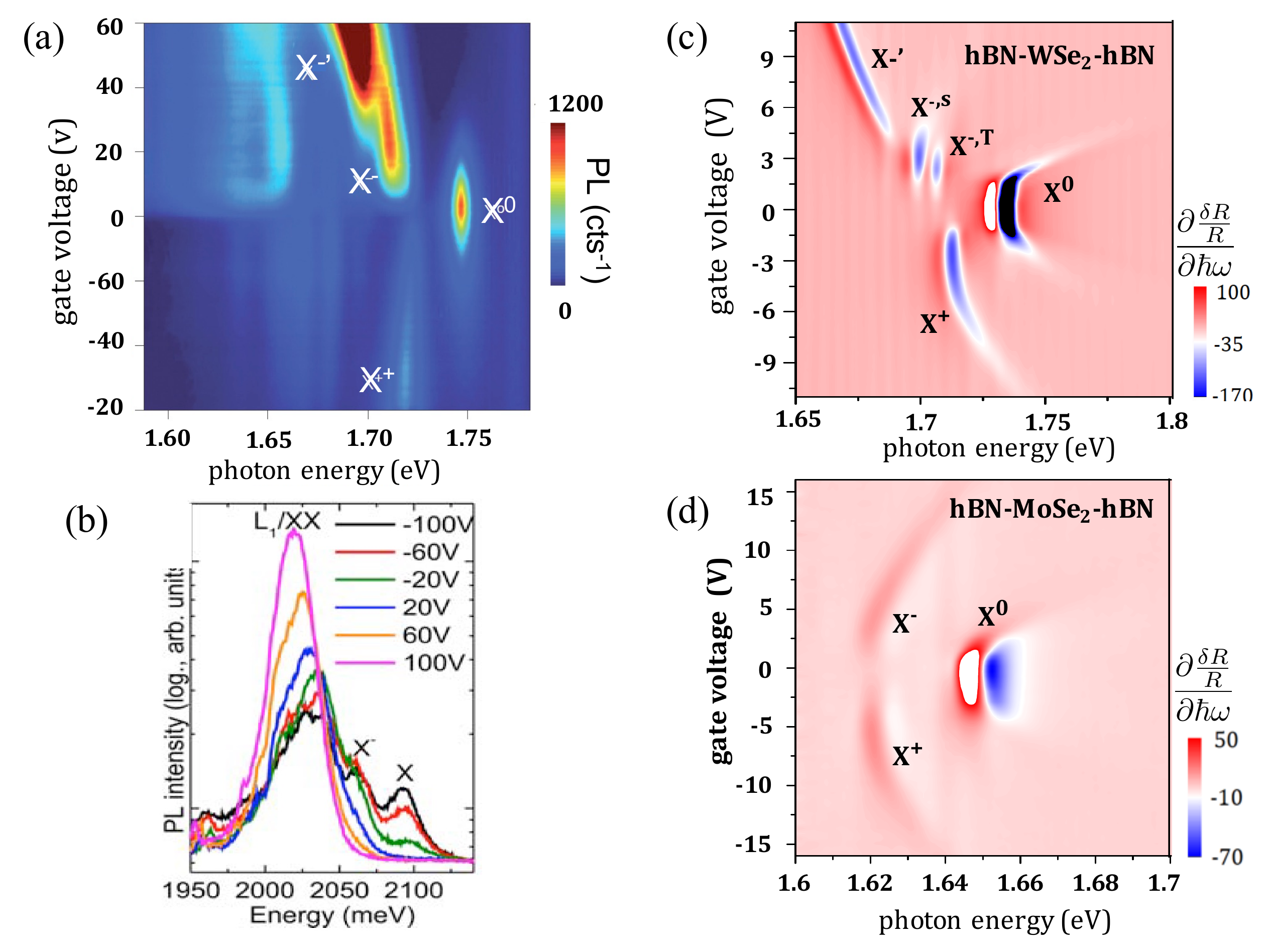}
\caption{(a) Photoluminescence intensity map of ML-WSe$_2$ on SiO$_2$ as a function of photon energy and gate voltage showing the neutral exciton ($X^0$), negative ($X^-$) and positive ($X^+$) trions as well as the optical sideband ($X-$'), taken from Ref.~\cite{Jones2013:NN}. (b) Photoluminescence spectra of ML-WS$_2$ as a function of photon energy for different gate voltages, taken from Ref.~\cite{Plechinger2015:PSS}. Derivative of the reflectance contrast spectra as a function of photon energy and gate voltage for (c) ML-WSe$_2$ and (d) ML-MoSe$_2$ encapsulated in hexagonal boron-nitride (hBN), taken from Ref.~\cite{Wang_NanoLett17}.}\label{fig:Gating}
\end{figure}

The 2D character and reduced screening in ML-TMDs lead to very large exciton binding energies up to $\sim$0.5 eV ~\cite{Mak2010:PRL,Chernikov2014:PRL,Amani2015:S}. Combined with their direct band gap, one can then study manifestations of many-body interactions through their optical spectrum across a wide range of background charge densities that are controlled by a gate voltage \cite{Dery_PRB16,Schmidt_NanoLett16,VanTuan_PRX17,Chaves_2DMater17,Steinhoff_NatCom17,Steinhoff_PRB18}. As illustrated in Fig.~\ref{fig:Gating}, the gate voltage leads to profound changes in the photoluminescence (PL) or differential reflectance measurements of ML-TMDs~\cite{Jones2013:NN,Plechinger2015:PSS,Wang_NanoLett17}. 
In the undoped case (zero gate voltage), one can clearly observe pronounced peaks associated with the neutral exciton, denoted by $X^0$ in Figs.~\ref{fig:Gating}(a), (c) and (d), or by $X$ in Fig.~\ref{fig:Gating}(b). The position of this peak remains almost unchanged for moderate gate voltages. It significantly loses spectral weight and decays only at high gate voltages/charge densities. As the charge carrier density increases, charged-excitons peaks emerge for both electron- and hole-doped conditions (trions). The resulting optical features are the ones denoted by $X^\pm$, or their singlet and triplet spin configurations, $X^{-,S}$ and $X^{-,T}$, in electron-doped ML-WSe$_2$ \cite{Jones_NatPhys16,Courtade_PRB17,Plechinger2016:NC}. Also shown is the optical sideband, $X-$', that we have recently associated to the unique coupling of neutral excitons and intervalley plasmons in W-based compounds ~\cite{VanTuan_PRX17,VanTuan_arXiv18}. Its signature is clear and resolved from that of the two negative trions in absorption-type experiments, as shown in Fig.~\ref{fig:Gating}(c). Its emission in PL experiments is very strong [$X-'$ in Fig.~\ref{fig:Gating}(a) and $L_1/XX$ in Fig.~\ref{fig:Gating}(b)].  

The unique optical sideband has been observed in various emission and absorption-type experiments of electron-doped WSe$_2$ and WS$_2$ MLs~\cite{VanTuan_PRX17,Jones2013:NN,Plechinger2015:PSS,Wang_NanoLett17,Plechinger2016:NC,Shang2015:ACS,Plechinger2016:NL,Wang_NatNano17,Koperski2017:NPhot}, but is conspicuously absent in Mo-based MLs [compare Figs.~\ref{fig:Gating}(c) and~(d)]. While this feature has variously been associated with the fine structure of negatively charged excitons, biexcitons, or defects in different experiments~\cite{Xu2014:NP,Jones2013:NN,Plechinger2015:PSS,Wang_NanoLett17,Plechinger2016:NC,Shang2015:ACS,Plechinger2016:NL,Wang_NatNano17,Koperski2017:NPhot,Molas_NanoS17}, it has recently been shown theoretically that its qualitative density-dependence, and absence in Mo-based compounds or hole-doped samples is actually consistent with an exciton-intervalley plasmon quasiparticle~\cite{VanTuan_PRX17}: A dynamical sideband due to this quasiparticle appears around one intervalley-plasmon energy below the indirect exciton. Since in ML-WX$_2$ the indirect exciton has lower energy than the direct exciton [recall Fig.~\ref{fig:Scheme}(b)], the dynamical sideband appears well below the direct exciton peak. In ML-MoX$_2$, where the indirect exciton is above the direct exciton [Fig.~\ref{fig:Scheme}(d)], on the other hand, the dynamical sideband coincides with the direct exciton, and thus cannot be spectrally resolved from this peak.

In this Review, we will overview the effect of dynamical screening on the self-energies of electrons and holes in ML-TMDs, followed by a short overview of how to include dynamical screening to compute neutral excitons $X^0$ in ML-TMDs using the Bethe-Salpeter Equation (BSE). A fully dynamical treatment of screening of $X^0$ by gate-induced background charge carriers will be presented, complemented with various approximations and a brief discussion of their limitations. While our focus will be on optical manifestations of many-body effects originating from changes in the carrier density, it is important to recognize that the BSE has found its use in the studies of many other systems and phenomena, including superconductivity and nonequilibrium transport~\cite{Vignale1985:PRB,Ferry2009,Tkachov2011:PRB}. Thus, the presented considerations for dynamical screening should have broader implications. 

The Review is organized as follows: We first give a short overview of the optical properties of ML-TMDs in Sec.~\ref{Sec:TMDMLs}, and introduce the bare Coulomb potential of 2D layers in Sec.~\ref{Sec:BareCoulomb}. We then introduce the dynamically-screened potential in Sec.~\ref{Sec:RPA}, studying both the long-wavelength and shortwave regimes, followed by an analysis of the resulting intravalley and intervalley plasmons in Sec.~\ref{Sec:Plasmons}. We then provide a quantitative analysis of the band-gap renormalization due to plasmons in Sec.~\ref{Sec:BGR}. Section~\ref{sec:general} deals with general properties of excitons in ML-TMDs. The computation of neutral excitons in the presence of a plasma of background charge carriers via the BSE is studied in Sec.~\ref{Sec:Excitons}, where we introduce several approximations to solve it, such as the quasistatic and Shindo approximations as well as a fully dynamical treatment of the BSE. Finally, we include an explicit coupling between excitons and shortwave plasmons in Sec.~\ref{sec:exciton_intervalley},  showing how this coupling  gives rise to an optical sideband that we identify with the $X-'$ peaks observed in experiments. A summary in Sec.~\ref{Sec:Con} concludes the Review, where we also provide an outlook for further needed investigations.

\section{Properties of monolayer transition-metal dichalcogenides}\label{Sec:TMDMLs}

ML-TMDs are direct band-gap semiconductors with the conduction band and valence band edges at the $\bm{K}$ and $\bm{K}'=-\bm{K}$ points \cite{Xiao2012:PRL}. These time-reversed points define two separate low-energy pockets/valleys in the Brillouin zone. Due to the lack of space inversion symmetry and strong spin-orbit coupling arising from the $d$ orbitals of the transition-metal atoms, the spin degeneracy in these valleys is lifted ~\cite{Song2013:PRL}, as shown in Figs.~\ref{fig:Scheme}(a,c). Whereas the spin splitting $\Delta_c$ in the conduction band is typically at most a few 10s of meV, the spin splitting in the valence band can be several 100s meV \cite{Song2013:PRL,Cheiwchanchamnangij_PRB13,Liu_PRB13,Kosmider_PRB13,Kormanyos_2DMater15}. There are two important aspects to this band structure: First, time-reversal symmetry results in a coupling between the spin and valley degrees of freedom, such that the spin ordering in opposite valleys is reversed as illustrated in Figs.~\ref{fig:Scheme}(a,c). Second, the spin ordering of the conduction band in Mo-based compounds is opposite to the one in W-based compounds, as can be seen by comparing Figs.~\ref{fig:Scheme}(a) and~(c) ~\cite{Song2013:PRL,Kormanyos_2DMater15,Dery2015:PRB}. This difference will be shown below to have profound consequences on their optical properties. Due to the large spin-splitting energy of the valence band, one can distinguish between two different series of neutral excitons in absorption experiments: One series involving the top valleys in the valence band, usually denoted as A excitons, and one involving its bottom valleys, usually denoted as B excitons \cite{Chernikov2014:PRL,He_PRL14}. The energies of type B excitons are larger by approximately the spin-splitting energy of the valence band, and therefore, they cannot be observed in the emission spectrum after energy relaxation. Below, we focus on the $A$ series [Figs.~\ref{fig:Scheme}(b,d)].

The spin-valley coupling in ML-TMDs gives rise to valley-dependent optical selection rules, where one can selectively address a given valley by circularly polarized light \cite{Xiao2012:PRL}. Most intriguingly, the 2D character and suppressed Coulomb screening lead to pronounced excitonic effects in ML-TMDs that also dominate their optical properties ~\cite{Mak2010:PRL}. The neutral excitons, bound electron-hole pairs, are either bright or dark depending on whether the optical transition between the electron and hole state is optically active or not. For an exciton to be bright (dark), the orbital transition between the electron and hole has to be dipole allowed (forbidden), whereas the spins of the electron in the conduction band and the missing electron in the valence band have to be parallel (antiparallel). Attempting to decipher many-body phenomena through optical spectroscopy measurements, we do not take into account dark excitons in the following. This approximation is reinforced by the fact that scattering between dark and bright excitons necessitates a spin-flip of the electron or hole, typically a much slower process than the lifetime of bright excitons~\cite{Lagarde2014:PRL,Wang2014:PRB,Yang2015:NP,Song2016:NL}, 

Apart from bright and dark excitons, we can also distinguish between direct and indirect excitons \cite{Dery2015:PRB}: A direct exciton is formed from an electron-hole pair within the same valley, which consequently results in a direct-gap optical transition. Indirect excitons, on the other hand, arise if the electron and hole reside in opposite valleys. For such indirect optical transitions to occur, the large momentum mismatch of the photon and indirect exciton in a perfect crystal has to be overcome by external agents such as shortwave phonons. Figures~\ref{fig:Scheme}(b,d) show that indirect excitons in ML-WX$_2$ have lower energy than direct ones, whereas direct excitons have lower energy in ML-MoX$_2$~\cite{Dery2015:PRB}. Direct and indirect excitons can be coupled via intervalley plasmons, which will be introduced in Sec.~\ref{Sec:Plasmons} along with their conventional intravalley counterparts. Before introducing them, however, we first look at the bare Coulomb interaction in ML-TMDs, responsible for the formation of the excitons in the undoped case.

\section{Bare Coulomb interaction in two-dimensional systems}\label{Sec:BareCoulomb}

\begin{figure}[htb!]
\centering
\includegraphics*[width=0.8\textwidth]{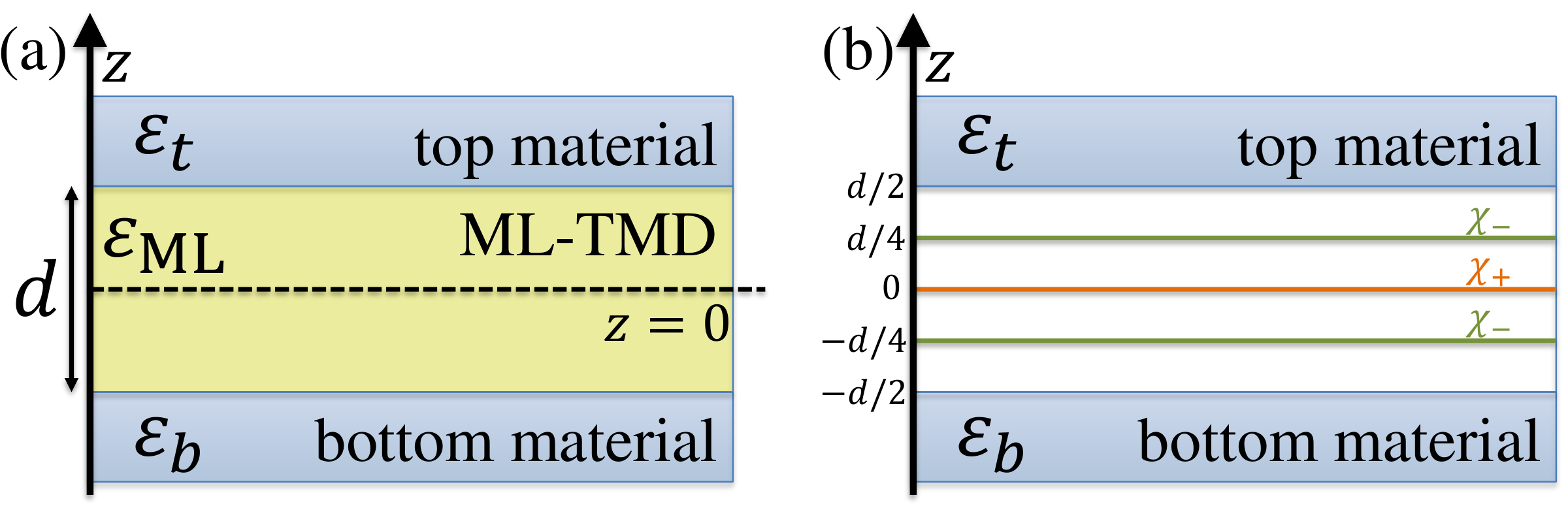}
\caption{Geometries considered to describe a ML-TMD sandwiched between two materials with the dielectric constants $\epsilon_t$ (top) and $\epsilon_b$ (bottom): (a) ML-TMD described as a dielectric medium of thickness $d$ with $\epsilon_\mathrm{ML}$. (b) ML-TMD described as three atomic layers with polarizabilities $\chi_+$ for the central layer of Mo/W atoms and $\chi_-$ for the top and bottom layers of S/Se/Te atoms [recall Fig.~\ref{fig:Scheme}(e)].}\label{fig:DielectricEnvironment}
\end{figure}

One of the main features of ML-TMDs is that they exhibit excitons with large binding energies and a Rydberg series that do not follow the 2D hydrogen model~\cite{Chernikov2014:PRL,He_PRL14}. This behavior originates from the wavevector dependence of the dielectric constant due to the environment (non-local effect), caused by  a dielectric contrast between the ML and the surrounding materials. If the dielectric constants of the top and bottom layers are small compared with the effective dielectric constant of the ML, $\epsilon_{\mathrm{ML}} > \epsilon_t,\,\epsilon_b$, then the Coulomb interaction is enhanced compared with conventional III-V semiconductor quantum well heterostructures wherein the non-local effect is weak because the well region has approximately the same dielectric constant as in the potential barrier regions. Here, we consider a ML-TMD sandwiched between two materials. Suspended MLs are modeled by assigning $\epsilon_t=\epsilon_b=1$, supported ones by $\epsilon_t=1$ and $\epsilon_b\neq 1$, and encapsulated ones by $
\epsilon_t \neq 1$ and $\epsilon_b\neq 1$. 

The bare Coulomb potential of a point charge $e$ in a 2D system, unscreened by free charge carriers, has the general form
\begin{equation}\label{Eq:BareCoulomb}
V(\bm{q})=\frac{2\pi e^2}{A\epsilon_d(q)q}\,\,,
\end{equation}
where $A$ is the sample area and $\epsilon_d(q)$ is the non-local dielectric function. Below, we present three approximated forms of $\epsilon_d(q)$ that behave similarly in the long-wavelength regime, $qd \ll 1$,  where $d$ is the thickness of the ML. The first approach is a step-like dielectric model, as shown in Fig.~\ref{fig:DielectricEnvironment}(a), similar to Refs.~\cite{Rytova_MSU67,Keldysh1979:JETP,Zhang2014:PRB,Trolle_SR17,Meckbach_PRB18}. We denote $\epsilon_d(q)$ by $\epsilon_{d,s}(q)$ in this case, and calculate the non-local dielectric function from the Poisson equation for a point charge located in $(\bm{\rho}',z')$, where $\left(\bm{\rho},z\right)$ are the cylindrical coordinates. We get,
\begin{equation}\label{Eq:Poisson}
\nabla\left[\kappa(z)\nabla\phi(\bm{\rho}-\bm{\rho}';z,z')\right]=-4\pi e\delta\left(\bm{\rho}-\bm{\rho}'\right)\delta\left(z-z'\right)\,,
\end{equation}
where the relative dielectric constant is given by
\begin{equation}\label{Eq:DielEnv1}
\kappa(z)=\left\{\begin{array}{ll}
 \epsilon_t & \mathrm{for}\quad z>d/2,\\
 \epsilon_\mathrm{ML} & \mathrm{for}\quad -d/2<z<d/2,\\
 \epsilon_b & \mathrm{for}\quad z<-d/2.\\
 \end{array}\right.
\end{equation}
Introducing the Fourier transform of $\phi$ with respect to the in-plane coordinates $\bm{\rho}$, Eq.~(\ref{Eq:Poisson}) becomes
\begin{equation}\label{Eq:PoissonMomentum}
\frac{\d}{\d z}\left[\kappa(z)\frac{\d\phi(\bm{q};z,z')}{\d z}\right]-\kappa(z)q^2\phi(\bm{q};z,z')=- \frac{4\pi e}{A}\delta\left(z-z'\right),
\end{equation}
where we require the global solution to be continuous and its derivative to be piecewise continuous with a jump of $-4\pi e/A$ at $z=z'$. This equation is then solved for $z=z'=0$, denoting the fact that electrons and holes in ML-TMDs are restricted to move in the mid-plane since they are governed by the $d$-orbitals of the transition-metal atoms. Having found the potential $\phi(\bm{q};0,0)$ in this way, one can use Eq.~(\ref{Eq:BareCoulomb}) to calculate the bare Coulomb interaction $V(\bm{q})=e\phi(\bm{q};0,0)$ between two electrons in the $xy$-plane ($z=z'=0$). The non-local dielectric function follows
\begin{equation}\label{Eq:InvDiFv1}
\epsilon_{d,s}(q)=  \frac{\left( 1-p_bp_t\e^{-2qd} \right)\epsilon_{\mathrm{ML}}}{1+(p_b+p_t)\e^{-qd}+p_bp_t\e^{-2qd}}.
\end{equation}
$p_{j}$ is  the polarization due to the dielectric contrast between the ML and the bottom/top layer ($j=b/t$),
\begin{equation}\label{Eq:pbt}
p_{j}=\frac{\epsilon_{\mathrm{ML}}-\epsilon_{j}}{\epsilon_{\mathrm{ML}}+\epsilon_{j}}.
\end{equation}

The second approximated form of the non-local dielectric function is achieved by taking the limit $qd\ll1$ in the step-like model. This approach was originally taken by Rytova and then by Keldysh to study thin semiconductor films ~\cite{Rytova_MSU67,Keldysh1979:JETP}. One can then simply expand the form of $\epsilon_s(q)$ in powers of $qd$, which to linear order yields 
\begin{equation}\label{Eq:epsgeo}
\epsilon_{RK}(q)\approx \frac{\epsilon_t+\epsilon_b}{2} + r_0q \,.
\end{equation}
The parameter $r_0$ describes a geometric correction due to the dielectric contrast between the ML and the surrounding materials, 
\begin{equation}\label{Eq:r0v1}
r_0=\frac{\epsilon_{ML}d}{2}\left(1-\frac{\epsilon_t^2+\epsilon_b^2}{2\epsilon_{ML}^2}\right)\,.
\end{equation}
It is this contrast that gives rise to a nonhydrogenic Rydberg series of neutral excitons. The dielectric function Eq.~(\ref{Eq:epsgeo}) has also been derived by Cudazzo et al. for a strictly 2D layer ($d=0$) with in-plane polarizability $\chi_\mathrm{2D}$, where $r_0=2\pi\chi_\mathrm{2D}$~\cite{Cudazzo2011:PRB}.  The advantage of using $\epsilon_{RK}(q)$ is that it yields a compact form for the real-space Coulomb potential in the mid-plane of the ML, 
\begin{eqnarray}
V(r) = \frac{A}{4\pi^2} \int \! \! d^2q \, V(q) e^{\i\mathbf{q}\cdot{\mathbf{r}}} = e^2 \int_0^{\infty} \! \! dq \frac{ J_0(qr)}{\epsilon_d(q)} \,\, , \label{eq:2D_potential_Fourier}
\end{eqnarray}
where $J_0$ is the zeroth-order Bessel function. Substituting Eq.~(\ref{Eq:epsgeo}) in this integral, one gets the celebrated Rytova-Keldysh potential \cite{Rytova_MSU67,Keldysh1979:JETP,Cudazzo2011:PRB}
\begin{eqnarray}
V_{RK}(r) = \frac{e^2}{r_0} \frac{\pi}{2} \left[ \mathbf{H}_0\left( \frac{\epsilon r}{r_0}\right) - Y_0\left( \frac{\epsilon r}{r_0}\right) \right ]\,,\,\,\,\,\, \label{eq:Keldysh}
\end{eqnarray}
where $\mathbf{H}_0$ and $Y_0$ are the zero-order Struve and Neumann special functions, and $\epsilon = (\epsilon_t+\epsilon_b)/2$.

In spite of the recent popularity of the Rytova-Keldysh dielectric function \cite{Chernikov2014:PRL,Courtade_PRB17,Berkelbach2013:PRB,Thilagam_JAP14,Wang_ADP14,Berghauser_PRB15,Zhang_NanoLett15,Ganchev_PRL15,Mayers_PRB15,Kylanpaa_PRB15,Velizhanin_PRB15,Latini_PRB15,Wu_PRB15,Qiu_PRB16,Kidd_PRB16,Kezerashvili_FBS17,Mostaani_PRB17,Szyniszewski_PRB17,Raja_NatComm17,Donck_PRB17,Stier_PRL18}, one should bear in mind that it is a correct description at macroscopic distances compared to the lattice constant ($r \gg a$). This problem is sharpened in ML-TMDs wherein the lattice constant is comparable to the thickness of the ML, and both are not excessively smaller than the Bohr radius. As a result, one has to introduce a short-range correction to the Coulomb potential in the range $r \sim d$, in which local exchange and correlation effects still do not play a significant role \cite{Courtade_PRB17,Mostaani_PRB17}. This correction is especially crucial for the description of few-body complexes due to their strong dependence on short-range interactions. For example, the binding energy of the trion (a three-body bound complex) is the difference in energy between that of the complex and that of an exciton plus a faraway third particle (electron or hole). At large distances, the interaction between the neutral exciton and the third charged particle is dipolar in nature and decays fast ($\sim$$1/r^2$). The contribution to the binding energy of trions and other few-body complexes is therefore governed by inter-particle interactions at short distances. Indeed, when using Eqs.~(\ref{Eq:InvDiFv1}) or~(\ref{Eq:epsgeo}) to calculate trion states, one finds a strong dependence of their binding energies on whether the ML is encapsulated in hexagonal boron nitride (hBN), supported on SiO$_2$, or suspended in air \cite{VanTuan_PRB18}. Empirically, however, the situation is reversed: The binding energy of trions in MoSe$_2$ and WSe$_2$, for example, have been repeatedly measured in various configurations showing that they are only slightly affected by the identity of the dielectric materials below and above the ML \cite{Wang_NanoLett17,Jones_NatPhys16,He_PRL14,Borghardt_PRM17,Branny_APL16,Ross_NatCommun12,Shepard_ACSnano17}. This behavior can only be reasoned if the binding of trions is governed by the Coulomb interaction inside the ML. 

The previous discussion suggests that the dependence of the non-local dielectric function on the environment, as provided by Eqs.~(\ref{Eq:InvDiFv1}) or~(\ref{Eq:epsgeo}), should be mitigated. The third approximated form of $\epsilon_d(q)$ is derived by taking strides in this direction, where the system is viewed as consisting of three atomic sheets that represent a ML-TMD embedded between top and bottom dielectric layers, as shown in Fig.~\ref{fig:DielectricEnvironment}(b). The central atomic sheet is that of Mo/W atoms while the top and bottom ones are of S/Se/Te atoms. In analogy to the approach by Cudazzo et al.~\cite{Cudazzo2011:PRB}, the central atomic sheet has an in-plane polarizability $\chi_+$ and the top and bottom ones are assigned an in-plane polarizability of $\chi_-$. These polarizabilities give rise to the effective dielectric constant of the ML. 
Writing the Poisson equation in cylindrical coordinates for the bare Coulomb potential of a point charge in this configuration yields
\begin{equation}\label{Eq:Poissonv2}
\nabla\left[\kappa(z)\nabla\phi(\bm{\rho}-\bm{\rho}';z,z')\right]=-4\pi e\delta\left(\bm{\rho}-\bm{\rho}'\right)\delta\left(z-z'\right)-4\pi n_\mathrm{ind}(\bm{\rho},z)
\end{equation}
with the relative dielectric constant
\begin{equation}\label{Eq:DielEnv2}
\kappa(z)=\left\{\begin{array}{ll}
 \epsilon_t & \mathrm{for}\quad z>d/2,\\
 1 & \mathrm{for}\quad -d/2<z<d/2,\\
 \epsilon_b & \mathrm{for}\quad z<-d/2.
 \end{array}\right.
\end{equation}
The induced charge density,
\begin{equation}\label{Eq:IndChargev2}
n_\mathrm{ind}(\bm{\rho},z)=\left[\delta\left(z\right)\chi_++\delta\left(z-\frac{d}{4}\right)\chi_-+\delta\left(z+\frac{d}{4}\right)\chi_-\right]\nabla^2_{\bm{\rho}}\phi\left(\bm{\rho};z,z'\right),\\
\end{equation}
is in turn related to the potential $\phi(\bm{\rho};z,z')$. After Fourier transformation with respect to $\bm{\rho}$, Eq.~(\ref{Eq:Poissonv2}) becomes
\begin{equation}\label{Eq:PoissonMomentumv2}
\begin{array}{l}
\frac{\d}{\d z}\left[\kappa(z)\frac{\d\phi(\bm{q};z,z')}{\d z}\right]-\kappa(z)q^2\phi(\bm{q};z,z')=-\frac{4\pi e}{A}\delta\left(z-z'\right)\quad\quad\\
\quad\quad\quad\quad+4\pi q^2\left[\delta\left(z\right)\chi_++\delta\left(z-\frac{d}{4}\right)\chi_-+\delta\left(z+\frac{d}{4}\right)\chi_-\right]\phi(\bm{q};z,z').
\end{array}
\end{equation}
Fixing $z'=0$, one can now solve Eq.~(\ref{Eq:PoissonMomentumv2}) for $\phi(\bm{q};z,0)$ with the boundary conditions that $\phi(\bm{q};z,0)$ is continuous and its derivative to be piecewise continuous with jumps of $4\pi[q^2\chi_+\phi(\bm{q};0,0) - e/A]$ at $z=z'=0$ and of $4\pi q^2\chi_-\phi(\bm{q};\pm d/4,0)$ at $z=\pm d/4$.

Then, $\phi(\bm{q};0,0)$ yields the Coulomb interaction given by Eq.~(\ref{Eq:BareCoulomb}), where the static dielectric function is now given by
\begin{equation}\label{Eq:DiFv2}
\epsilon_{3\chi}(q)=\frac{1}{2}\left[\frac{N_t(q)}{D_t(q)}+\frac{N_b(q)}{D_b(q)}\right]
\end{equation}
with
\begin{equation}\label{Eq:DiFv2def}
\begin{array}{l}
D_j(q)=1+q\ell_- -q\ell_-\left(1+\scriptstyle{\mathcal{P}}_j\right)\e^{-\frac{qd}{2}} - \left(1-q\ell_-\right)\scriptstyle{\mathcal{P}}_j\e^{-qd},\\
N_j(q)=\left(1+q\ell_-\right)\left(1+q\ell_+\right) + \left[\left(1-\scriptstyle{\mathcal{P}}_j\right)-\left(1+\scriptstyle{\mathcal{P}}_j\right)q\ell_+\right]q\ell_-\e^{-\frac{qd}{2}}\\
\quad\quad\quad\quad + \left(1-q\ell_-\right)\left(1-q\ell_+\right)\scriptstyle{\mathcal{P}}_j\e^{-qd},
\end{array}
\end{equation}
where $\ell_\pm=2\pi\chi_\pm$ and $\scriptstyle{\mathcal{P}}$$_j=(\epsilon_j-1)/(\epsilon_j+1)$ for $j=b/t$. In the strict 2D limit ($d=0$), Eq.~(\ref{Eq:DiFv2}) reduces to Eq.~(\ref{Eq:epsgeo}) where the polarizability parameter follows $r_0=\ell_++2\ell_-$. The use of the dielectric function in Eq.~(\ref{Eq:DiFv2}) is more accurate when describing few body systems such as trions \cite{VanTuan_PRB18}. For neutral excitons, however, the use of any of the other forms for the dielectric function is also acceptable since the main correction introduced by Eq.~(\ref{Eq:DiFv2}) is for the short-range part of the potential \cite{VanTuan_PRB18}. 

The results in the following parts use Eqs.~(\ref{Eq:DiFv2})-(\ref{Eq:DiFv2def}) for the non-local dielectric function. We use the parameters in Ref.~\cite{VanTuan_PRB18}, where $d=0.6$~nm for all ML-TMDs,  $\ell_{\pm}=5.4d$ for ML-WS$_2$, $\ell_{\pm}=5.9d$ for ML-WSe$_2$, and $\ell_{\pm}=7.1d$ for ML-MoSe$_2$. We simulate three configurations in which the ML is encapsulated in hBN ($\epsilon_{b}=\epsilon_{t}=3.8$ leading to $\scriptstyle{\mathcal{P}}$$_b=\,\scriptstyle{\mathcal{P}}$$_t=14/19$), supported on SiO$_2$ ($\epsilon_{b}=2.1$ and $\epsilon_{t}=1$ leading to $\scriptstyle{\mathcal{P}}$$_b=11/31$ and $\scriptstyle{\mathcal{P}}$$_t=0$), or suspended in air ($\epsilon_{b}=\epsilon_{t}=1$ leading to $\scriptstyle{\mathcal{P}}$$_b=\,\scriptstyle{\mathcal{P}}$$_t=0$). The chosen values of all parameters are explained in Ref.~\cite{VanTuan_PRB18}. Next, we discuss how the bare Coulomb interaction changes by the presence of other charge carriers.


\section{Dynamical screening in monolayer transition-metal dichalcogenides}\label{Sec:RPA}

In the presence of a background of charge carriers (for example, induced by a gate voltage), the bare interaction is replaced with the dynamically-screened potential \cite{HaugKoch1994,Haug1984:PQE}
\begin{equation}\label{Eq:potdyn}
W(\bm{q},\omega)= \frac{2\pi e^2}{A\epsilon_d(q)q} \cdot \frac{1}{\epsilon(\bm{q},\omega)},
\end{equation}
where $\omega$ is the angular frequency and $\epsilon(\bm{q},\omega)$ is dynamical dielectric function. Note that $\epsilon_d(q)$ is not related and should not be confused with the static limit $\omega \rightarrow 0$ of the dynamical dielectric function. The role of the non-local dielectric function, $\epsilon_d(q)$, is to capture the $q$-dependence of the effective dielectric constant due to material parameters of the ML and its surrounding \cite{Latini_PRB15,Qiu_PRB16}. The dynamical dielectric function, on the other hand, describes the response of the delocalized electrons (or holes) in the ML to a test charge, and in the limit of zero charge density we get $\epsilon(\bm{q},\omega) \rightarrow 1$.

We discuss the forms of the dynamical dielectric function in the long-wavelength and shortwave limits. The former limit corresponds to intravalley charge excitations whose wavelength is much longer than the lattice constant, governed by electrons that cross the Fermi surface with small crystal momentum transfer and thus occur within the same valley ($qa\ll1$). The shortwave limit, on the other hand, corresponds to intervalley charge excitations  through which electrons  are scattered between a populated state in one valley and an empty state in another. Hereafter, we assign $\bm{q} = \bm{K}_0 + \bar{\mathbf{q}}$ in Eq.~(\ref{eq:eta_RPA}) where $\bm{K}_0$ is the wavevector that connects the valley centers, implying that $K_0=0$ in the long-wavelength limit, while $K_0 \gg \bar{q}$  in the shortwave limit. Using the random-phase approximation (RPA) \cite{Bohm_PR51,Pines_PR52,Bohm_PR53,Pines_PR53}, the dynamical dielectric function can be expressed as \cite{VanTuan_arXiv19}
\begin{eqnarray}
\epsilon(\bm{q},\omega) \!=\!  1 -  \frac{2\pi e^2}{A\epsilon_d(q)q} \cdot \frac{\chi(\bar{\mathbf{q}},\omega)}{\eta}   \,\,, \,\, \label{eq:eta_RPA} 
\end{eqnarray}
where $\chi(\bar{\mathbf{q}},\omega)$ is the density response function and $\eta$ is a scalar that lumps together local-field effects (umklapp processes) \cite{Adler1962:PR,Wiser1963:PR}. Starting with the long-wavelength limit, the density response function has the standard RPA form
\begin{eqnarray}
\chi(\mathbf{q},\omega) =  2  \sum_{\mathbf{k}} \!\frac{ f(\varepsilon_{\mathbf{k}})-f(\varepsilon_{\mathbf{k}+{\mathbf{q}}})}{ \hbar\omega -   \varepsilon_{\mathbf{k}+ {\mathbf{q}}} +  \varepsilon_{\mathbf{k}} + i\delta }  \,.\,\,\,\,\, \label{eq:chi_l}
\end{eqnarray}
The pre-factor 2 corresponds to spin-conserving transitions between spin-up or spin-down states. The $f$-terms in the numerator are Fermi-Dirac distribution functions. Assuming parabolic energy dispersion relation, $\varepsilon_{\bm{k}}=\hbar^2k^2/2m_e$, the only dependence of the density response  function on the underlying band structure is through the effective mass, $m_e$. 

\begin{figure}
\centering
\includegraphics*[width=0.95\textwidth]{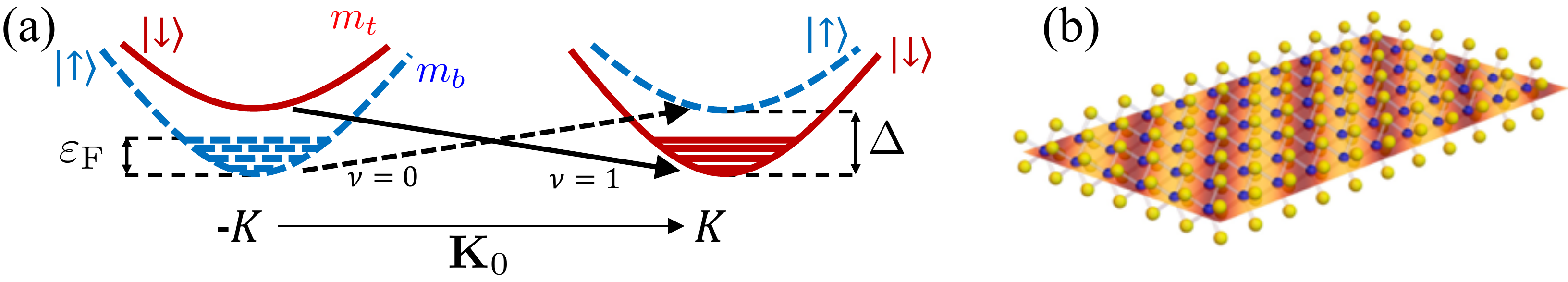}
\caption{The intervalley Coulomb interaction in ML-TMDs. (a) Spin-conserving charge excitations from the $-K$ to the $K$ valleys. The excitation energy is finite because of the spin-splitting energy gap, indicated by $\Delta$. (b) The resulting shortwave charge fluctuations in the monolayer.}\label{fig:InterValley}
\end{figure}

The shortwave limit in ML-TMDs is modeled as  a two-valley system, as shown in Fig.~\ref{fig:InterValley}.  The valleys are centered around distinct time-reversed points in the Brillouin zone (BZ), marked by $\bm{K}$ and $-\bm{K}$. The amplitude of the wavevector that connects the valley centers is ${K}_0=4\pi/3a$, where $a \simeq 3.2~\AA$ is the lattice constant (distance between neighboring transition-metal atoms). Each of the valleys is spin-split, where $\Delta$ is the spin splitting energy at the valley center. We assume parabolic energy dispersion for electronic states, $\varepsilon_{b,\bm{k}} = \hbar^2 k^2 / 2m_{b}$ and $\varepsilon_{t,\bm{k}} = \hbar^2 k^2 / 2m_{t}$,  where $m_b$ and $m_t$ are the effective masses in the bottom and top valleys, respectively. The RPA expression of the density response function in this two-valley system reads \cite{VanTuan_arXiv19}
\begin{eqnarray}
\chi(\bar{\mathbf{q}},\omega) =  \sum_{\mathbf{k},\nu} \!\frac{ f(\varepsilon_{b,\mathbf{k}})-f(\varepsilon_{t,\mathbf{k}+\bar{\mathbf{q}}}\!+\!\Delta)}{ (-1)^{\nu}(\hbar\omega + i\delta) - \!(\Delta  + \varepsilon_{t,\mathbf{k}\!+\!\bar{\mathbf{q}}} - \varepsilon_{b,\mathbf{k}}) }  \,,\,\,\,\,\, \label{eq:chi_s}
\end{eqnarray}
$\nu=\{0,1\}$ are the two spin configurations that contribute to intervalley excitations (see arrows in Fig.~\ref{fig:InterValley}).

Next, we discuss the local-field effect parameter $\eta$ in Eq.~(\ref{eq:eta_RPA}). Its general form follows \cite{VanTuan_arXiv19}
\begin{eqnarray}
\frac{1}{\eta} =   \sum_{\mathbf{G}}  \frac{ V_{\mathbf{q}+\mathbf{G}} }{  V_{\mathbf{q}}  }   |\mathcal{F}(\mathbf{q}+\bm{G})|^2 \,,\,\,\,\,\, \label{eq:eta}
\end{eqnarray}
where the sum runs over reciprocal lattice vectors ($\bm{G}$), and 
\begin{eqnarray}
\mathcal{F}(\bm{q}+\bm{G}) =  |\langle \mathbf{K}_f | e^{i(\mathbf{q} + \bm{G})\mathbf{r}} | \mathbf{K}_i \rangle|^2. \,\,\,\,\, \label{eq:Fq}
\end{eqnarray}
$| \mathbf{K}_i\rangle$ and $|\mathbf{K}_f\rangle$ are the valley-center states. When dealing with intravalley excitations, $\mathbf{K}_i = \mathbf{K}_f = \pm \bm{K}$, we get $\eta \sim 1$ because the ratio $V_{\mathbf{q}+\mathbf{G}} /  V_{\mathbf{q}}$ in Eq.~(\ref{eq:eta}) is negligible when $\bm{G} \neq 0$. That is, local-field effects do not play a significant role when the excitation wavelength is much longer than the lattice constant ($qa\ll1$).  

The behavior changes in the shortwave limit for which $\bm{q} \sim \bm{K}_0$, $\mathbf{K}_i = -\bm{K}$ and $\mathbf{K}_f = \bm{K}$. In this limit, the ratio $V_{\mathbf{q}+\mathbf{G}} /  V_{\mathbf{q}}$ in Eq.~(\ref{eq:eta}) is no longer negligible because $K_0$ is comparable to $| \mathbf{K}_0+\mathbf{G} |$ for the first few umklapp processes (when the amplitude of $G$ is comparable to that of the reciprocal lattice basis vectors). As shown in Ref.~\cite{VanTuan_arXiv19}, $\eta \sim 0.2$ in electron-doped ML-TMDs and $\eta \sim 0.47$ in hole-doped ML-TMDs. The reason for the difference stems from the orbital composition of states in the conduction and valence bands. The orbital $d_{z^2}$ of the transition-metal atom is dominant in the bottom of the conduction band, while the orbital $d_{(x \pm iy)^2}$ is dominant in the top of the valence band.

Finally, we discuss the form of the non-local dielectric function in Eq.~(\ref{eq:eta_RPA}).  In the long-wavelength limit,  $\epsilon_d(q)$ can be described by Eqs.~(\ref{Eq:InvDiFv1}), (\ref{Eq:epsgeo}), or (\ref{Eq:DiFv2}) since all of these forms converge to the same value when $q \rightarrow 0$. In the shortwave limit, we approximate the 
non-local dielectric function by 
\begin{eqnarray}
\epsilon_d(|\mathbf{K}_0+\mathbf{G}|) \simeq    1 + (\epsilon_d(K_0) - 1) \left(\frac{K_0}{|\mathbf{K}_0+\mathbf{G}|} \right)^P .   \,\,\,\,\, \label{eq:eps_ratio}
\end{eqnarray}
The power-law parameter $P$ denotes how fast the non-local dielectric function decays to unity because of the vanishing induced non-local potential when $G \rightarrow \infty$. The form in Eq.~(\ref{eq:eps_ratio}) is extracted from the results of DFT calculations in ML-MoS$_2$ \cite{Latini_PRB15,Qiu_PRB16}, where $\epsilon_d(K_0 \simeq1.3\, \AA^{-1}) \sim 2.5 $ and the power-law is about quadratic ($P \approx 2$). Note that unlike the long-wavelength case,  the materials below and above the ML do not affect  the value of $\epsilon_d(K_0)$  because of the shortwave nature of $K_0$ (e.g., $K_0d \sim 7.8$ where $d \sim 0.6$~nm is the thickness of the ML). 

\section{Plasmons in monolayer transition-metal dichalcogenides}\label{Sec:Plasmons}

Plasmons, collective charge-density oscillations, can arise as the electromagnetic response in solid-state systems because of the Coulomb interaction between electrons \cite{Bohm_PR51,Pines_PR52,Bohm_PR53,Pines_PR53,Pines_RMP56,Ferrell_PR56,Ritchie_PR57,Nozieres_PR58,Bozhevolnyi_Nat06,Tudorovskiy_PRB10,Berini_NatPhys12,Merano_OL16}. The energy dispersion of plasmons is found by looking for values $\omega$ as a function of ${\mathbf{q}}$ for which the dynamically-screened potential diverges (i.e., $\epsilon(\bm{q},\omega(\bm{q}))=0$).
 
\subsection{Intravalley plasmons}
Long-wavelength/intravalley plasmons are studied through Eq.~(\ref{eq:eta_RPA}) with $\eta=1$ and using Eq.~(\ref{eq:chi_l}) for the density response function. Focusing on the zero-temperature case, the sum over $\bm{k}$ in Eq.~(\ref{eq:chi_l}) can be handled analytically because the Fermi-Dirac distributions become step functions. The condition $\epsilon(\bar{\bm{q}},\omega({\bar{q}}))=0$ yields the well-known energy dispersion relation of 2D plasmons in the long-wavelength limit \cite{HaugKoch1994},
\begin{eqnarray}\label{Eq:lwplasmon}
\hbar\omega_{\ell}(q)=\sqrt{\frac{2e^2\varepsilon_F q}{\epsilon_d(q)}}\,\,,
\end{eqnarray}
where $\varepsilon_F$ is the Fermi energy. The energy dispersion relation of intravalley plasmons is gapless, $\hbar\omega_{\ell}(q \rightarrow 0) = 0$, which is different from that of a 3D system wherein the plasmon energy is finite when $q \rightarrow 0$. This difference can be understood by considering the nature of the charge excitation. A charge-wave profile in a 2D (3D) electron gas can be viewed as parallel wires (planes) with alternating charges ($...+-+-+-...$). The restoring force between charged wires depend on their distance, while the restoring force between infinite planes does not. This difference leads to the $q$-dependent ($q$-independent) plasmon frequency in a 2D (3D) system. 

In the context of optical transitions in semiconductors, long-wavelength plasmons screen the electron-hole attraction and reduce the band-gap energy by assisting electrons (or holes) of similar spins to avoid each other. These effects will be further explained in Sec.~\ref{Sec:BGR}. 

\subsection{Intervalley plasmons}\label{sec:intervally}

Whereas the intravalley plasmons introduced above and their effects on optical properties of ML-TMDs are conceptually similar to other conventional semiconductors \cite{Haug1984:PQE}, the additional valley degree of freedom gives rise to a second species of plasmons, shortwave/intervalley plasmons \cite{Dery_PRB16}. These plasmons originate from the short-range Coulomb potential through which electrons transition between valleys. 

\begin{figure}
\centering
\includegraphics*[width=0.95\textwidth]{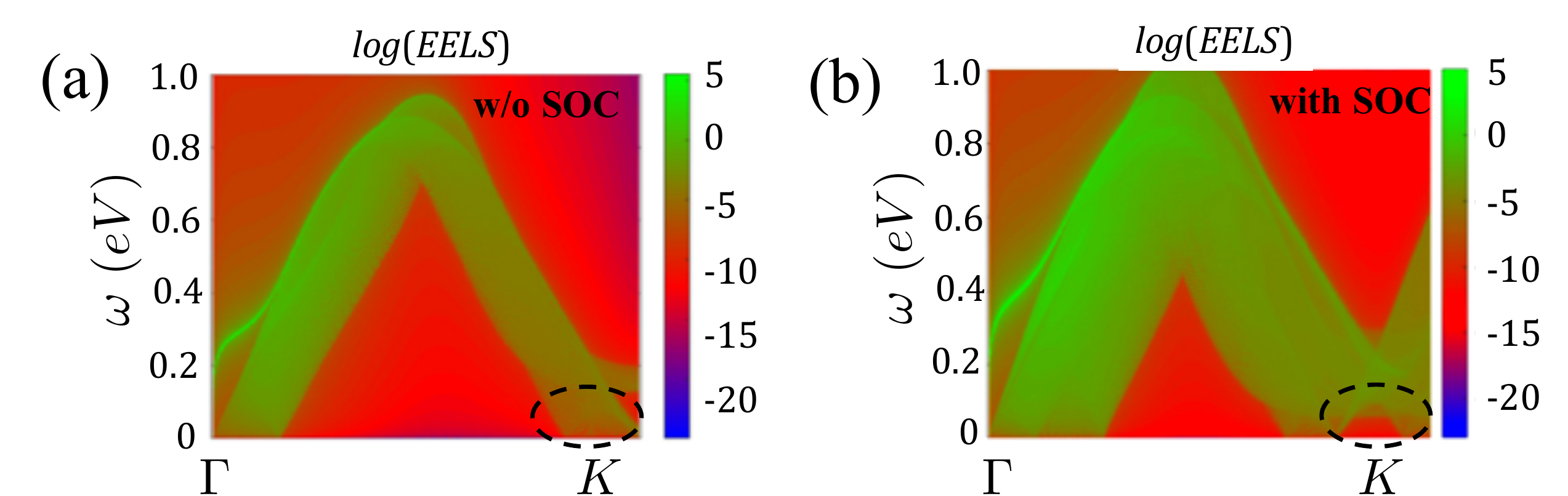}
\caption{The intervalley Coulomb interaction in ML-TMDs. (a) and (b) Model calculations of the electron energy loss spectroscopy (EELS) spectra for hole-doped MoS$_2$ without and with SOC, respectively, taken from Ref.~\cite{Groenewald2016:PRB}.}\label{fig:EELS}
\end{figure}

Figures~\ref{fig:EELS} shows \textit{ab initio}-based model calculations of the electron energy loss spectroscopy (EELS) spectrum, measuring the imaginary part of the inverse dielectric function, for hole-doped MoS$_2$, without and with SOC~\cite{Groenewald2016:PRB}. Here, the momenta are along the $\bm{\Gamma}-\bm{K}$ path in the Brillouin zone and the pronounced maxima in the EELS spectrum correspond to plasmon modes. Due to the hole doping, the mechanism described above now involves the valence band instead of the conduction band, but remains otherwise the same: For finite SOC, one can clearly distinguish between the gapless long-wavelength intravalley mode at $\bm{\Gamma}$ and the gapped shortwave intervalley mode at $\bm{K}$, corresponding to the spin-splitting energy of the valence band. Note that in the absence of SOC, the intervalley mode at $\bm{K}$ remains gapless. In contrast to the long-wavelength plasmons, shortwave plasmons can give rise to optical signatures absent in conventional semiconductors as will be discussed in Sec.~\ref{sec:exciton_intervalley}.

A compact way to model the energy dispersion relation of intervalley plasmons is to assume parabolic energy bands, and use Eq.~(\ref{eq:eta_RPA}) for the dynamical dielectric function and Eq.~(\ref{eq:chi_s}) for the density response function. One can then find that intervalley plasmons play an important role in ML-TMDs when the parameter 
\begin{eqnarray}
\alpha_0 = \frac{1}{\eta K_0} \frac{m_{b} e^2}{  \epsilon_d(K_0)\hbar^2}\, \label{eq:alpha0}
\end{eqnarray}
is not much smaller than unity \cite{VanTuan_arXiv19}. As we will show, the reason is that intervalley plasmons can propagate without damping in the range $\bar{q} \leq q_{\mathrm{max}}$, where $q_{\mathrm{max}}$ is commensurate with both $\alpha_0$ and the charge density. The damping-free propagation range is defined by the solution of $\epsilon(\bar{\bm{q}},\omega({\bar{q}}))=0$ for which $\omega({\bar{q}})$ is a real-value plasmon frequency. 

To derive the dispersion of intervalley plasmons and to find their damping-free propagation range, we focus on the zero-temperature case, in which the sum over $\bm{k}$ in Eq.~(\ref{eq:chi_s}) can be handled straightforwardly (the Fermi-Dirac distributions become step functions).  The solution for  $\epsilon(\bar{\bm{q}},\omega({\bar{q}}))=0$ yields compact analytical results in two cases \cite{VanTuan_arXiv19}. The first one is for the mode $\bar{q}=0$, corresponding to intervalley transitions with wavenumber $K_0$, 
\begin{eqnarray}
\hbar \omega_\mathrm{s}(\bar{q}\!=\!0)  = \!\sqrt{ \Delta^2 +  \frac{2(\gamma \!-\! c_0) \beta}{\gamma - 1}\Delta \varepsilon_F   +  \frac{(\gamma \!-\! c_0^2)  \beta^2}{\gamma-1} \varepsilon_F^2}\,,\,\,\,\,\,\label{eq:plasmon_energy_0}
  \end{eqnarray}
where $\beta = m_{b}/m_{t}-1$ is the valley mass asymmetry, $\gamma = \exp(\beta/\alpha_0)$, and $c_0 = 0$ when $\varepsilon_F < \Delta$ or $c_0= (\varepsilon_F - \Delta)/(\beta+1)\varepsilon_F$ when $\varepsilon_F > \Delta$. In mass symmetric valleys where $m_b=m_t$ leads to $\beta = 0$ and $\gamma = 1$, we get that $\beta/(\gamma-1) \rightarrow \alpha_0$,  and accordingly $\hbar \omega_\mathrm{s}(\bar{q}\!=\!0) \rightarrow \sqrt{\Delta^2 +  2(1 - c_0) \alpha_0 \Delta \varepsilon_F}$.  

The second case where we can find a compact analytical result is when $\varepsilon_F \ll \Delta$.  In this regime, the contribution from the term $\nu=1$ in Eq.~(\ref{eq:chi_s})  can be neglected because it is far off the plasmon resonance. The energy dispersion relation of intervalley plasmons in this case reads
\begin{eqnarray}
\hbar \omega_\mathrm{s}(\bar{\bm{q}}) \approx \Delta + \frac{\gamma\beta}{\gamma-1}\varepsilon_\mathrm{F} +\left( \frac{\gamma-1}{\beta} + \gamma \right)\varepsilon_{t,\bar{\bm{q}}}  \,, \,\,\,\,\,\,\,\,\,\,\, \label{eq:dispersion}
\end{eqnarray}
and the damping-free propagation range is limited to
\begin{eqnarray}
\bar{q} \,\,  \lesssim  \,\, \frac{\beta}{\gamma-1} \frac{k_F}{\beta+1}  \,\,,\label{eq:propagation} 
\end{eqnarray}
where $k_F$ is the Fermi wavenumber. In mass symmetric valleys, where $\beta = 0$ and $\gamma = 1$, we get that $\bar{q} \lesssim \alpha_0 k_F$. The validity of Eqs.~(\ref{eq:dispersion})-(\ref{eq:propagation}) degrades as $\varepsilon_\mathrm{F}$ continues to grow  because the effect of the spin-splitting energy is slowly washed-out. The main changes are somewhat a smaller plasmon frequency and a narrower damping-free propagation range \cite{VanTuan_arXiv19}. 

\subsection{Single-plasmon pole (SPP) approximation} \label{sec:spp}
The SPP approximation is a compact way to replace the relatively cumbersome RPA excitation spectrum by a single collective mode, $\omega_{\bm{q}}$.  As will be shown in the next section, the SPP approximation allows one to obtain analytical results for the self-energies of electrons or holes through which the band-gap renormalization is calculated.

The dynamically screened potential in the SPP approximation reads, 
\begin{equation}
\widetilde{W}_i(\bm{q},\omega)= \frac{V_{\bm{q}}}{\tilde{\epsilon}_i(\bar{\bm{q}},\omega)} =  \frac{2\pi e^2}{A\epsilon_d(q)q} \left( 1 + \frac{r_{i}(\bar{\bm{q}})}{\omega^2-\omega^2_{i,\bar{\bm{q}}} } \right)  . \label{eq:spp}
\end{equation}  
The index $i=\ell$ or $i=s$ denotes whether we deal with the long-wavelength limit where $\bm{q}=\bar{\bm{q}}$ or the shortwave one where $\bm{q}=\bm{K_0}+\bar{\bm{q}}$. The plasmon-pole frequency is denoted by $\omega_{i,\bar{\bm{q}}}$, and $r_{i}(\bar{\bm{q}})$ is the residue that represents the weight carried by the summation over $\bm{k}$ in the density response function. The residue  can be found from the asymptotic behavior of the RPA dielectric function at high-frequencies,
   \begin{equation}
\mathrm{Re} \left\{ \epsilon(\bm{q},\omega \rightarrow \infty) \right\} =  1 -  \frac{2\pi e^2}{A\epsilon_d(q)q} \cdot \frac{\chi(\bar{\mathbf{q}},\omega \rightarrow \infty)}{\eta} =  1 - \frac{r_{i}(\bar{\bm{q}})}{\omega^2} \label{eq:cond_real}
  \, .  \,\,\,\,\,\, \label{eq:csumrule}
   \end{equation} 
Alternatively, employing the Kramers-Kronig relation, the residue can also be extracted from the conductivity sum-rule, 
   \begin{equation}
r_{i}(\bar{\bm{q}}) = 2   \int_0^\infty \!\! d \omega  \,\omega \mathrm{Im} \left\{ \epsilon( \mathbf{q}, \omega) \right\}  .  \,\,\,\,\,\, \label{eq:csumrule}
   \end{equation} 
Starting with the long-wavelength limit ($\bm{q}=\bar{\bm{q}}$), we assign $\eta=1$ and use Eq.~(\ref{eq:chi_l}) for the density response function. We then get that 
 \begin{eqnarray}
r_{\ell}(\bar{\bm{q}})&=& \omega_{\ell}^2(q) = \frac{2e^2\varepsilon_F q}{\hbar^2 \epsilon_d(q)}\,\,.
  \label{eq:rl}
  \end{eqnarray}
Repeating the analysis for the shortwave limit ($\bm{q}=\bm{K_0}+\bar{\bm{q}}$) by  using Eq.~(\ref{eq:chi_s}) for the density response function, we get that \cite{VanTuan_arXiv19}
 \begin{eqnarray}
r_{s}(\bar{\bm{q}})&=& \frac{2\alpha_0  \varepsilon_F}{\hbar^2} \left( (1-c_0)\Delta + \left( 1 + \frac{c_0}{1+\beta} \right) \varepsilon_{t,\bar{\bm{q}}}  + \frac{\beta(1-c_0^2)}{2} \varepsilon_F \right) .\,
  \label{eq:rs}
  \end{eqnarray}
where $c_0$ and $\beta$ were defined after Eq.~(\ref{eq:plasmon_energy_0}). Local field effects are lumped together in the parameter $ \alpha_0 \propto \eta^{-1}$.   

The single collective mode, $\omega_{i,\bar{\bm{q}}}$, is found from comparing the static limits of the RPA and SPP dielectric functions,
   \begin{equation}
  1 -  \frac{2\pi e^2}{A\epsilon_d(q)q} \cdot \frac{\chi(\bar{\mathbf{q}},\omega =0)}{\eta} =  1 + \frac{r_{i}(\bar{\bm{q}})}{\omega^2_{i,\bar{\bm{q}}} - r_{i}(\bar{\bm{q}})}  \,\,. \label{eq:compress_sum_rule}
   \end{equation} 
In the long-wavelength case ($\bm{q}=\bar{\bm{q}}$), a straightforward calculation yields \cite{Lundquist_PMK67,Overhauser_PRB71,Rice_NC74,Zimmermann_PSS76}
   \begin{eqnarray}
  \omega^2_{\ell,\bar{\bm{q}}}  &=&  \left( 1 + \frac{q}{\kappa(q)}\right)\omega_{\ell}^2(q) + C_\mathrm{eff}\varepsilon_{b,{\bm{q}}}^2 \,\,,\,\,\,\,\,\label{eq:w_q_l} 
   \end{eqnarray} 
where $C_\mathrm{eff}$ is a constant of the order of unity needed to compensate for the fact that the static approximation typically overestimates the screening effect. $\kappa(q)$ is the screening length \cite{Ando1982:RMP}, 
\begin{equation}\label{eq:kappa_RPA}
\kappa(q)={g_sg_v} \frac{e^2 m_\mathrm{b}}{\hbar^2\epsilon_d(q)} \left[1-\sqrt{1-\left( \frac{2k_\mathrm{F}}{q}\right)^2}\Theta(q-2k_\mathrm{F})\right]\,.\,\,\,\,\,\,
\end{equation}
$g_s=1$ and $g_v=2$ are the spin and valley degeneracies, respectively. $m_b$ is the effective mass of an electron in the conduction-band bottom valley. In hole-doped systems, this mass is replaced with that of the valence-band top valley.  Omitting the term in square brackets on the right-hand side of Eq.~(\ref{eq:kappa_RPA}) amounts to the Thomas-Fermi approximation of an ideal 2D system. However, this approximation overestimates the screening, leading to unphysical results where the screening length is independent of the charge density. The static RPA model resolves this problem,  where the screening length is suppressed when the wavenumber is larger than $2k_F$. 

Repeating the calculation of Eq.~(\ref{eq:compress_sum_rule}) in the shortwave limit ($\bm{q}=\bm{K_0}+\bar{\bm{q}}$), we get that \cite{VanTuan_arXiv19}
  \begin{eqnarray}
  \omega^2_{s,\bar{\bm{q}}}  =  r_{i}(\bar{\bm{q}}) \left[ 1 + \frac{|\beta|}{2\alpha_0}\left( \ln \frac{1 + |\beta|\mathcal{R}(\varepsilon_{t,\bar{\bm{q}}},(1+\beta)\varepsilon_{t,\bar{\bm{q}}},\varepsilon_F) }{1 + |\beta|\Theta(\varepsilon_F-\Delta)\mathcal{R}(-\varepsilon_{b,\bar{\bm{q}}},\varepsilon_{b,\bar{\bm{q}}},c_0\varepsilon_F) }\right)^{-1}\right]  ,\label{eq:wqs}
   \end{eqnarray} 
where
  \begin{eqnarray}
\mathcal{R}(\varepsilon_1,\varepsilon_2,\varepsilon_3) = \frac{\sqrt{(\Delta+\varepsilon_1+\beta\varepsilon_3)^2 - 4\varepsilon_2\varepsilon_3} - (\Delta+\varepsilon_1-|\beta|\varepsilon_3)}{(\Delta+\varepsilon_1)(|\beta|+\beta) - 2\varepsilon_2}   \,\,.\,\,\,\,\,\label{eq:R}
   \end{eqnarray} 
In mass symmetric or nearly symmetric systems ($\beta \rightarrow 0$ and $\varepsilon_{\bar{\bm{q}}} = \varepsilon_{b,\bar{\bm{q}}} = \varepsilon_{t,\bar{\bm{q}}} $), we can simplify Eq.~(\ref{eq:wqs}) when $\Delta \gg \varepsilon_F, \varepsilon_{\bar{\bm{q}}}$,  and write
 \begin{eqnarray}
 \!\!\!\!\!\!\!\!\! \hbar^2\omega^2_{s,\bar{\bm{q}}}  =  \hbar^2 \omega^2_s({\bar{q}=0}) + 2(1+c_0) \left( \alpha_0\varepsilon_F + \frac{\Delta}{(1-c_0)\Delta - (1+c_0)\varepsilon_{\bar{\bm{q}}} } \right)\varepsilon_{\bar{\bm{q}}}\,.\label{eq:wqs2}
   \end{eqnarray} 
$\hbar \omega_s({\bar{q}=0}) = \sqrt{\Delta^2 + 2(1-c_0)\alpha_0\Delta\varepsilon_F}$ is the original plasmon frequency  [see discussion after Eq.~(\ref{eq:plasmon_energy_0})]. Similar to the long-wavelength case \cite{Overhauser_PRB71}, the pole and plasmon frequencies coincide when $\bar{q}=0$ (or $\varepsilon_{\bar{\bm{q}}}=0$).

\section{Band-gap renormalization}\label{Sec:BGR}
The band-gap renormalization (BGR) describes how the single-particle band structure is renormalized by the background charge plasma.  The BGR has contributions from intravalley and intervalley charge excitations in the long-wavelength and shortwave regimes, respectively. 

Using the finite-temperature Green's function formalism \cite{Mahan_book}, we quantify the BGR through the self-energy under the $GW$-approximation
\begin{equation}\label{Eq:SEgen}
\Sigma_{i,j}(\bm{k},z)=-k_BT \sum\limits_{\bm{q},z'} G^0_{j'}(\bm{k}-\bar{\bm{q}},z') \widetilde{W}_i(\bm{q},z-z') \,.
\end{equation}
$G^0$ is the unperturbed propagator (Green's function) of the charged particle, and $\widetilde{W}$ is the dynamically-screened potential.  As before, the index $i=\ell$ or $i=s$ denotes whether we consider the potential in the long-wavelength limit where $\bm{q}=\bar{\bm{q}}$ or the shortwave limit where $\bm{q}=\bm{K_0}+\bar{\bm{q}}$. The index $j=b$ or $j=t$ denotes whether we evaluate the self-energy of states in the bottom or top valleys. When the self-energy is due to intravalley virtual transitions, $i=\ell$, we assign $j=j'$.  On the other hand, we assign $j=b$ and $j'=t$ or vice versa when  $i=s$ because  the self-energy of a state in the bottom (top) valley of the $K$-point is affected by intervalley virtual transitions to the top (bottom) valley of the $-K$ point (Fig.~\ref{fig:InterValley}). 

Other parameters in Eq.~(\ref{Eq:SEgen}) are $z$ and $z'$, which denote odd (Fermion) imaginary Matsubara energies, $(2\ell+1)\pi\i k_BT$, where $\ell$ is an integer and $k_BT$ is the thermal energy. The self-energy can be evaluated by making use of the following substitutions. The sum over $z'$ is transformed into a contour integration in the complex plane by using the identity 
\begin{equation}\label{eq:identity}
k_BT\sum_{z'} A(z') = \oint_C \frac{dz'}{2\pi i} \cdot \frac{A(z')}{\exp(z'/k_BT)+1}  \,.
\end{equation}
The unperturbed Green's function is
\begin{equation}\label{Eq:freeG}
G^0_{j'}(\bm{k}-\bar{\bm{q}},z')=\frac{1}{z'-\varepsilon_{j',\bm{k}-\bar{\bm{q}}} - \Delta \delta_{j',t} + \mu},
\end{equation} 
where $\mu = \mu_{0}+\Sigma_{i,j}(\bm{k}_F,\varepsilon_F-\mu_{0})$ is the renormalized chemical potential, and  $\mu_{0}$ is the temperature- and density-dependent chemical potential of the non-interacting electron or hole gas. Finally, we employ the SPP approximation [Eq.~(\ref{eq:spp})] for the screened potential and make use of its spectral representation,
\begin{equation}\label{eq:spectral_representation}
\widetilde{W}_i(\bm{q},z-z') = V_{\bm{q}}\left( 1 - \int_{-\infty}^{\infty} \frac{d\omega }{\pi} \mathrm{Im} \left\{ \frac{1 }{\tilde{\epsilon}_i(\bar{\bm{q}},\omega)} \right\} \cdot \frac{\hbar}{z-z' -\hbar\omega}\right)  \,.
\end{equation}

\subsection{BGR due to intravalley Coulomb excitations} \label{sec:bgr_intra}

Making use of Eqs.~(\ref{eq:identity})-(\ref{eq:spectral_representation}) and the residue theorem, we quantify the self-energy in Eq.~(\ref{Eq:SEgen}). In the long-wavelength limit ($\bm{q}=\bar{\bm{q}}$), we can break the resulting BGR into contributions from screened-exchange and Coulomb-hole energies \cite{Haug1984:PQE}.

The former is denoted by
\begin{equation}\label{eq:sx}
\!\!\!\!\!\!\!\!\!\!\!\!\!  \Sigma_{\ell,j}^{\mathrm{sx}}(\bm{k}, z \rightarrow \varepsilon_{j,\bm{k}}-\mu+ i\delta ) = - \sum_{\bm{q}} \widetilde{W}_{\ell}(\bm{q}, \varepsilon_{j,\bm{k}-\bm{q}}-\varepsilon_{j,\bm{k}}) f_j( \varepsilon_{j,\bm{k}-\bm{q}} + \Delta \delta_{j,t}),
\end{equation}
and the latter by
\begin{equation}\label{eq:ch}
\!\!\!\!\!\!\!\!\!\!\!\!\!\!\!\!\!\!\!\!\!\!\!\! \!\!\!\!\!\!\!\!\! \Sigma_{\ell,j}^{\mathrm{ch}}(\bm{k}, z \rightarrow \varepsilon_{j,\bm{k}}-\mu+ i\delta ) = \sum_{\bm{q}} \int_{-\infty}^{+\infty} \frac{d\omega}{\pi} \, \mathrm{Im}\left\{ \frac{1}{\tilde\epsilon_{\ell}(\bm{q}, \omega)}\right\} \cdot \frac{\hbar V_{\bm{q}} g(-\omega)}{\varepsilon_{j,\bm{k}-\bm{q}}\!-\!\varepsilon_{j,\bm{k}} \!-\! \hbar\omega \!+\! i\delta}\,,
\end{equation}
where $z$ is analytically continued from the imaginary (Matsubara) energy axis to the real one.  $f(x)$ and $g(x)$ are Fermi-Dirac and Bose-Einstein distributions, respectively. $\bm{k}$ and $\bm{q}$ are measured with respect to the valley center. 

Starting with the screened-exchange energy when $T \rightarrow 0$, we substitute Eq.~(\ref{eq:spp}) into Eq.~(\ref{eq:sx}) and get that
\begin{equation}\label{eq:sx_2}
\!\!\!\!\!\!\!\!\!\!\!\!\! \!\!\!\!\!\!\!\!\!\!\!\!\! \!\!\!\! \Sigma_{\ell,j}^{\mathrm{sx}}(\bm{k}, \varepsilon_{j,\bm{k}}-\mu+ i\delta ) = \frac{2\pi e^2}{A } \sum_{\bm{q}} \frac{\Theta(k_{F,j}-|\bm{k}-\bm{q}|)}{q \epsilon_d(q) } \! \left( 1 \!+\! \frac{\hbar^2r_{\ell}(\bm{q})}{ (\varepsilon_{j,\bm{k}-\bm{q}}-\varepsilon_{j,\bm{k}})^2 - \hbar^2\omega^2_{\ell,\bm{q}}} \right).
\end{equation}
$k_{F,j}$ is the Fermi wavenumber in the top ($j=t$) or bottom ($j=b$) valleys. The Fermi wavenumber  is nonzero only for populated valleys, where electron populated conduction-band valleys shift down in energy and hole populated valence-band valleys shift up.  Substituting Eqs.~(\ref{eq:rl}) and (\ref{eq:w_q_l}) for the residue and plasmon-pole frequency, and considering the limit $q\rightarrow 0$, we can neglect the recoil energy term $(\varepsilon_{j,\bm{k}-\bm{q}}-\varepsilon_{j,\bm{k}})^2$ in Eq.~(\ref{eq:sx_2}) and the last term in Eq.~(\ref{eq:w_q_l}) due to their quartic wavevector dependence. Converting the sum to an integral, we get that
\begin{equation}\label{Eq:sx_3}
\Sigma_{j,\mathrm{sx}} \equiv   \Sigma_{\ell,j}^{\mathrm{sx}}(\bm{k}, \varepsilon_{j,\bm{k}}-\mu + i\delta )   \approx - \frac{\pi \hbar^2 n_j}{2m_j} = - \frac{1}{2}\varepsilon_{F,j} \,.
\end{equation}
The $k$-dependence of the screened exchange energy is neglected because the Fermi wavenumber is typically much smaller than the screening length in 2D systems [Eq.~(\ref{eq:kappa_RPA})]. We can then assume a rigid energy shift, so that  $\Sigma_{j,\mathrm{sx}} \simeq - \varepsilon_{F,j}/2$ for all low-energy states in the $j$ valley. Unpopulated valleys for which $\varepsilon_{F,j}=0$ are not affected by the screened exchange. 


Next, we evaluate the Coulomb-hole energy due to long-wavelength plasma excitations [Eq.~(\ref{eq:ch})]. It is by far the dominant contribution to the BGR. The term Coulomb-hole refers to the lack of charge next to a charged particle due to Pauli exclusion principle, and should not be confused with valence-band holes. Using Dirac's identity theorem, 
\begin{eqnarray}
\int_a^b \frac{f(x)}{x+i0^+} = -i\pi f(0) + \mathrm{P}\int_a^b \frac{f(x)}{x}  \,\,,\,\,\,\,\,\,\, \,\,\,\,\,\,\,\label{eq:SP_identity}
\end{eqnarray}
the imaginary part of the screened-potential in Eq.~(\ref{eq:spp}) is  the sum of two delta functions at $\pm  \omega_{\ell,\bm{q}}$, and we get 
\begin{eqnarray}\label{eq:ch2}
\!\!\!\!\!\!\!\!\!\!\!\!\!  \Sigma_{\ell,j}^{\mathrm{ch}}(\bm{k}, \varepsilon_{j,\bm{k}}-\mu+ i\delta )&=& \frac{\pi  e^2 }{A} \sum_{\bm{q}} \frac{1}{q\epsilon_d(q)} \cdot \frac{\hbar r_{\ell}(\bm{q})}{\omega_{\ell,\bm{q}}} \times \nonumber \\
&\,& \left[ \frac{g(\hbar \omega_{\ell,\bm{q}})}{ \varepsilon_{j,\bm{k}}-\varepsilon_{j,\bm{k}-\bm{q}} +\hbar\omega_{\ell,\bm{q}}} -\frac{g(-\hbar \omega_{\ell,\bm{q}})}{ \varepsilon_{j,\bm{k}}-\varepsilon_{j,\bm{k}-\bm{q}} - \hbar\omega_{\ell,\bm{q}}} \right]. 
\end{eqnarray}
This result is further simplified at the band edge, $\Sigma_{j,\mathrm{ch}} \equiv \Sigma_{\ell,j}^{\mathrm{ch}}(\bm{k}=0,  \varepsilon_{j,\bm{k}=0}-\mu+ i\delta )$, 
\begin{eqnarray}\label{eq:ch3}
\!\!\!\!\!\!\!\!\!\!\!\!\!  \Sigma_{j,\mathrm{ch}} = - \frac{ e^2 }{2} \int_0^{q_c} dq \, \frac{1}{\epsilon_d(q)} \cdot \frac{\hbar r_{\ell}(\bm{q})}{\omega_{\ell,\bm{q}}}  \times \left[ \frac{1+g(\hbar \omega_{\ell,\bm{q}})}{ \hbar\omega_{\ell,\bm{q}} + \varepsilon_{j,\bm{q}}} - \frac{g(\hbar \omega_{\ell,\bm{q}})}{ \hbar\omega_{\ell,\bm{q}} - \varepsilon_{j,\bm{q}}} \right] \,\,,
\end{eqnarray}
where we have used the identity $g(-x) = -1 - g(x)$, and introduced an integration cutoff ($q_c$). The latter denotes the fact that plasmons whose energy is much larger than the Fermi energy experience Landau damping due to single-particle excitations. Unlike the screened-exchange contribution which depends on the Fermi-Dirac distribution, the Coulomb-hole energy is largely the same for populated and unpopulated valleys. Conduction bands shift down in energy while valence bands shift up, and the shift has similar magnitude in ML-TMDs. The shift of different valleys may be slightly different if their effective masses are not the same. Nonetheless, the difference is small because $\hbar\omega_{\ell,\bm{q}} \gg |\varepsilon_{j,\bm{k}-\bm{q}}-\varepsilon_{j,\bm{k}}|$ when $q, k\rightarrow 0$. As a result, the Coulomb-hole energy in Eq.~(\ref{eq:ch2}) is largely independent of $\bm{k}$, 
\begin{eqnarray}\label{eq:ch4}
\!\!\!\!\!\!\!\!\!\!\!\!\!  \!\!\!\!\!\!\!\!\!\!\!\!\!  \!\!\!\!\!\!   \Sigma_{\ell,j}^{\mathrm{ch}}(\bm{k}, \varepsilon_{j,\bm{k}}-\mu+ i\delta ) \approx \Sigma_{\mathrm{ch}} &\equiv&  \frac{1}{2} \sum_{\bm{q}} \left[ \widetilde{W}_\ell(\bm{q},0) -V_{\bm{q}} \right] = - \frac{ e^2 }{2} \int_0^{q_c} dq \, \frac{1}{\epsilon_d(q)} \cdot \frac{r_{\ell}(\bm{q})}{\omega_{\ell,\bm{q}}^2}  \nonumber \\ &=&  - \frac{ e^2 }{2} \int_0^{q_c} dq \, \frac{1}{\epsilon_d(q)} \cdot \left[ 1 + \frac{q}{\kappa(q)} + C_\mathrm{eff} \left( \frac{\varepsilon_{b,\bm{q}}}{\omega_{\ell}(q)}\right)^2  \right]^{-1} ,
\end{eqnarray}
where we have substituted Eqs.~(\ref{eq:rl}) and (\ref{eq:w_q_l}) for the residue and plasmon-pole frequency. Employing Eq.~(\ref{Eq:InvDiFv1}), (\ref{Eq:epsgeo}), or (\ref{Eq:DiFv2}) for the non-local dielectric function, $\epsilon_d(q)$, yields similar results  since all of these expressions are similar when $q \rightarrow 0$.

\begin{figure}[t!]
\centering
\includegraphics*[width=0.9\textwidth]{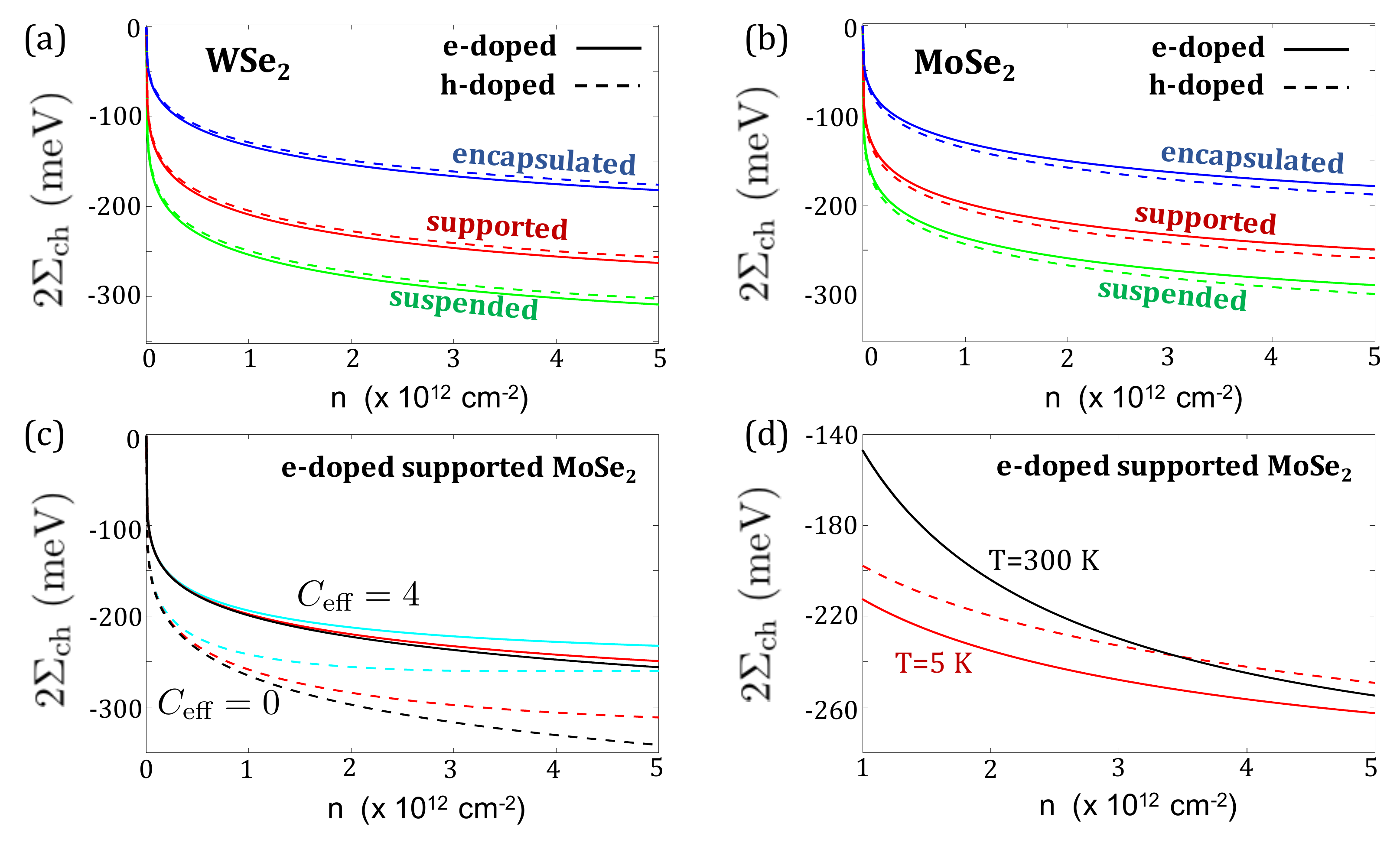}
\caption{The Coulomb-hole contribution to the band-gap renormalization. (a) and (b) show the results for ML-WSe$_2$ and ML-MoSe$_2$, respectively. Solid (dashed) lines correspond to electron- (hole-) doped samples. The results are shown for MLs encapsulated in hBN, supported on SiO$_2$, and suspended in air.  (c) The effect of the correction parameter and cutoff energies on the Coulomb-hole energy. The results are shown for ML-MoSe$_2$ supported on SiO$_2$. The solid (dashed) lines correspond to  $C_\mathrm{eff}=4$ ($C_\mathrm{eff}=0$) in the plasmon-pole frequency [Eq.~(\ref{eq:w_q_l})]. Each case shows three results that correspond to different cutoff energies in Eq.~(\ref{eq:ch4}): $\hbar^2q_c^2/2m_b=$0.05, 0.12 and 1~ eV from top to bottom (cyan, red, and black lines). (d) The effect of temperature in ML-MoSe$_2$ supported on SiO$_2$. The dashed line is calculated through Eq.~(\ref{eq:ch4}) and the solid lines through Eq.~(\ref{eq:ch3}). See text for details.}\label{fig:ch}
\end{figure}

Figure~\ref{fig:ch} shows the Coulomb-hole contribution to the band-gap renormalization, $2\Sigma_{\mathrm{ch}}$,  where the factor of 2  comes from the simultaneous energy downshift and upshift of the conduction and valence bands, respectively.  Figures~\ref{fig:ch}(a) and (b) show the results for ML-WSe$_2$ and MoSe$_2$, respectively. We have used $C_\mathrm{eff}=4$ and $\hbar^2q_c^2/2m_b=0.12$~eV in all of the calculations. In addition, we have employed Eq.~(\ref{Eq:DiFv2}) for the non-local dielectric function with the parameters values after Eq~(\ref{Eq:DiFv2def}). The effective mass parameters include the polaron effect (interaction of the charged particles with the lattice through the Fr\"{o}hlich interaction) \cite{VanTuan_PRB18}. Namely, $m_{b(t)}=(1+\delta_P)m_{b(t),0}$ where $m_{b(t),0}$ is the bare band-edge effective mass and $\delta_P$ is the polaron parameter.  The values of $m_{b(t),0}$  follow the DFT calculations in Ref.~\cite{Kormanyos_2DMater15}: the effective masses in the top and bottom valleys of the conduction band in ML-WSe$_2$ (ML-MoSe$_2$) are $0.29m_0$ and $0.4m_0$ ($0.58m_0$ and $0.5m_0$), respectively. The effective masses in the top and bottom valleys of the valence band in ML-WSe$_2$ (ML-MoSe$_2$) are $0.36m_0$ and $0.54m_0$ ($0.6m_0$ and $0.7m_0$), respectively. The polaron parameters are $\delta_P=0.17$ in ML-WSe$_2$ and $\delta_P=0.25$ in ML-MoSe$_2$ \cite{VanTuan_PRB18}.  The amplitude of $\delta_P$ is commensurate with the Born effective charge, which sets the amplitude of the Fr\"{o}hlich interaction in the long-wavelength limit \cite{Sohier_PRB16}.

The main trend in Figs.~\ref{fig:ch}(a) and (b) is that the BGR in ML-TMDs is significant when the charge density increases from zero to about 10$^{12}$~cm$^{-2}$, and it starts to saturate at larger densities. This behavior does not change qualitatively when we use different values for the integration cutoff  and correction to the plasmon-pole energy [$q_c$ and $C_\mathrm{eff}$ in Eq.~(\ref{eq:ch4})]. Figure~\ref{fig:ch}(c) shows the Coulomb-hole contribution to the BGR when choosing different values for these parameters in electron-doped MoSe$_2$ supported on SiO$_2$. The solid (dashed) lines correspond to $C_\mathrm{eff}=4$ ($C_\mathrm{eff}=0$), where the results are shown for three cutoff energies: $\hbar^2q_c^2/2m_b=0.05$, 0.12, and 1~eV.  The term associated with $C_\mathrm{eff}$ in Eq.~(\ref{eq:ch4}) corresponds to the correction one has to introduce at large energies. The BGR is overestimated when $C_\mathrm{eff}=0$, especially if the integration cutoff energy is large. 

Finally, Fig.~\ref{fig:ch}(d) shows the BGR at room and cryogenic temperatures in electron-doped MoSe$_2$ supported on SiO$_2$. Here, we have used Eq.~(\ref{eq:ch3}) to account for the temperature effect. The dashed line follows Eq.~(\ref{eq:ch4}),  corresponding to the zero-temperature and static-limit calculation that we have used in Figs~\ref{fig:ch}(a)-(c).  Note that Fig.~\ref{fig:ch}(d) shows the results for $n > 10^{12}$~cm$^{-2}$ because the Landau damping of plasmons is  not adequately described by the above formalism when $\varepsilon_F \ll k_BT$. The main effect of the temperature is to mitigate the BGR, rendering it more gradual. The BGR at room- and low-temperature  is similar when $\hbar\omega_{\ell,q} \gg k_BT$.

\subsection{BGR due to intervalley Coulomb excitations} \label{sec:bgr_inter}
 In the shortwave limit, $\bm{q}=\bm{K}_0+\bar{\bm{q}}$, we make use of the fact that $\bar{q}  \ll K_0$ and rewrite Eq.~(\ref{eq:spectral_representation}) after using Eq.~(\ref{eq:SP_identity}),  
 \begin{equation}\label{eq:spectral_representation_s}
\widetilde{W}_s(\bm{q},z-z') \simeq V_{\bm{K}_0} \left( 1 +  \frac{\hbar^2r_s(\bar{\bm{q}})}{ (z-z')^2 - \hbar^2 \omega_{s,\bar{q}}^2} \right)  \,,
\end{equation}
where $V_{\bm{K}_0} = 2\pi e^2/A\epsilon_d(K_0)K_0$. Making use of this SPP form, we break the self-energy in Eq.~(\ref{Eq:SEgen}) into contributions from exchange and correlation \cite{VanTuan_arXiv19}. The exchange contribution comes from the bare potential
\begin{equation}
\Sigma_{s,j}^{x} \simeq -k_BT \cdot V_{\bm{K}_0} \sum_{\bar{\mathbf{q}},z'} G_{j'}(\bm{k}-\bar{\bm{q}}, z' ), \label{eq:sigma_x1}
\end{equation} 
and the correlation contribution comes from its dynamical part,
\begin{eqnarray}
\Sigma_{s,j}^{c}(\bm{k},z) & \simeq & -k_BT \cdot V_{\bm{K}_0}  \sum_{\bar{\mathbf{q}},z'}   \frac{\hbar^2r_s(\bar{\bm{q}}) G_{j'} (\bm{k}-\bar{\mathbf{q}}, z' )}{(z-z')^2 - \hbar^2 \omega_{s,\bar{q}}^2} . \,\,\,\,\,\,\, \,\,\,\,\,\,\,\label{eq:sigma_c1}
\end{eqnarray} 
Recall that $j=b$ and $j'=t$ or vice versa in the shortwave limit because  the self-energy of a state in the bottom (top) valley is affected by intervalley virtual transitions to the top (bottom) valley, as shown in Fig.~\ref{fig:InterValley}. 

\begin{figure}[t!]
\centering
\includegraphics*[width=0.9\textwidth]{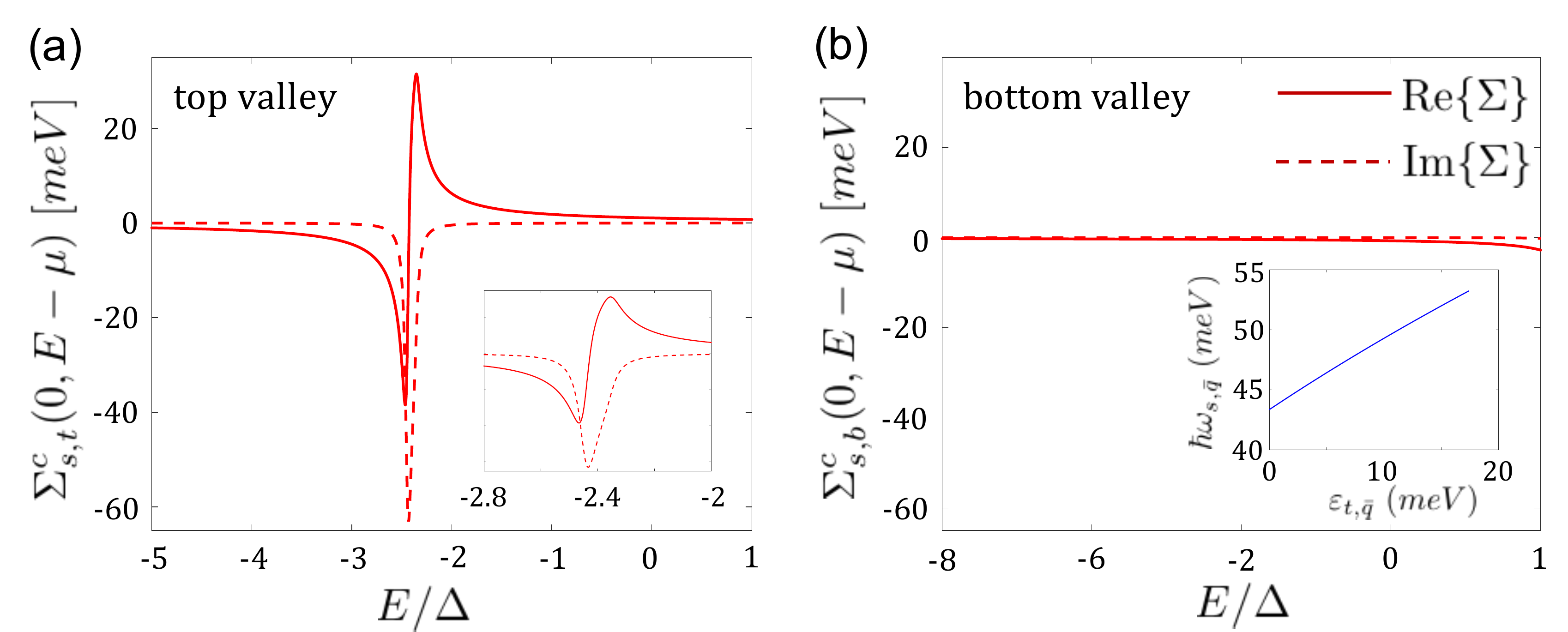}
 \caption{ Intervalley correlation self-energy at the band edge ($k=0$) of the top (a) and bottom (b) valleys. The inset in (a) is a zoom-in of the singular region. The inset in (b) shows the plasmon-pole energy in the damping-free propagation range. See text for the parameters. Taken from \cite{VanTuan_arXiv19}.} \label{fig:correlation}
\end{figure}

Replacing the sum over Matsubara frequencies with contour integration [Eq.~(\ref{eq:identity})], and the Green's function with Eq.~(\ref{Eq:freeG}), we get
\begin{eqnarray}
&&\!\!\!\!\!\!\!\!\!\!\!\!\! \Sigma_{s,j}^{x}  \simeq  -\eta\alpha_0\varepsilon_F \left(\delta_{i,t} + c_0 \delta_{i,b} \right) \,\,, \label{eq:sigma_x2}
\label{eq:sigma_x2} 
\end{eqnarray} 
and
\begin{eqnarray}
&&\!\!\!\!\!\!\!\!\!\!\!\!\!\!\!\!\!\!\!\!\!\!\!\!\!\!\!\!\!\!\!\!\!\!\! \Sigma_{s,j}^{c}(\bm{k},z\!-\!\mu) \!  \simeq  \!  \frac{ V_{\bm{K}_0} }{2} \! \sum_{\bar{\mathbf{q}}} \frac{\hbar r_{iv}(\bar{\bm{q}})}{  \omega_{\mathrm{s,\bar{\bm{q}}}}} \!  \left[ \frac{ f(\varepsilon_{j',\bf k-\bar{q}} \!+\! \Delta \!\cdot\! \delta_{j',t}) + g(\hbar \omega_{\mathrm{s,\bar{\bm{q}}}})}{z- \varepsilon_{j',\bf k-\bar{q}} - \Delta \cdot \delta_{j',t} +  \hbar \omega_{\mathrm{s,\bar{\bm{q}}}}} \,\,- \,\, ( \omega_{\mathrm{s,\bar{\bm{q}}}} \,\,\rightarrow \,\,- \omega_{\mathrm{s,\bar{\bm{q}}}}) \right]\!\!. \label{eq:sigma_c2} 
\end{eqnarray} 
In ML-TMDs, the exchange-driven  redshift of the top valley is about $1$~meV per electron density of $n \sim 10^{12}$ cm$^{-2}$ in the bottom valley. For the correlation term, the sum 
is limited to the damping-free propagation range ($q \leq q_{_\mathrm{max}}$). Figure~\ref{fig:correlation} shows the correlation contribution to the self-energy at the band edge ($k=0$) for a system with the following parameters: $T=10$~K, $\varepsilon_F=20$~meV, $\Delta=30$~meV, $\alpha_0=1.35$, $\eta=0.2$, and the conduction-band effective masses of WSe$_2$ (see discussion of Fig.~\ref{fig:ch}). The residue and plasmon-pole frequency, $r_{iv}(\bar{\bm{q}})$ and $\omega_{\mathrm{s},\bar{\bm{q}}}$, were calculated from Eqs.~(\ref{eq:rs}) and (\ref{eq:wqs}). In addition, we have used the thermal energy for broadening, $z = E + ik_BT$. The self-energy of an electron in the top valley, $\Sigma_{s,t}^{c}(\bm{k},E\!-\!\mu)$, includes resonance features below the continuum whose magnitude is about a few tens meV, as shown in Fig.~\ref{fig:correlation}(a). The singular region  lies at the interval $ -(\Delta - \varepsilon_{b,q_{_\mathrm{max}}} + \hbar \omega_{\mathrm{s,q_{\mathrm{max}}}}) < E < -(\Delta + \hbar \omega_\mathrm{s,\bar{q}=0})$, where $q_{_\mathrm{max}}$ is the largest possible value for damping-free plasmon propagation. The singularity arises from the Fermi-distribution term in the first expression of the square brackets in Eq.~(\ref{eq:sigma_c2}). The second term in square brackets does not lead to a resonance feature because $q_{_\mathrm{max}} < k_F$ in this case, and hence $f(\varepsilon_{b,\bar{q}} ) + g(-\hbar \omega_\mathrm{s}(\bar{\bm{q}})) \simeq 0$ for the entire integration range. 

\subsection{Total BGR} \label{sec:bgr_total}
Having all of the self-energy components and assuming a rigid energy shift, we evaluate the BGR from the continuum edge state, $k=0$ and $\varepsilon_{b(t),\bm{k}}=0$. The dominant contribution comes from the Coulomb-hole energy, whereas the correlation energy due to intervalley plasmons is not effective at the band edge. The latter is denoted by $E=0$ in Fig.~\ref{fig:correlation},  showing that the large  resonance is only effective in the singular energy interval below the continuum edge.  While this resonance does not lead to BGR, we will explain in Sec.~\ref{sec:exciton_intervalley} how this singularity is manifested in the optical spectrum.  

The charge-density-dependent energy difference between the edges of the bottommost valley in the conduction band and topmost valley in the valence band reads
\begin{equation}\label{eq:bgr_1}
E_{g,1} \approx E_{g,0} + 2\Sigma_{\mathrm{ch}} -  \left( \frac{1}{2}  + c_0\eta \alpha_0 \right) \varepsilon_F  \,,
\end{equation}
where $E_{g,0}$ is the fundamental band-gap energy at zero density. $E_{g,1}$ is relevant for direct-exciton optical transitions in Mo-based ML-TMDs, as well as for indirect and dark excitons in W-based ones (Fig.~\ref{fig:Scheme}). The energy difference between the top conduction-band valley and topmost valence-band valley reads
\begin{equation}\label{eq:bgr_2}
\!\!\!\!\!\!\!\!\!\!\!\!\!\!\!\!\!\!\!\!E_{g,2} \approx E_{g,0} + |\Delta_{c,0}| + 2\Sigma_{\mathrm{ch}} - \delta_v \left( \frac{1}{2}  + c_0\eta \alpha_0 \right) \varepsilon_F - (1-\delta_v)  \left( \frac{c_0}{2}  + \eta \alpha_0 \right) \varepsilon_F \,,
\end{equation}
where $\Delta_{c,0}$ is the spin-splitting energy in the conduction band due to spin-orbit coupling, and $\delta_v=1\,(0)$ for hole- (electron-) doped conditions. $E_{g,2}$ is relevant for direct-exciton optical transitions in W-based ML-TMDs, as well as for indirect and dark excitons in Mo-based ones (Fig.~\ref{fig:Scheme}). The energy splitting between the top and bottom valleys in the  conduction band can now be written as 
\begin{equation}\label{Eq:Delta_c}
\Delta_c \approx E_{g,2} - E_{g,1} =  |\Delta_{c,0}| + (1 - \delta_v)(1-c_0)\left( \frac{1}{2} - \eta \alpha_0 \right)\varepsilon_F \,.
\end{equation}
In hole-doped samples, we exchange the indices $c \leftrightarrow v$ and recall that the value of $\eta$ differs in these two doping cases \cite{VanTuan_arXiv19}.

The density-induced change in the spin-splitting energy may also be affected by the small difference between the Coulomb-hole energies of the top and bottom valleys because their effective masses differ. That is, one can employ Eq.~(\ref{eq:ch3}) instead of (\ref{eq:ch4}) to calculate the Coulomb hole term, and accordingly replacing $\Sigma_{\mathrm{ch}}$ with $\Sigma_{b,\mathrm{ch}}$ in Eq.~(\ref{eq:bgr_1}) and with $\Sigma_{t,\mathrm{ch}}$ in Eq.~(\ref{eq:bgr_2}). As a result, a small component, $2(\Sigma_{t,\mathrm{ch}}-\Sigma_{b,\mathrm{ch}})$, should also be added to Eq.~(\ref{Eq:Delta_c}).

\section{Excitons in monolayer transition-metal dichalcogenides}\label{sec:general}
One of the most attractive properties of ML-TMDs is the large binding energy of their excitons, stemming from the reduced dielectric screening when the ML is surrounded by low dielectric materials and from the relatively heavy masses of both electrons and holes (compared with the case of III-V semiconductors where the electron mass is small, $m_e \lesssim 0.1m_0$). When electrons or holes are added to the ML through electrostatic doping (gate voltage), we have two competing effects because of the dynamically-screened Coulomb interaction in the long-wavelength limit. On the one hand, we have seen the large signature of the Coulomb-hole energy on the self-energies of holes and electrons (Fig.~\ref{fig:ch}). The outcome of such effect would be to redshift the overall optical spectrum because the band-gap energy shrinks.  On the other hand, the attraction between the electron and hole is weaker due to screening, resulting in reduced binding energy of the exciton. The outcome of such effect would be to blueshift the neutral-exciton peak ($X^0$)  towards the continuum of free electron-hole pairs. The two effects almost completely compensate each other because long-wavelength charge excitations (intravalley plasmons) correspond to both screening and BGR. The overall outcome is nearly a fixed spectral position of $X^0$. 

In addition to the fact that the spectral position of $X^0$ is hardly affected by the background charge density,  there are two other reasons that render it difficult to resolve the contribution of long-wavelength plasmons in the exciton spectrum. First, the plasmon interaction with the electron component is offset by its interaction with the hole if $qa_X \ll 1$, where  $q$ is the plasmon wavenumber and $a_X$ is the exciton Bohr radius. Second, the gapless plasmon dispersion does not impart spectrally resolved peaks. As we will show later, these problems are not relevant when the exciton interacts with shortwave plasmons. Before providing a quantitative analysis of the interaction between excitons and plasmons, we first outline the general properties of excitons in the presence of charge carriers.  

\begin{figure}[t!]
\centering
\includegraphics*[width=0.7\textwidth]{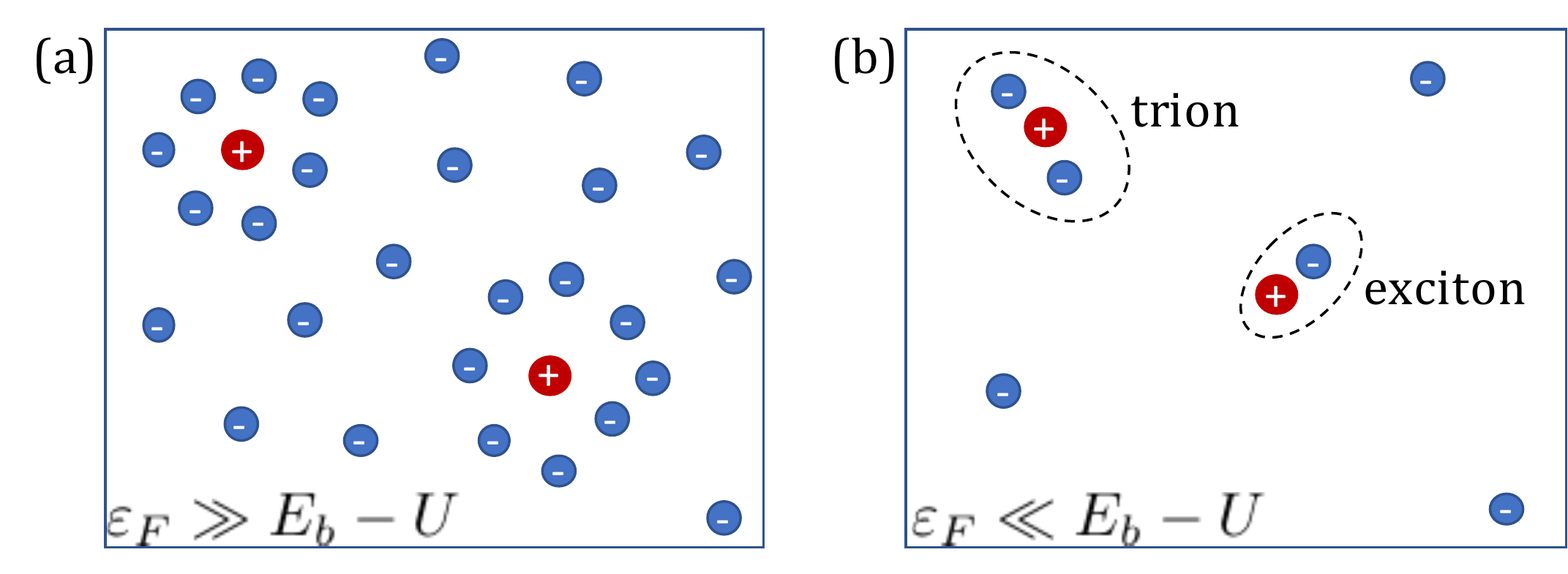}
 \caption{ Cartoon of a semiconductor following photoexcitation. (a) The high-density regime where the hole is surrounded by electrons. In this case, the Fermi energy exceeds the difference between the electron-hole binding energy and electron-electron repulsion energy $\varepsilon_F \gg E_b-U$. (b) The low-density regime where the ground state is the trion, the excited state is the exciton, and both reside below the continuum edge. The energy of the trion is $2E_b-U$, and its binding energy with respect to the exciton is $E_b-U$.} \label{fig:hole_in_a_sea}
\end{figure} 

\subsection{Excitons in doped semiconductors}

A photon absorption in a semiconductor creates an electron in the conduction band and a hole in the valence band. We assume an electron-doped semiconductor in the discussion below, bearing in mind that similar picture can be drawn for a hole-doped semiconductor. Assuming that our system contains $N$ delocalized electrons in the conduction band prior to the photon absorption, the post-excitation  state of the system involves correlation of the hole with $N+1$ electrons \cite{Mahan_PR67a,Hawrylak_PRB91}. Figure~\ref{fig:hole_in_a_sea}(a) shows a cartoon of the post-excitation state with two (uncoupled) holes  in the Fermi sea. This problem is analogous to the famous X-ray catastrophe in metals \cite{Mahan_PR67a,Hawrylak_PRB91,Mahan_PR67b,Nozieres_PR69,Schotte_PR69,Combescot_JdP71,Skolnick_PRL87,Kane_PRB94}, where semiconductors present the advantage that the carrier density can be tuned through a gate voltage.  When the average distance between electrons is much larger than the Bohr radii of electrons and holes, the ground state of the system is typically the trion (two-electrons and a hole), and the excited state is the neutral exciton (bound electron-hole pair). This behavior is shown in Fig.~\ref{fig:hole_in_a_sea}(b). As a result, the absorption spectrum shows two distinct peaks below the continuum.  The trion energy is lowered by the energy $2E_b-U$ with respect to the continuum where $E_b$ is the binding energy of the neutral electron-hole pair and $U$ is the Coulomb interaction energy between two electrons in the trion complex. The binding energies of configurations that include more than two electrons with the hole are smaller than those of trions and neutral excitons due to the strong Coulomb repulsion between electrons in such complexes. It is emphasized that experiments measure the binding energy of trions with respect to that of neutral excitons, which is $E_b-U$ in this example. However,  the total energy of a trion with respect to the continuum is $2E_b-U$, and the trion size is comparable to that of an exciton \cite{Donck_PRB17,VanTuan_PRB18}. 

\subsection{The absorption spectra of electrostatically-doped ML-TMDs} \label{sec:exciton_behavior}
The absorption process reveals information on intrinsic material and many-body properties, not subjected to energy relaxation and localization effects that often dominate the emission process \cite{Godde_PRB16,VanTuan_arXiv18b}. Focusing on the absorption process, the following behavior seem to be universal in ML-TMDs \cite{Wang_NanoLett17,Courtade_PRB17}:
\begin{enumerate}
\item Increasing the charge density through application of a gate voltage is initially accompanied by decay of the neutral exciton peak and the amplification of the trion peak. Figure~\ref{fig:blueshift_low} shows this behavior in electron and hole doped ML-TMDs. Notably, the strong decay of the neutral-exciton peak is not commensurate with the rise of the trion peak. 

\item The spectral position of the neutral-exciton peak does not shift during the decay when holes are added to the system [Fig.~\ref{fig:blueshift_low}(b) and (d)]. On the other hand, the position of this peak blueshifts when electrons are added to the system [Fig.~\ref{fig:blueshift_low}(a) and (c)]. This behavior can be seen by tracing the peak position of $X^0$ in Figure~\ref{fig:blueshift_low}, marked by the solid diamond symbols.

\begin{figure}[t!]
\centering
\includegraphics*[width=0.8\textwidth]{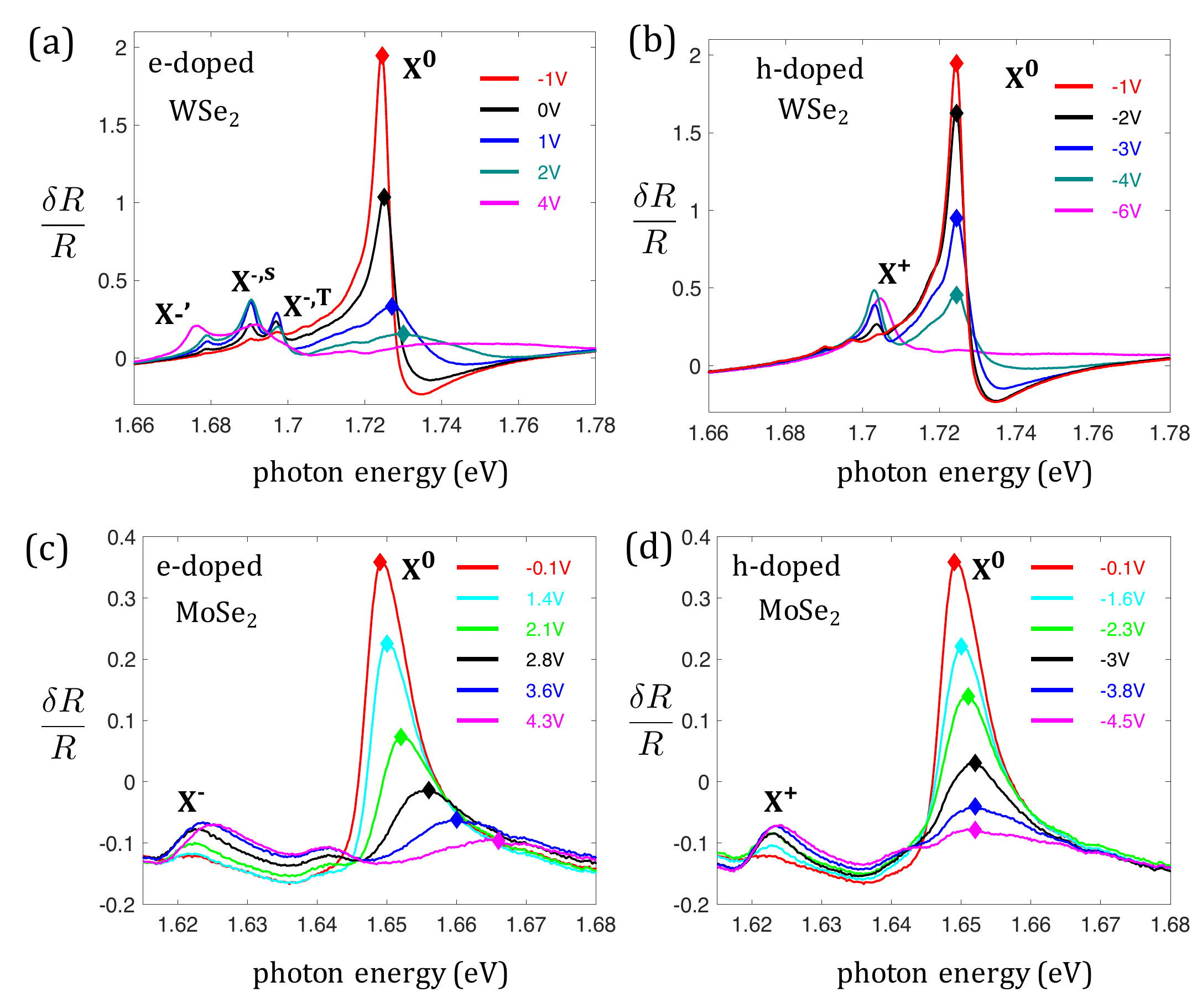}
 \caption{ The decay of the neutral exciton and rise of the trion in the low and moderate electrostatic doping. (a) and (b) show the reflectance contrast spectra ($\delta R/R$) at 4K in electron and hole-doped ML-WSe$_2$ encapsulated in hBN. (c) and (d) show the respective cases in encapsulated ML-MoSe$_2$. The diamond symbols show the peak position of $X^0$. Charge neutrality in ML-WSe$_2$ is achieved when the gate voltage is $\sim -1$~V because the sample is unintentionally electron-doped. Every increase (decrease) of 1~V amounts to adding an electron (hole) density of $\sim10^{12}$~cm$^{-2}$.  Taken from Ref.~\cite{VanTuan_arXiv18}.} \label{fig:blueshift_low}
\end{figure}

\item Continuing to increase the charge density eventually leads to disappearance of the neutral exciton peak. This behavior is shown in Figure~\ref{fig:blueshift_high} for ML-WSe$_2$, where noticeably, the trion peak starts to experience a significant blueshift, decay and broadening after the exciton peak disappears. Figure~\ref{fig:blueshift_low}, on the other hand, shows that when the neutral-exciton peak can still be observed in the spectrum the trion peaks increase in amplitude when the magnitude of the gate voltage increases. 

\item  An optical sideband, marked by $X-'$ in Fig.~\ref{fig:blueshift_high}, emerges in electron-doped WSe$_2$ (a similar behavior is expected in electron-doped WS$_2$). Whereas the trion peaks blueshift, broaden and decay at elevated charge densities (open circle symbols in Fig.~\ref{fig:blueshift_high}), the optical sideband redshifts and its magnitude increases until it saturates (solid hexagon symbols).
\end{enumerate}

\begin{figure}[t!]
\centering
\includegraphics*[width=0.7\textwidth]{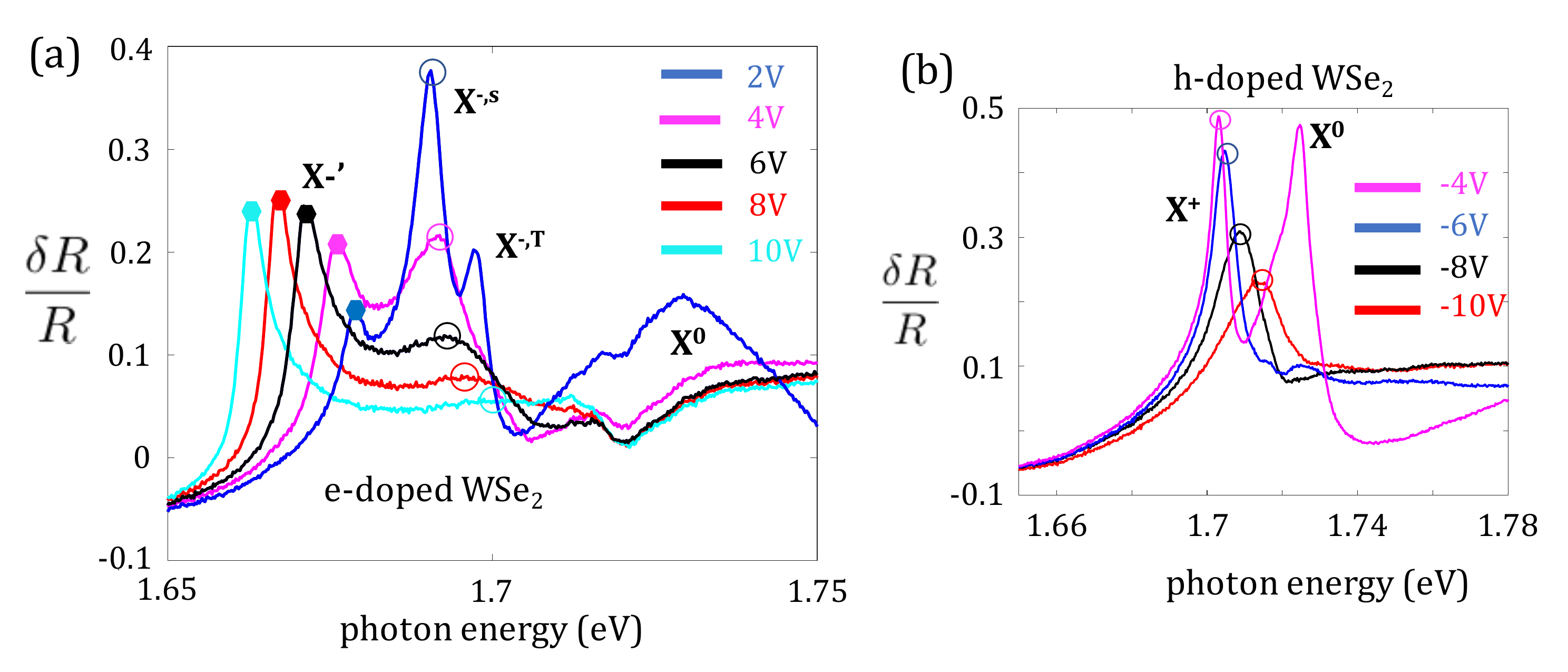}
 \caption{ The decay of the trion in large electrostatic doping. (a) and (b) show the reflectance contrast spectra ($\delta R/R$) at 4K in electron and hole-doped ML-WSe$_2$ encapsulated in hBN. Note that the $y$-scale amplitude is smaller than in Figs.~\ref{fig:blueshift_low}(a) and (b). The solid hexagon symbols in (a) trace the peak of $X-'$, while the open circle symbols in (a) and (b) trace the singlet trion peaks ($X^{-,s}$ and $X^+$). We are grateful to Jie Shan and Kin Fai Mak for providing  these results (from the raw data of Ref.~\cite{Wang_NanoLett17}). } \label{fig:blueshift_high}
\end{figure} 

We do not address the behavior of trions in a plasma in this review. However, we make use of what is known in conventional semiconductor quantum wells \cite{Suris_PSSB01,Ramon_PRB03,Kossacki_JPCM03,PortellaOberli_PRB04,BarJoseph_SST05,Keller_PRB05}, and suggest an explanation of their absorption behavior in ML-TMDs. Continuing the example of a hole in a Fermi sea of electrons, we believe that the empirical behavior in absorption-type experiments can be explained as a smooth transition of the ground-state between that of isolated three-body complexes (trions) at the low density regime and that of a hole in a Fermi sea of electrons at elevated densities (Fig.~\ref{fig:hole_in_a_sea}). As Hawrylak suggested, the absorption process in a 2D semiconductor is governed by trions and neutral excitons at low densities and by Fermi-edge singularities at elevated densities if the hole's mass is infinite \cite{Hawrylak_PRB91,Skolnick_PRL87}. On the other hand, when the hole's mass is comparable to that of the electrons in the Fermi sea (e.g., in ML-TMDs), one should expect a smooth transition between trion states and a `muted' Fermi-edge singularity  \cite{Hawrylak_PRB91}. This smooth transition can be thought of as a trion that manages to keep both its electrons when the surrounding electron density is small, as shown in Fig.~\ref{fig:hole_in_a_sea}(b), and that this physical picture gradually turns into the one of a hole in a sea of electrons when their density increases, as shown in Fig.~\ref{fig:hole_in_a_sea}(a). This gradual change is accompanied by a blueshift, decay and broadening of the trion peak, and it starts when the neutral exciton peak disappears from the spectrum (merges to the redshifting continuum), as shown in Fig.~\ref{fig:blueshift_high}. 

The above view is different from the one that considers the neutral and charged excitons in ML-TMDs as repulsive and attractive Fermi-polarons \cite{Sidler_NatPhys17,Efimkin_PRB17}. In the Fermi-polaron picture, the electron-hole pair does not break and it interacts with the Fermi sea around it through van-der-Waals-type interaction due to its charge neutrality. However, the fact that the exciton disappears from the absorption spectrum at densities around a few 10$^{12}$~cm$^{-2}$ means that the exciton size is no longer small compared with the average distance between electrons or holes in the surrounding plasma (the exciton radius in ML-TMDs  is $\sim$1~nm when the binding energy is a few hundreds meV \cite{Stier_PRL18}, and this picture is valid only when the BGR is negligible; i.e., when the free-charge density is negligible). Therefore, our view is that  the electron-hole pair picture breaks at large densities and the system is better described as a hole in a sea of electrons rather than  a neutral  exciton that interacts with the surrounding plasma. Further studies are needed to address this problem and see if the observed behavior of the trion peak can indeed be described  as a muted Fermi-edge singularity. 


 

\section{Excitons in the presence of long-wavelength charge excitations}\label{Sec:Excitons}

In the rest of this review, we focus on the behavior of neutral excitons in the presence of dynamic screening. In this section, we outline the formalism for calculating the exciton states with  the dynamically-screened potential in the long-wavelength limit. Three levels of approximations will be presented, where the well-studied quasistatic limit of the screened potential is employed first  \cite{HaugKoch1994}. The Shindo approximation is presented in the next step, where the dynamical part of the potential is kept but the calculation is simplified by approximating the form of the electron-hole pair Green's function \cite{Haug1984:PQE}. We then present a fully dynamical scheme  \cite{VanTuan_PRX17}, showing that even in this case, all that the long-wavelength charge excitations can explain is the BGR and eventual merging of the exciton into the redshifting continuum. Noticeably, these calculations reveal that long-wavelength charge excitations cannot explain two critical experimental observations in ML-TMDs. The first one is the enhanced blueshift of the neutral exciton peak in electron-doped but not in hole-doped ML-TMDs (Fig.~\ref{fig:blueshift_low}), and the second one is the emergence of the optical sideband in electron-doped  W-based compounds ($X-'$ in Figs.~\ref{fig:Gating} and \ref{fig:blueshift_high}). We will present a compact analytical model in Sec.~\ref{sec:exciton_intervalley} that shows how the coupling of neutral excitons to shortwave plasmons gives rise to both observed phenomena in ML-TMDs \cite{VanTuan_arXiv18}. 

The absorption spectrum of excitons is studied through the Green's function of the electron-hole pair. Focusing on photon absorption without the assistance of shortwave plasmons or phonons in this section, we consider only direct excitons in which the electron and hole reside in the same valley [Fig.~\ref{fig:Scheme}]. Starting with the case that the electron and hole do not interact with each other (i.e., an unbound exciton), the pair function is the product of the electron and hole Green's functions 
\begin{equation}\label{Eq:freePairG}
G_p^0(\bm{q},\bm{k}_i,\bm{k}_f,z,\Omega)=G_{e}(\bm{k}_i+\bm{q},\Omega-z)G_{h}(-\bm{k}_i,z)\delta_{\bm{k}_i,\bm{k}_f}\,\,.
\end{equation} 
$z=(2\ell+1)\pi\i k_BT$ and $\Omega=2\ell\pi\i k_BT$ are odd (Fermion) and even (Boson) imaginary Matsubara energies. $\bm{k}_{i/f}$ and $\bm{q}$ denote 2D crystal wavevectors, where $\bm{q}$  is the exciton's center-of-mass component. The restriction $\bm{k}_i = \bm{k}_f$ will be removed when we consider the interaction between the electron and hole through the dynamically-screened potential. 

Previously, we have only seen the effect of the dynamically-screened potential on the self-energy of electrons and holes, $\Sigma_{e/h}(\bm{k},z)$. Their renormalized Green's functions become
\begin{equation}\label{Eq:freeG2}
G_{e/h}(\bm{k},z)=\frac{1}{z-\varepsilon_{e/h}(\bm{k})-\Sigma_{e/h}(\bm{k},z)+\mu_{e/h}},
\end{equation} 
where $\mu_{e/h} =\mu_{e/h,0}+\Sigma_{e/h}(\bm{k}_F,E_F-\mu_{e/h,0})$ are the quasichemical potentials of electrons and holes. In the long-wavelength limit, the contributions to the self energy come from screened-exchange and Coulomb-hole energies (Sec.~\ref{sec:bgr_intra}). We focus on type-A excitons, so that $G_{h}(\bm{k},z)$  denotes states in the top valley of the valence band. The distinction between top and bottom valleys in the conduction band is implied when we choose the electron's effective mass in the kinetic energy, $\varepsilon_{e}(\bm{k})$, and when the screened-exchange contribution to the self-energy is evaluated [recall that the screened-exchange energy mostly affects the bottom populated valleys; see discussion after Eq.~(\ref{Eq:sx_3})]. Finally, when we consider electron states in the top valley (e.g., direct excitons in ML-WSe$_2$, as shown in Fig.~\ref{fig:Scheme}), the spin-splitting energy added to the top valley can be absorbed in the  energy reference level for the quasichemical potential.

\subsection{Bethe-Salpeter Equation}
\begin{figure}[t!]
\centering
\includegraphics*[width=0.9\textwidth]{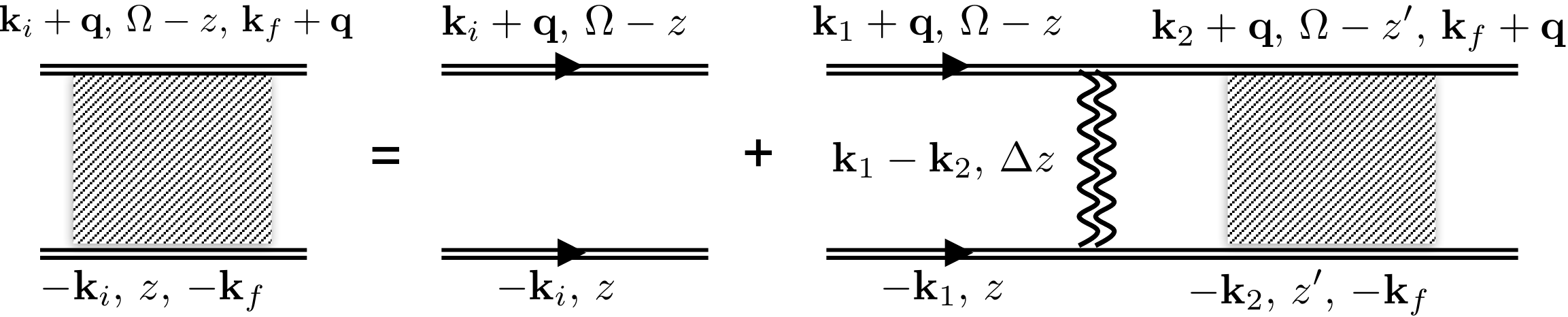}
\caption{Feynman diagram of the Bethe-Salpeter Equation in the screened-ladder approximation.}\label{fig:BSE}
\end{figure}

The equation of motion of bound excitons is studied by the so-called Bethe-Salpeter Equation (BSE) under the screened-ladder approximation (also referred to as shielded potential/$GW$ approximation) ~\cite{Haug1984:PQE,SchmittRink1986:PRB,Rohlfing2000:PRB}. This formalism considers the repeated interaction between the electron and hole through the screened potential but it neglects exciton-exciton interactions. Accordingly, the screened-ladder approximation is a valid approach to model the physics when the ML is not subjected to intense photoexcitation. Figure~\ref{fig:BSE} shows the Feynman diagram of the BSE, where the top and bottom horizontal propagators are the electron and hole Green's Functions (after considering the effect of the dynamically-screened potential on their self-energies). The first term on the right-hand side of this diagrammatic equation is the free-pair function, Eq.~(\ref{Eq:freePairG}). The vertical wiggly double-line in the second term denotes the Coulomb interaction between the electron and hole components of the exction through the dynamically-screened potential.  The BSE is formally written as
\begin{eqnarray}\label{Eq:BSEgeneral}
\!\!\!\!\!\!\!\!\!\!\!\!\!\!\!\!\!\!\!\!\!\!\!\!\!\!\!\!\!\!\!\!G_{p}(\bm{q},\bm{k}_i,\bm{k}_f,z,\Omega) &=&G_p^0(\bm{q},\bm{k}_i,\bm{k}_f,z,\Omega) \\
&\!\!\!\!\!\!\!\!\!\!\!\!\!\!\!\!+\!\!\!\!\!\!\!\!\!\!\!\!\!\!\!\!& \!\!\!\!\!\!\!\!\!\!\!\!\!\!\!\! k_BT\sum\limits_{\bm{k}_1,\bm{k}_2,z'}G_p^0(\bm{q},\bm{k}_i,\bm{k}_1,z,\Omega)W_{\ell}(\bm{k}_1-\bm{k}_2,z-z')G_{p}(\bm{q},\bm{k}_2,\bm{k}_f,z',\Omega), \nonumber
\end{eqnarray}
where $W_{\ell}(\bm{k},z)$ is the dynamically screened potential in the long-wavelength limit. Solving Eq.~(\ref{Eq:BSEgeneral}) is computationally expensive and often additional approximations are invoked to obtain the solution, as we discuss below. The solution of the BSE can then be compared with the observed absorption spectrum by ~\cite{SchmittRink1986:PRB}
\begin{equation}\label{Eq:abs}
\!\!\!\!\!\!\!\!\!\!\!\!\!\!\!\! A(\omega)\propto k_BT\sum\limits_{\bm{k}_i,\bm{k}_f,z}\mathrm{Im}\left[G_p(\bm{q}\rightarrow0,\bm{k}_i,\bm{k}_f,z,\Omega\rightarrow\hbar\omega-\mu_{e}-\mu_{h}+\i0^+)\right],
\end{equation}
 where $\hbar \omega$ is the photon energy. The sum integrates out the fermion degrees of freedom in the pair function, leaving only the boson degrees of freedom, $\bm{q}$ and $\Omega$. The former is considered in the limit $\bm{q} \rightarrow 0$ to denote the negligible momentum of direct excitons in the light cone (these excitons are coupled to photons). The boson Matsubara energy is related to the photon energy by analytical continuation to the real energy axis. 

\subsection{Quasistatic approximation}\label{Sec:QuasiStatic}
One procedure commonly employed to greatly simplify Eq.~(\ref{Eq:BSEgeneral}) is the (quasi)static approximation, where $W_{\ell}(\bm{k},z)$ is replaced by its static value $W_{\mathrm{\ell}}(\bm{k})\equiv W_{\ell}(\bm{k},0)$.\footnote{Note that the procedure described in Sec.~\ref{Sec:QuasiStatic} is also employed in the case of zero density because the potential $W_{\ell}(\bm{k},z)$ does not have any $z$-dependence in this case.} In this case, one can first contract the non-interacting and interacting pair functions independently,
\begin{equation}\label{Eq:ContractedFreeGP}
G_p^0(\bm{q},\bm{k}_i,\bm{k}_f,\Omega)=- k_BT\sum\limits_zG_p^0(\bm{q},\bm{k}_i,\bm{k}_f,z,\Omega),
\end{equation}
\begin{equation}\label{Eq:ContractedIGP}
G_p(\bm{q},\bm{k}_i,\bm{k}_f,\Omega)=-  k_BT \sum\limits_zG_p(\bm{q},\bm{k}_i,\bm{k}_f,z,\Omega),
\end{equation}
and Eq.~(\ref{Eq:BSEgeneral}) reduces to~\cite{SchmittRink1986:PRB}
\begin{eqnarray}\label{Eq:BSEqs}
 G_{p}(\bm{q},\bm{k}_i,\bm{k}_f,\Omega) &=&G_p^0(\bm{q},\bm{k}_i,\bm{k}_f,\Omega) \\
&-&\sum\limits_{\bm{k}_1,\bm{k}_2}G_p^0(\bm{q},\bm{k}_i,\bm{k}_1,\Omega)W_{\mathrm{\ell}}(\bm{k}_1-\bm{k}_2)G_{p}(\bm{q},\bm{k}_2,\bm{k}_f,\Omega). \nonumber
\end{eqnarray}
According to Eq.~(\ref{Eq:abs}), the absorption is described by the pair functions after the analytical continuation $\Omega\rightarrow\hbar\omega-\mu_{e}-\mu_{h}+\i0^+$. For now, we are only interested in direct excitons with $\bm{q}=\bm{0}$, and define
\begin{equation}\label{Eq:ContractedFreeGPq0}
G_p^0(\bm{k}_i,\bm{k}_f,\hbar\omega) \equiv G_p^0(\bm{0},\bm{k}_i,\bm{k}_f,\Omega\rightarrow\hbar\omega-\mu_{e}-\mu_{h}+\i0^+),
\end{equation}
\begin{equation}\label{Eq:ContractedIGPq0}
G_p(\bm{k}_i,\bm{k}_f,\hbar\omega) \equiv G_p(\bm{0},\bm{k}_i,\bm{k}_f,\Omega\rightarrow\hbar\omega-\mu_{e}-\mu_{h}+\i0^+).
\end{equation}
Within the quasistatic approximation and by using Eqs.~(\ref{eq:identity}), (\ref{Eq:freePairG}) and (\ref{Eq:ContractedFreeGPq0}) to evaluate Eq.~(\ref{Eq:ContractedFreeGP}), the free-pair function becomes 
\begin{equation}\label{Eq:ContractedFreeGPqs}
\begin{array}{ll}
\!\!\!\!\!\!\!\!\!\!\!\!\!\!\! G_p^0(\bm{k}_i,\bm{k}_f,\hbar\omega)=F_{\bm{k}_i} \Big[ \hbar\omega+\i0^+ -\varepsilon_e(\bm{k}_i)- \Sigma_{eh} - \varepsilon_h(-\bm{k}_i) \Big]^{-1} \delta_{\bm{k}_i,\bm{k}_f}
\end{array}
\end{equation}
with $F_{\bm{k}}=1-f_e(\varepsilon_{e}(\bm{k}))-f_h(\varepsilon_{h}(-\bm{k}))$.  The derivation of Eq.~(\ref{Eq:ContractedFreeGPqs}) was made under the quasistatic approximation, by which the $z$-dependence of $\Sigma_{e/h}(\bm{k},z)$ is replaced by $z\to\varepsilon_{e/h}(\bm{k})+\Sigma_{e/h}-\mu_{e/h}+\i0^+$ and recoil effects are neglected [see derivations of Eqs.~(\ref{Eq:sx_3}) and (\ref{eq:ch4})]. 
The total self-energy largely yields a rigid energy shift, $\Sigma_{eh} = \Sigma_{e} + \Sigma_h = 2\Sigma_{\mathrm{ch}} + \Sigma_{\mathrm{sx}}$, which has contributions from the Coulomb-hole and screened-exchange energies in the long-wavelength limit (Sec.~\ref{sec:bgr_intra}). 

Next, we solve the BSE in the quasistatic limit by using Eq.~(\ref{Eq:ContractedFreeGPqs}) and the ansatz
\begin{equation}\label{Eq:IntGPqsAnsatz}
\begin{array}{ll}
\!\!\!\!\!\!\!\!\!\!\!\!\!\!\! \!\!\!\!\!\!\!\!\!\!\!\! \!\!\!\! \!\!\!\!G_p(\bm{k}_i,\bm{k}_f,\hbar\omega)=\sum\limits_{S_x} \sqrt{|F_{\bm{k}_i}|}\,\mathcal{A}^{S_x}_{\bm{k}_i}\,\sqrt{|F_{\bm{k}_f}|}\,\mathcal{A}^{S_x}_{\bm{k}_f}   \Big[ \hbar\omega+\i0^+ \! -\! \Omega_{S_x}\Big]^{-1} \!\!\! \mathrm{sgn}(\Omega_{S_x}\!-\!\mu_e\!-\!\mu_h) \,.
\end{array}
\end{equation}
The exciton/excitation is described by its wave function in reciprocal space, $\mathcal{A}^{S_x}_{\bm{k}}$, and its energy $\Omega_{S_x}$ where $S_x$ is the discrete energy level.  With this ansatz, one can recast Eq.~(\ref{Eq:BSEqs}) into a non-Hermitian eigenvalue problem,
\begin{eqnarray}\label{Eq:BSEev}
\!\!\!\!\!\!\!\!\!\!\!\!\!\!\! \!\!\!\!\!\!\!\!\!\!\!\! \!\!\!\! \!\!\!\! \sum\limits_{\bm{k}' }  \left( \left[\varepsilon_e(\bm{k})\!+\!\varepsilon_h(-\bm{k})\!+\!\Sigma_{eh} \right]\delta_{\bm{k},\bm{k}'} - \mathrm{sgn}(F_{\bm{k}})\! \sqrt{\left| F_{\bm{k}} \right|}W_{\mathrm{\ell}}\left(\bm{k}\!-\!\bm{k}'\right)\!\sqrt{\left|F_{\bm{k}'}\right|}  \right)  \! \mathcal{A}^{S_x}_{\bm{k}'} = \Omega_{S_x}\mathcal{A}^{S_x}_{\bm{k}}.
\end{eqnarray}
Having solved Eq.~(\ref{Eq:BSEev}) for $\Omega_{S_x}$ and $\mathcal{A}^{S_x}_{\bm{k}}$, the optical absorption can be determined from Fermi's Golden Rule or the Kubo formalism as
\begin{eqnarray}\label{Eq:absd}
\!\!\!\!\!\!\!\!\!\!\!\!\!\!\!\!\!\!\!\!\!\!\!\!\!\!\!\!\!\!\!\! A(\omega)=\frac{4e^2\pi^2g_sg_v}{A c\omega} \sum\limits_{S_x} \! \left|\sum\limits_{\bm{k}}\!\sqrt{\left|F(\bm{k})\right|}d_\mathrm{vc}(\bm{k})\mathcal{A}^{S_x}_{\bm{k}}\right|^2 \!\!\! \mathcal{L}(\hbar\omega;\Omega_{S_x},\Gamma) \!\;\! \mathrm{sgn}(\Omega_{S_x}\!-\!\mu_e\!-\!\mu_h).
\end{eqnarray}
Here, $c$ is the speed of light and $d_\mathrm{vc}(\bm{k})$ is the $\bm{k}$-dependent single-particle dipole-matrix element for transitions between the valence and conduction bands. Due to the smallness of the Fermi energy with respect to the band gap, $d_\mathrm{vc}(\bm{k})$ can often be considered $\bm{k}$-independent~\cite{HaugKoch1994}, in which case Eq.~(\ref{Eq:absd}) reduces to Eq.~(\ref{Eq:abs}). The Lorentzian function in Eq.~(\ref{Eq:absd}), $\mathcal{L}(\hbar\omega;\Omega_{S_x},\Gamma)$, accounts for homogenous broadening due to electron-electron and electron-lattice interactions~\cite{Huang1988:PRB,Honold1989:PRB,Moody2015:NC}. Additionally, we model the broadening $\Gamma$ to be energy dependent \cite{HaugKoch1994},
\begin{equation}\label{Eq:broadening}
\Gamma\equiv\Gamma(\hbar\omega)=\Gamma_1+\frac{\Gamma_2}{1+\exp\left[\left(E_\mathrm{cont}-\hbar\omega\right)/\Gamma_3\right]}.
\end{equation}
Here, $\Gamma_1$ describes the homogeneous broadening due to radiative decay time as well as disorder-induced inhomogeneous broadening, while $\Gamma_2$ and $\Gamma_3$ describe the enhanced homogenous broadening when the energy $\hbar\omega$ enters the continuum, $\hbar\omega>E_\mathrm{cont}$.

\begin{figure}[htb!]
\centering
\includegraphics*[width=0.95\textwidth]{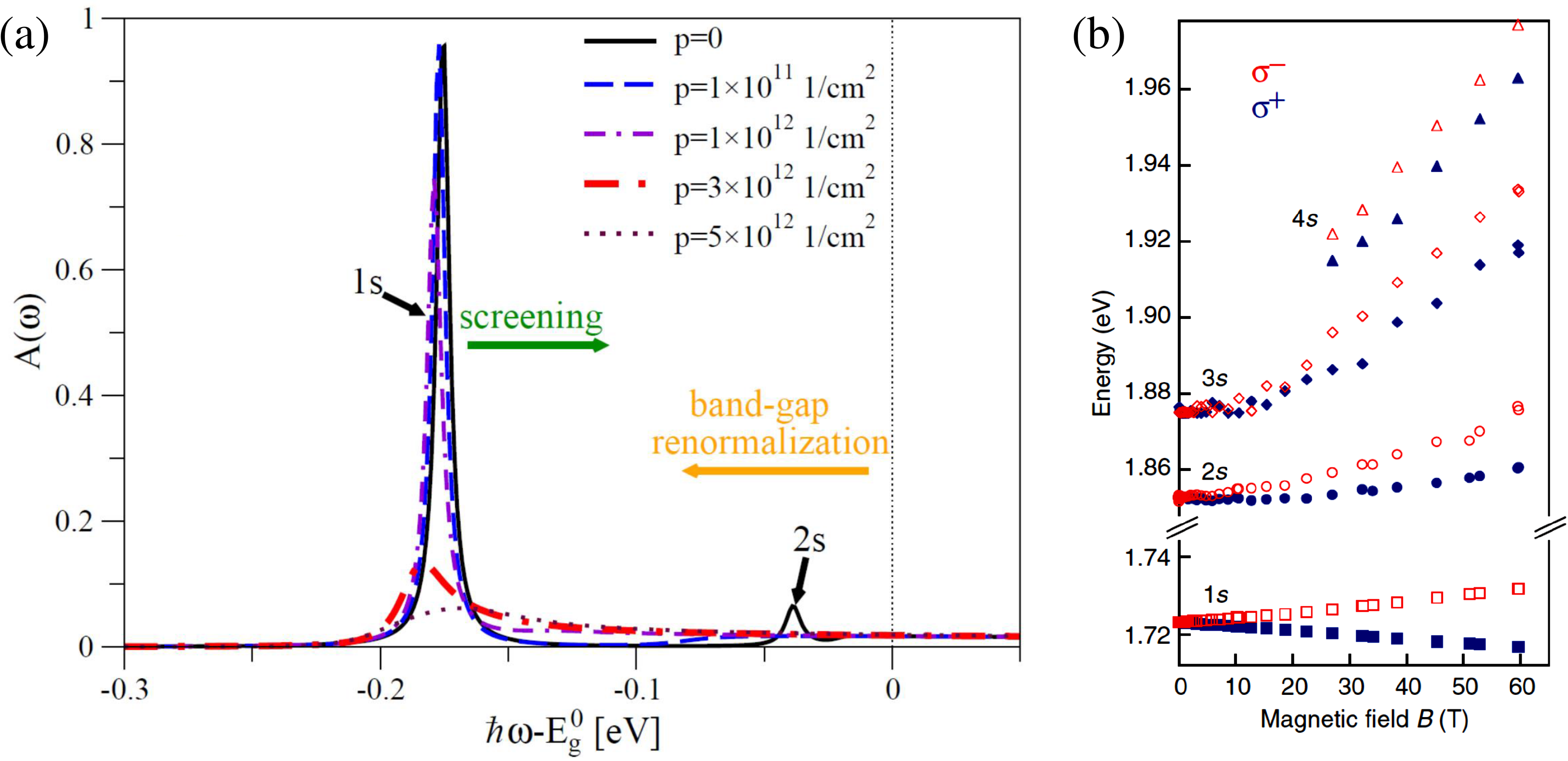}
\caption{(a) Using the quasistatic approach to calculate the absorption spectrum of the ground-state $X^0$ in ML-WSe$_2$ embedded in hBN for different background hole densities. The photon energy $\hbar\omega$ is measured with respect to the bare band gap $E_g^0$. (b) Measured $s$-states exciton energies versus magnetic field $B$ for $\sigma_+$ and $\sigma_-$ polarizations, where the free particle band-gap at $B=0$ is around  $E_g^0 \sim1.89$~eV; taken from Ref.~\cite{Stier_PRL18}. The measured exciton binding energy of the $1s$ is around 175~meV, and the difference between $1s$ and $2s$ is about 130~meV. Both values are reproduced by the modeled absorption spectrum in (a).}\label{fig:qsAbsorption}
\end{figure}

We now have all of the ingredients needed to solve the eigenvalue problem~(\ref{Eq:BSEev}) and to calculate the ensuing absorption spectrum from Eq.~(\ref{Eq:absd}). Numerically, Eq.~(\ref{Eq:BSEev}) can be diagonalized on a coarse uniform $N\times N$ $k$-grid with a spacing of $\Delta k=2\pi/(Na_0)$ and $a_0=3.2$ {\AA} in each direction. The Coulomb matrix elements $W_{\mathrm{\ell}}\left(\bm{k}-\bm{k}'\right)$, however, are not evaluated at the grid points of the $N\times N$ $k$-grid, but are instead averaged over a square centered around the coarse-grid point $\bm{k}-\bm{k}'$ with side widths $\Delta k$ on a fine $N_\mathrm{int}\times N_\mathrm{int}$ grid [with a corresponding spacing of $\Delta k_\mathrm{int}=\Delta k/N_\mathrm{int}=2\pi/(NN_\mathrm{int}a_0)$]. Our numerical calculations have been performed with $N=200$ (and an energy cutoff 1.2~eV above the band gap) and $N_\mathrm{int}=100$. 

Figure~\ref{fig:qsAbsorption}(a) shows the zero-temperature absorption spectrum computed in the quasistatic approximation with parameters corresponding to moderately hole-doped ML-WSe$_2$ embedded in hBN. The parameters used to model the non-local dielectric function, $\epsilon_d(q)$, are mentioned after Eq.~(\ref{Eq:DiFv2def}). The effective masses of the hole and electron are $m_h=0.36m_0$ and $m_e=0.29m_0$, respectively ~\cite{Kormanyos_2DMater15}. We have used the parameters $C_\mathrm{eff}=4$ and $\hbar^2q_c^2/2m_b=0.12$~eV in the calculations of the screened interaction and Coulomb-hole energy [Eqs.~(\ref{eq:spp}), (\ref{eq:w_q_l}), and (\ref{eq:ch4})]. The broadening of the spectral lines were modeled by using $\Gamma_1=3$~meV, $\Gamma_2=30$~meV and $\Gamma_3=10$~meV in Eq~(\ref{Eq:broadening}). The dipole matrix elements $d_\mathrm{vc}(\bm{k})=5\times10^5$ m/s are chosen to be constant. Figure~\ref{fig:qsAbsorption}(a) shows that two peaks emerge when the background hole density vanishes ($p=0$). These peaks correspond to the $1s$ state of $X^0$ and its first excited Rydberg state ($2s$). The binding energy of the $1s$ state is around 175~meV and the difference between the $1s$ and $2s$ exciton energies is around 130~meV. These values are in very good agreement with the experimental findings \cite{Stier_PRL18}, shown in Fig.~\ref{fig:qsAbsorption}(b) for WSe$_2$ sandwiched between hBN~\footnote{Due to broadening, only the $1s$ and $2s$ excitons are clearly visible in Fig.~\ref{fig:qsAbsorption}(a). An inspection of the energies of the higher Rydberg states computed with the bare potential~(\ref{Eq:DiFv2})-(\ref{Eq:DiFv2def}) reveals that a good agreement with experimental data from Fig.~\ref{fig:qsAbsorption}(b) also extends to these states for $p=0$.}. 

\begin{figure}[htb!]
\centering
\includegraphics*[width=0.8\textwidth]{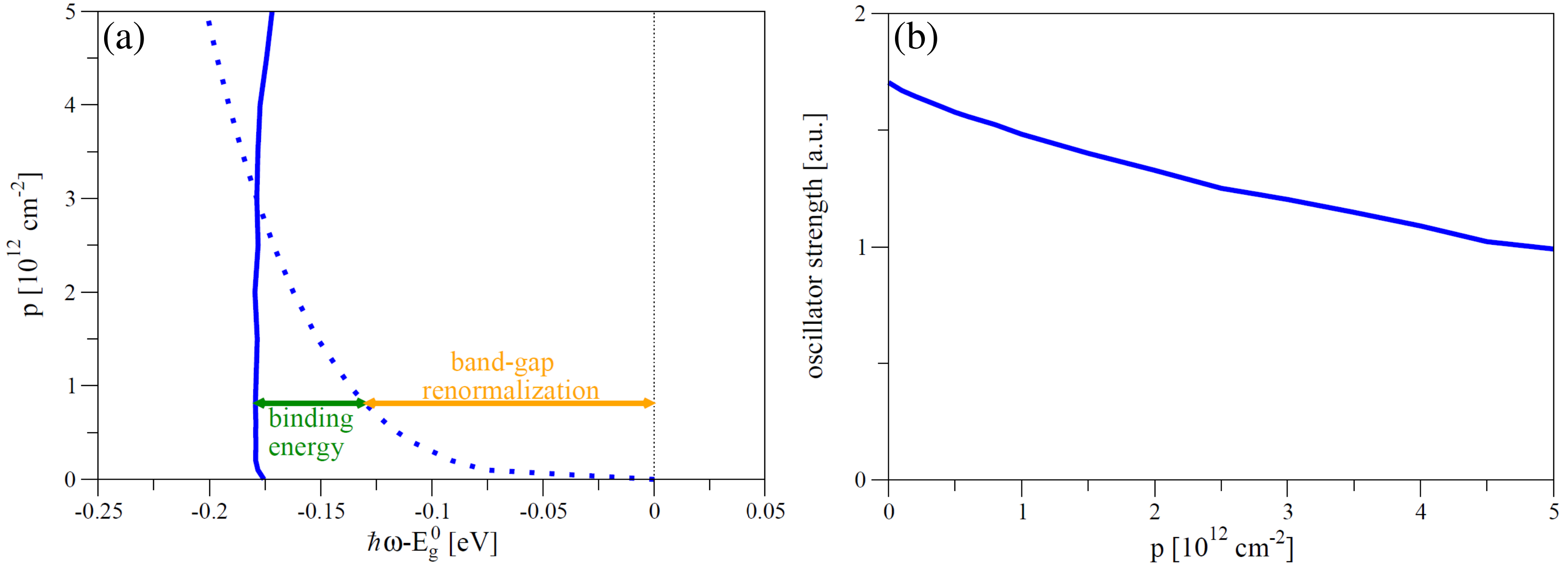}
\caption{(a) Calculated position of the $X^0$ peak ($1s$) and BGR. The solid line shows the peak position, and the dotted line is the redshifting continuum. (b) oscillator strength in ML-WSe$_2$. The parameters are the same as in Fig.~\ref{fig:qsAbsorption}(a).}\label{fig:qsBGR}
\end{figure}

Figure~\ref{fig:qsAbsorption}(a) shows that the absorption of the $1s$ state is attenuated  when $p$ is increased and its position slightly redshifts. The calculated dependence of the peak position on charge density is governed by the chosen values of $C_\mathrm{eff}$ and the integration cutoff energy ($\hbar^2q_c^2/2m_b$). As we have seen in Fig.~\ref{fig:ch}(c), choosing a smaller (larger) integration cutoff energy leads to a smaller (larger) BGR. As a result, the calculated peak position blueshifts  (redshifts) when the charge density is ramped up by decreasing (increasing) the integration cutoff energy.  That is, we can lessen/amplify the BGR compared with screening, whose competing effects are illustrated by the arrows in Fig.~\ref{fig:qsAbsorption}(a). These competing effects are also illustrated by Fig.~\ref{fig:qsBGR}(a), which shows the position of the $1s$ peak from Fig.~\ref{fig:qsAbsorption}(a) versus $p$ as well as the position of the continuum edge determined by $E_g^0+\Sigma_{eh}$. One can see that the continuum edge rapidly redshifts with $p$. On the other hand, as $W_{\ell}(\bm{k})$ decreases with $p$, the binding energy of the exciton also decreases, which roughly compensates the redshift due to the BGR. Moreover, Fig.~\ref{fig:qsBGR}(b) shows that although the oscillator strength of the $1s$ exciton decreases with $p$, it still carries significant oscillator strength/spectral weight even at densities of $p=5\times10^{12}$ cm$^{-2}$. The decrease of the height of the $1s$ peak with $p$ in Fig.~\ref{fig:qsAbsorption}(a) is then due to two effects: the loss of oscillator strength as depicted in Fig.~\ref{fig:qsBGR}(b) and an increase in the broadening $\Gamma(\hbar\omega)$ as the continuum edge moves closer to the $1s$ peak position. While the $1s$ peak can thus be still clearly seen when $p > 10^{12}$ cm$^{-2}$, the $2s$ peak quickly disappears as it merges into the continuum.

\begin{figure}[htb!]
\centering
\includegraphics*[width=0.95\textwidth]{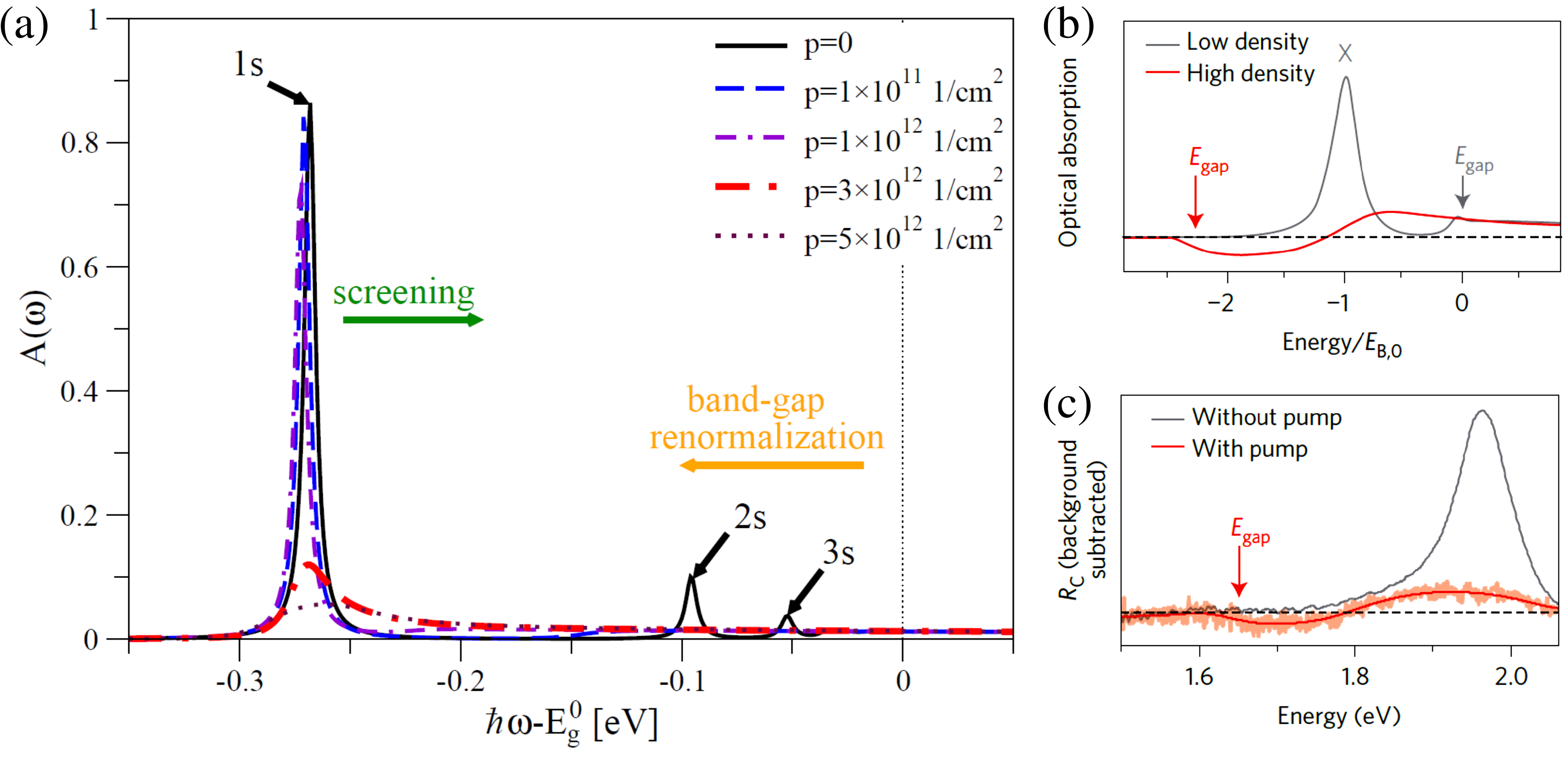}
\caption{(a) Absorption spectrum of the ground-state $X^0$ in ML-WS$_2$ on SiO$_2$ for different hole dopings as calculated from the quasistatic approach. The photon energy $\hbar\omega$ is measured with respect to the bare band gap $E_g^0$. (b) and (c) Scheme of the optical absorption of a 2D semiconductor (in units of the binding energy $E_{B,0}$) with BGR at high densities and measured reflectance contrast in a WS$_2$ bilayer supported on SiO$_2$, taken from Ref.~\cite{Chernikov2015:NPhot}.}\label{fig:qsBGRcomp}
\end{figure}

Similar to Fig.~\ref{fig:qsAbsorption}(a), Fig.~\ref{fig:qsBGRcomp}(a) shows the zero-temperature absorption spectrum computed in the quasistatic approximation, but now for hole-doped ML-WS$_2$ supported on SiO$_2$. The parameters used to model the non-local dielectric function, $\epsilon_d(q)$, are mentioned after Eq.~(\ref{Eq:DiFv2def}). Following Ref.~\cite{Kormanyos_2DMater15}, we have taken the effective masses of the hole and electron as $m_h=0.36m_0$ and $m_e=0.27m_0$, respectively. At zero doping $p=0$, the binding energy of the $1s$ state of $X^0$ is around 270~meV, which decreases to a binding energy of around 90~meV at $p=1\times10^{12}$ 1/cm$^2$. The BGR is around 180~meV when $p=10^{12}$~cm$^{-2}$, and it increases to around 250 meV at $p=5\times10^{12}$~cm$^{-2}$. This BGR is around half of the value estimated for the BGR in WS$_2$ following excitation by intense optical pump pulses (with an estimated carrier density per layer of $1.1\times10^{14}$~cm$^{-2}$)~\cite{Chernikov2015:NPhot}, also shown in Figs.~\ref{fig:qsBGRcomp}(b,c). Such a giant BGR has also been observed in ML-MoSe$_2$~\cite{Ugeda2014:NM}.

The results presented in Figs.~\ref{fig:qsAbsorption}(a),~\ref{fig:qsBGR} and~\ref{fig:qsBGRcomp}(a) have been obtained for hole-doped WSe$_2$ embedded in hBN (or WS$_2$ supported on SiO$_2$), but are also representative for the results one would obtain for electron- and hole-doped MoX$_2$ and WX$_2$ in the quasistatic approximation. 
In the following, we will go beyond the quasistatic approximation and take into account dynamical effects at different approximation levels.

\subsection{Shindo approximation}\label{Sec:Shindo}
In the previous section, we have not taken into account dynamical effects and have replaced the fully dynamical potential $W_{\ell}(\bm{q},z)$ in Eq.~(\ref{Eq:BSEgeneral}) by its static limit ($z=0$). The objective in this part  is to include dynamical effects in an approximate way before showing a fully dynamical treatment in Sec.~\ref{Sec:ExcitonsDyn}. For this purpose, we employ the so-called Shindo approximation~\cite{Shindo1970:JPSJ} to solve Eq.~(\ref{Eq:BSEgeneral}). 

First, we use the identity
\begin{equation}\label{Eq:FreeGPrewritten}
\begin{array}{l}
\!\!\!\!\!\!\!\!\!\!\!\!\!\!\!\!\!\!\!\!\! G_{e}(\bm{k}_i+\bm{q},\Omega-z)G_{h}(-\bm{k}_i,z)=\frac{G_{e}(\bm{k}_i+\bm{q},\Omega-z)+G_{h}(-\bm{k}_i,z)}{\Omega-\varepsilon_{e}(\bm{k}+\bm{q})-\varepsilon_{h}(-\bm{k})-\Sigma_e(\bm{k}+\bm{q},\Omega-z)-\Sigma_h(-\bm{k},z)+\mu_e+\mu_h}
\end{array}
\end{equation}
to rewrite Eq.~(\ref{Eq:BSEgeneral}) as
\begin{eqnarray}\label{Eq:BSEgeneralrewritten}
\!\!\!\!\!\!\!\!\!\!\!\!\!\!\!\!\!\!\!\!\!\!\!\!\!\!\!\!\!\!\!\!\!\!\!\!\! &\!\!\!&\!\! \Big[{\Omega\!-\!\varepsilon_{e}(\bm{k}_i\!+\!\bm{q})\!-\varepsilon_{h}(-\bm{k}_i)\!-\!\Sigma_e(\bm{k}_i\!+\!\bm{q},\Omega\!-\!z)\!-\!\Sigma_h(-\bm{k}_i,z) \!+\!\mu_e\!+\!\mu_h}\Big]\! G_{p}(\bm{q},\bm{k}_i,\bm{k}_f,z,\Omega)= \nonumber \\ 
\!\!\!\!\!\!\!\!\!\!\!\!\!\!\!\!\!\!\!\!\!\!\!\!\!\!\!\!\!\!\!\! &\!\!\!&\!\! \Big[\delta_{\bm{k}_i,\bm{k}_f}\!+\!k_BT\!\sum\limits_{\bm{k},z'}W_{\ell}(\bm{k}_i\!-\!\bm{k},z\!-\!z')G_{p}(\bm{q},\bm{k},\bm{k}_f,z',\Omega)\Big]\!\Big[ \! G_{e}(\bm{k}_i\!+\!\bm{q},\Omega\!-\!z)\!+\!G_{h}(-\bm{k}_i,z)\Big].
\end{eqnarray}
For the absorption calculated via Eq.~(\ref{Eq:abs}), we need the contracted pair Green's function $G_p(\bm{k}_i,\bm{k}_f,\hbar\omega)$ given by Eq.~(\ref{Eq:ContractedIGPq0}). Retardation effects in Eq.~(\ref{Eq:BSEgeneralrewritten}), however, prevent the derivation of a closed equation for $G_p(\bm{k}_i,\bm{k}_f,\hbar\omega)$ as in Sec.~\ref{Sec:QuasiStatic}, where dynamic effects in $W_{\ell}(\bm{k},z)$ have been neglected. The Shindo approximation provides a way to simplify this problem by replacing the $z$-dependence of $G_{p}(\bm{q},\bm{k}_i,\bm{k}_f,z,\Omega)$ with that of $G_{e}(\bm{k}_i+\bm{q},\Omega-z)+G_{h}(-\bm{k}_i,z)$~\cite{Shindo1970:JPSJ,Haug1984:PQE}:
\begin{equation}\label{Eq:ShindoApprox}
G_{p}(\bm{q},\bm{k}_i,\bm{k}_f,z,\Omega)\approx\frac{G_{e}(\bm{k}_i+\bm{q},\Omega-z)+G_{h}(-\bm{k}_i,z)}{1-f_e(\varepsilon_{e}(\bm{k}+\bm{q}))-f_h(\varepsilon_{h}(-\bm{k}))}G_p(\bm{q},\bm{k}_i,\bm{k}_f,\Omega),
\end{equation}
where $G_p(\bm{q},\bm{k}_i,\bm{k}_f,\Omega)$ has been defined in Eq.~(\ref{Eq:ContractedIGP}).

Inserting Eq.~(\ref{Eq:ShindoApprox}) into Eq.~(\ref{Eq:BSEgeneralrewritten}) and summing over $z$ then yields a closed equation, which in the case of direct excitons in the light cone ($q=0$) reads \cite{Haug1984:PQE}
\begin{equation}\label{Eq:BSEShindo}
\begin{array}{l}
\left[\hbar\omega+\i0^+-\varepsilon_{e}(\bm{k}_i)-\varepsilon_{h}(-\bm{k}_i)-\Delta_{eh}(\bm{k}_i,\hbar\omega)\right]G_p(\bm{k}_i,\bm{k}_f,\hbar\omega)=\\
\quad\quad\quad\quad\quad\quad\quad\quad F(\bm{k}_i)\left[\delta_{\bm{k}_i,\bm{k}_f}+\sum\limits_{\bm{k}}I_{eh}(\bm{k}_i,\bm{k},\hbar\omega)G_p(\bm{k}_i,\bm{k}_f,\hbar\omega)\right].
\end{array}
\end{equation}
Equation~(\ref{Eq:BSEShindo}) needs to be solved self-consistently for the exciton/excitation energies $\Omega_{S_x}$ and is equivalent to the self-consistent wave equation
\begin{equation}\label{Eq:BSEShindoWE}
\quad\quad\sum\limits_{\bm{k}'}\mathcal{H}_{\bm{k},\bm{k}'}(\Omega_{S_x})\mathcal{A}^{S_x}_{\bm{k}'}(\Omega_{S_x})=\Omega_{S_x}\mathcal{A}^{S_x}_{\bm{k}}(\Omega_{S_x}),
\end{equation}
\begin{displaymath}
\mathcal{H}_{\bm{k},\bm{k}'}(\Omega)=\left[\varepsilon_e(\bm{k})+\varepsilon_h(-\bm{k})+\Delta_{eh}(\bm{k},\Omega)\right]\delta_{\bm{k},\bm{k}'}- F(\bm{k})\,I_{eh}(\bm{k},\bm{k}',\Omega).
\end{displaymath}
Using the SPP approximation, the interaction kernel and BGR read
\begin{eqnarray}\label{Eq:BSEShindoIntKernel_BGR}
&&\!\! \!\! \!\! \!\! \!\! \!\! \!\! \!\! \!\! \!\! \!\! \!\! \!\! \!\! \!\! \!\! \!\! I_{eh}(\bm{k},\bm{k}',\Omega) \!=\! V(\!\bm{k}\!-\!\bm{k}') \!+\!\frac{\omega_\mathrm{pl}^2\,g\!\!\left(\hbar\omega_{|\bm{k}-\bm{k}'|}\right)}{2\omega_{|\bm{k}-\bm{k}'|}}\!\left[\frac{V(\bm{k}-\bm{k}')}{\Omega\!+\!\i0^+\!-\!\varepsilon_{e}(\bm{k})\!-\!\varepsilon_{h}(-\bm{k})\!+\!\hbar\omega_{|\bm{k}-\bm{k}'|}}\!+\!(\bm{k}\leftrightarrow\bm{k}')\right] \nonumber \\ \!\! \!\! \!\! \!\! \!\! \!\! \!\! \!\! && \!\! \!\! \!\! \!\! \!\! \!\! \!\! \!\! \!\! + \frac{\omega_\mathrm{pl}^2\,\left[1+g\left(\hbar\omega_{|\bm{k}-\bm{k}'|}\right)\right]}{2\omega_{|\bm{k}-\bm{k}'|}}\left[\frac{V(\bm{k}-\bm{k}')}{\Omega+\i0^+-\varepsilon_{e}(\bm{k})-\varepsilon_{h}(-\bm{k})-\hbar\omega_{|\bm{k}-\bm{k}'|}}+(\bm{k}\leftrightarrow\bm{k}')\right] ,\nonumber \\
&& \!\! \!\! \!\! \!\! \!\! \!\! \!\! \!\! \!\! \!\! \!\! \!\! \!\! \!\! \!\! \!\! \!\! \Delta_{eh}(\bm{k},\Omega)=\sum\limits_{\bm{k}'}\left[F(\bm{k}')\,I_{eh}(\bm{k},\bm{k}',\Omega)-V(\bm{k}-\bm{k}')\right].
\end{eqnarray}

\begin{figure}[h!]
\centering
\includegraphics*[width=0.7\textwidth]{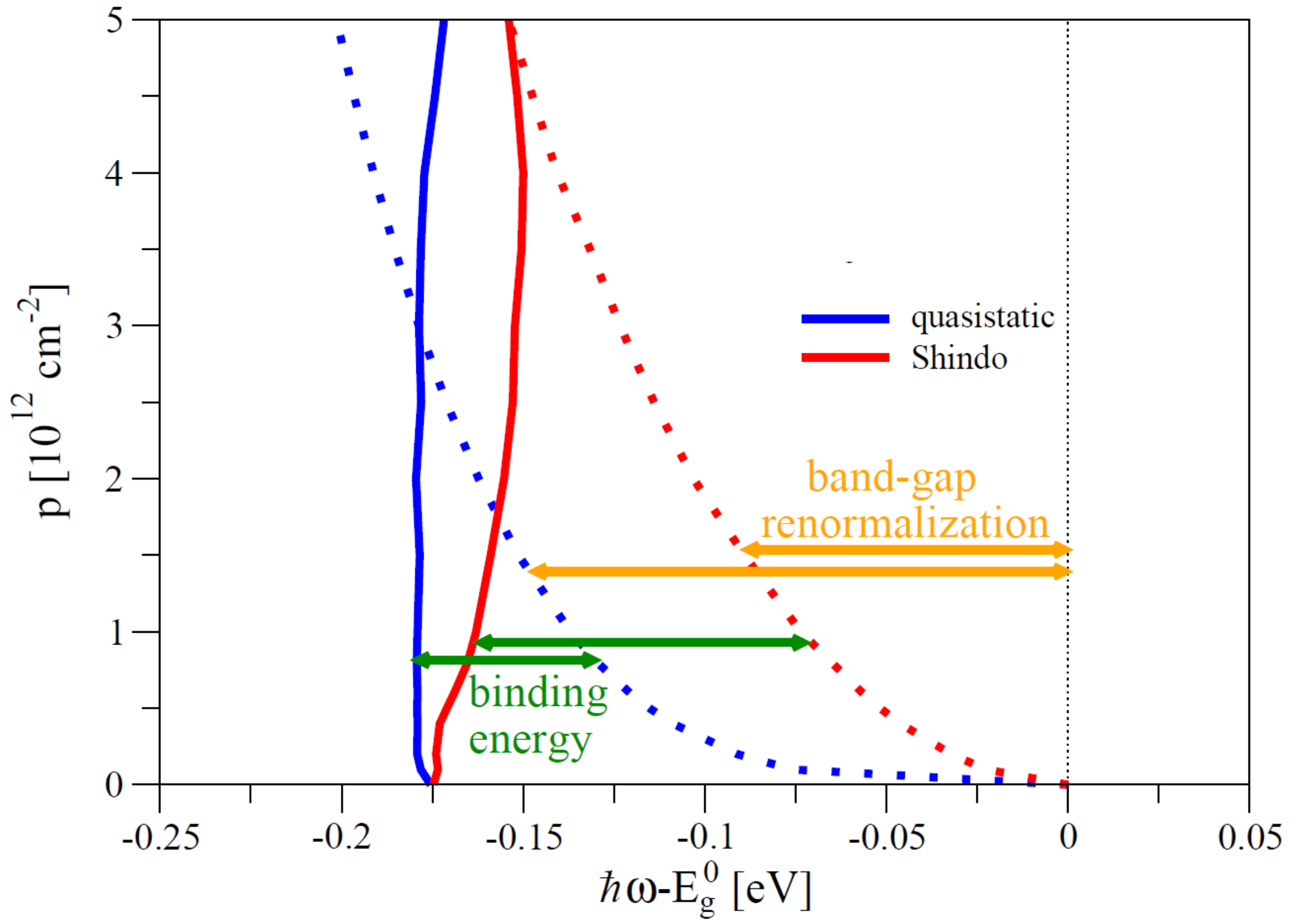}
\caption{Comparison between the calculated positions of the $X^0$ ($1s$) peak and the BGR in ML-WSe$_2$ embedded in hBN as computed with the Shindo and quasistatic approximations. The parameters are the same as in Fig.~\ref{fig:qsAbsorption}(a) for both approximations.}\label{fig:ShindoVsqsBGR}
\end{figure}

As before, there is a competition between screening of the electron-hole interaction [described by $I_{eh}(\bm{k},\bm{k}',\Omega)$] and BGR [described by $\Delta_{eh}(\bm{k},\Omega)$]. Figure~\ref{fig:ShindoVsqsBGR} compares the positions of the $1s$ $X^0$ peak computed self-consistently within the Shindo approximation, Eqs.~(\ref{Eq:BSEShindoWE})-(\ref{Eq:BSEShindoIntKernel_BGR}), with those obtained from the quasistatic approximation (see also Fig.~\ref{fig:qsBGR}).\footnote{Equation~(\ref{Eq:BSEShindoWE}) has been diagonalized on a coarse $100\times100$ $k$-grid as the one used to compute Figs.~\ref{fig:qsAbsorption}(a) and~\ref{fig:qsBGR}. Likewise, the interaction $I_{eh}(\bm{k},\bm{k}',\Omega)$ has additionally been averaged over a fine $100\times100$ $k$-grid.} The results obtained from the Shindo treatment again show the $X^0$ peak position does not change significantly for densities up to around $10^{12}$ cm$^{-2}$. For larger densities, there is a slight blueshift compared to the quasistatic result. The inclusion of dynamic effects leads to a weaker screening and BGR than in the quasistatic treatment. The net result of these two effects is a slight blueshift for the parameters in Fig.~\ref{fig:ShindoVsqsBGR}. As mentioned in the discussion of Fig.~\ref{fig:qsAbsorption}(a), the blueshift can be mitigated by choosing a larger integration cutoff energy or a smaller value for $C_{\mathrm{eff}}$.

\subsection{Fully dynamical treatment}\label{Sec:ExcitonsDyn}
The main complication in solving the dynamical BSE, Eq.~(\ref{Eq:BSEgeneral}),  is that it has to be solved for momentum and frequency variables simultaneously. In contrast to the Shindo approximation, we contract the pair Green's function~(\ref{Eq:BSEgeneral}) in momentum rather than Matsubara space. This approach allows one to integrate out the final wavevector $k_f$ in Eq.~(\ref{Eq:BSEgeneral}), and the contracted Green's function $G_{p}(\bm{q},\bm{k}_i,z,\Omega)$ can be determined from
\begin{eqnarray}\label{Eq:DynPairGFContractedMomentum}
 G_{p}(\bm{q},\bm{k},z,\Omega)&=&G_p^0(\bm{q},\bm{k},z,\Omega) \\
&+&k_BT\sum\limits_{\bm{k}',z'}G_p^0(\bm{q},\bm{k},z,\Omega)W_{\ell}(\bm{k}-\bm{k}',z-z')G_{p}(\bm{q},\bm{k}',z',\Omega), \nonumber
\end{eqnarray}
where the contracted non-interacting Green's function is given by
\begin{equation}\label{Eq:freePairGv2}
G_p^0(\bm{q},\bm{k},z,\Omega)=G_{e}(\bm{k}+\bm{q},\Omega-z)G_{h}(-\bm{k},z).
\end{equation}
While changing the contracting parameter from $z$ to $\mathbf{k}_f$ seems trivial, it has the advantage of retaining the dynamical effect by keeping both even and odd Matsubara energies. Importantly, it does not introduce any further approximations beyond the screened-ladder approximation: If the single-particle dipole matrix elements $d_\mathrm{vc}(\bm{k})$ are assumed to be independent of $\bm{k}$, then one can model the absorption spectrum by using Eq.~(\ref{Eq:abs}). Finally, the self-energy of electrons or holes is calculated by considering the spectral representation of the interaction [Eq.~(\ref{eq:spectral_representation})],
\begin{eqnarray} \label{eq:sigma_kz}
  \!\!\!\!\!\!\!\!\!\!\!\!\!\! \!\!\!\!\!\!\!  \Sigma_{i}({\bf k},z) &=& -  \frac{e^2}{\pi }  \int_0^\infty \frac{dq}{\epsilon_d(q)} \int_0^{\pi} d\theta  f_{i}({\bf k-q})   \,\, + \,\, \frac{e^2}{2\pi}  \int_0^{q_c} \frac{dq}{\epsilon_d(q)} \int_0^{\pi}  d\theta \,\,   \frac{\hbar \omega_{\ell}^2(q) }{\omega_{\ell,\bar{\bm{q}}}} \times \nonumber \\
  &   &  \left\{\frac{f_{i}({\bf k-q})+g(\hbar \omega_{\ell,\bar{\bm{q}}} )}{ z  -\epsilon_{i}({\bf k-q}) +\mu_{i} +\hbar \omega_{\ell,\bar{\bm{q}}} } -\left(\hbar \omega_{\ell,\bar{\bm{q}}} \leftrightarrow-\hbar \omega_{\ell,\bar{\bm{q}}} \right)\right\} ,
\end{eqnarray}
where $i=\,$e or h, and $\theta$ is the angle between $\bm{k}$ and $\bm{q}$. The first and second integrals denote contributions to the BGR from exchange and correlation energies, respectively. This partition is an alternative to the one that considers contributions from screened-exchange and Coulomb-hole energies.

%
%


\subsubsection{Numerical details:} The Green's function in Eq.~(\ref{Eq:DynPairGFContractedMomentum}) is obtained by matrix inversion. Each row (or column) in the matrix is indexed via $\{k_j,z_n\}$ in $G_{p}(\bm{q},\bm{k}_j,z_n,\Omega)$ where $q$ and $\Omega$ are treated as parameters. Since we are interested first in direct bright exctions, we assign $q=0$ due to the negligible photon momentum. The angular dependence of ${\bf k}-{\bf k}'$ in $W_{\ell}({\bf k}-{\bf k}',z-z')$ is averaged out, and then all the 2D wavevectors in Eq.~(\ref{Eq:DynPairGFContractedMomentum}) are treated as scalars. For the Fermion Matsubara energies, $z_n=i\pi (2n_z+1)k_BT$, $n_z$ runs between $-N_z$ and $+N_z$. We solve Eq.~(\ref{Eq:DynPairGFContractedMomentum}) for each Bosonic frequency $\Omega_n=i2\pi n_\Omega k_BT$ in the range $n_\Omega\in[-N_\Omega,N_\Omega]$. $N_\Omega$, $N_z$, and the cut-off momentum $k_{Max}$ are chosen such that the maximum energies $2\pi N_\Omega k_BT$, $2\pi N_z k_BT$, and $\varepsilon_{\bm{k}_{\mathrm{Max}}}$ are much greater than the exciton binding energy whereas the energy resolution, $2\pi k_BT$, should be smaller than the energy of interest (e.g., the binding energy of 2s state). The momenta of small energies play an more important role in the convergence of the calculated binding energies. Therefore, we use the nonuniform grid $k_{j}=j^2 \Delta k$ since it yields better results than the uniform one when the number of $k_j$ points, $N_k=\sqrt{k_{Max}/\Delta k}$, is relatively small. The calculations below were performed with $N_k=50$, $N_z=160$, $N_\Omega=90$, and $\varepsilon_{\bm{k}_{\mathrm{Max}}}=1$~eV. 

\subsubsection{Analytical continuation:} After obtaining the pair Green's function, it is further contracted by integrating out the Fermion degrees of freedom ($\bm{k}$ and odd Matsubara energies $z$),
\begin{eqnarray}
 G_p(\mathbf{q}, \Omega) =\sum_{\mathbf{k},z}G_p(\mathbf{q}, \mathbf{k}, z, \Omega).
\end{eqnarray}
Evaluating the pair Green's function at real photon energies is then obtained by analytical continuation using Pad\'{e} approximants~\cite{Vidberg1977:JLTP,Osolin2012:PRB}. Specifically, knowing the values of a general analytic complex function, $F_N(\Omega)$, at $N$ points $\{\Omega_i\}$ in the complex plane, 
\begin{equation}
 F_N(\Omega_i)=F_i; \quad i=1,\cdots,N,
 \label{Pade:Condition}
 \end{equation}
we look for a function $F_N(\Omega)$ in the form of continued fractions
\begin{equation}
F_N(\Omega)=\frac{a_1}{1+\frac{a_2(\Omega-\Omega_1)}{1+\frac{a_3(\Omega-\Omega_2)}{1+\dots \frac{\vdots}{ \frac{\vdots}{1+\frac{a_N(\Omega-\Omega_{N-1})}{1}} } } } },
\label{Pade1977}
\end{equation}
where the coefficients $a_i$ are to be determined so that Eq.~(\ref{Pade1977}) fulfills Eq.~(\ref{Pade:Condition}). 
The coefficients $a_i$ are then given by 
\begin{equation}
 a_i=f_i(\Omega_i), \quad i=1,\cdots,N,
\end{equation}
where $f_n(\Omega)$ follows the recurrence relation
\begin{equation}
 f_n(\Omega)=\frac{f_{n-1}(\Omega_{n-1})-f_{n-1}(\Omega)}{(\Omega-\Omega_{n-1})f_{n-1}(\Omega)}
\end{equation}
and
\begin{equation}
 f_1(\Omega_i)=F_i; \quad i=1,\cdots,N.
\end{equation}
After finding the analytic function of the electron-hole pair Green function in the upper complex plane, the connection to the absorption spectrum is then established via its imaginary part, 
\begin{eqnarray}
A(\omega) \propto \,\, k_BT\cdot  \mathrm{Im} \left\{G_p(\mathbf{q} \rightarrow 0, \Omega = \hbar\omega-\mu_{e}-\mu_{h}+\i \Gamma) \right\}.
\end{eqnarray}

\begin{figure}[htb!]
\centering
\includegraphics*[width=\textwidth]{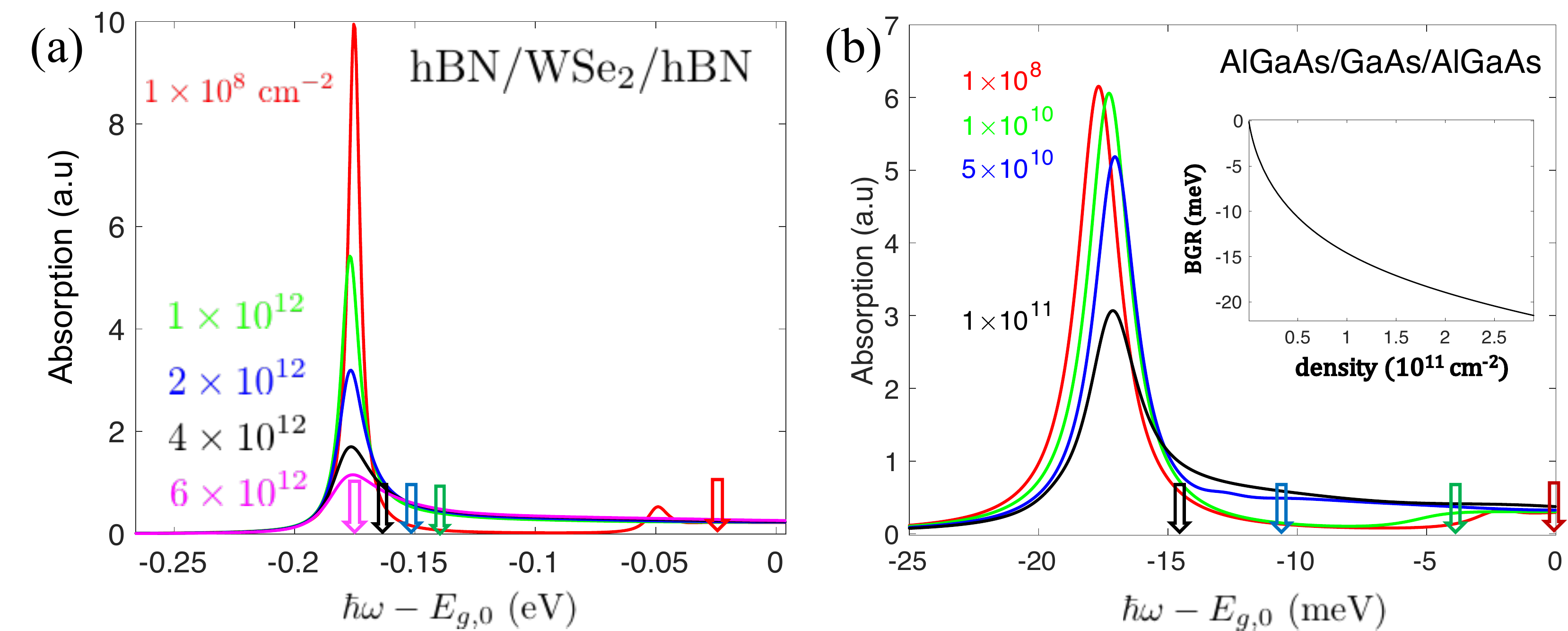}
\caption{Calculated absorption spectrum of neutral excitons in ML-WSe$_2$ (left) and infinite GaAs quantum well (right) using the fully dynamical treatment. $E_{g,0}$ is the optical band-gap for electron-hole pairs in the continuum at zero charge density. Increasing the background charge density leads to a decay of the exciton peak. The band-gap renormalization is indicated by the arrows on the $x$-axis, showing the redshift for the onset of optical transitions in the continuum (noninteracting electron-hole pairs). The inset shows the BGR as a function of electron densities.}\label{fig:DynAbsorptBefore}
\end{figure}




\subsubsection{Results:}  Figures~\ref{fig:DynAbsorptBefore}(a) and (b) show the absorption spectrum of ML-WSe$_2$ embedded in hBN at $T=20$~K, and GaAs quantum well at $T=10$~K. For the case of ML-WSe$_2$, $C_{\mathrm{eff}}=0$ and the integration cutoff energy is 40~meV. All other parameters are similar to the ones used in the quasistatic and Shindo approximations. Because the screening is overestimated when using the quasistatic approximation, we had to use smaller values for $C_{\mathrm{eff}}$ and the integration cutoff energy in order to achieve agreement with experiment. Figure~\ref{fig:DynAbsorptBefore}(a) shows that the behavior is qualitatively similar to the one modeled by using the Shindo and quasistatic approximations. The arrows on the $x$-axes indicate the continuum redshift due to BGR. Above this energy the excitons are no longer bound. The $2s$ bound state merges into the continuum already at relatively small densities due to its smaller binding energy. Note the BGR arrow for the lowest density in ML-WSe$_2$ is positioned about 26 meV below the zero energy level. The reason is that we made a distinction in this simulation between the optical band-gap and the fundamental band-gap energies, which in ML-WSe$_2$ involve the top and bottom valleys in the conduction band, respectively. 

Figure~\ref{fig:DynAbsorptBefore}(b) simulates the case of an infinite GaAs quantum-well with the conventional 2D Coulomb potential ($\epsilon = 12.9$). The electron and hole masses are $m_e=0.063 m_0$ and $m_h=0.51m_0$, respectively, and the integration energy cutoffs for electron and hole self-energies are $160$~meV and $20$~meV, respectively (due to the large difference in electron and hole masses in this material). These cutoff energies are the same for all densities, and as before, larger (smaller) cutoff energies induce a redshift (blueshift) of the peak position when the charge density increases. The broadening parameters in the GaAs quantum well simulation are $\Gamma_1=1$~ meV, $\Gamma_2$=5~meV and $\Gamma_3=1$~meV.  Comparing Figs.~\ref{fig:DynAbsorptBefore}(a) and (b), the change between a fully-dynamical BSE model of ML-TMDs and conventional semiconductor quantum wells is quantitative in nature, where the binding energies and involved densities are much larger in ML-TMDs. 



\subsection{Shortcomings of the dynamical BSE}

So far we have only considered the long-wavelength part of the dynamical potential in the BSE, from which we were able to calculate the density-dependent BGR and exciton binding energy. However, this theoretical framework does not capture noticeable empirical properties that are unique to ML-TMDs compared with conventional semiconductor quantum-well heterostructures. For example, the experimental results in Fig.~\ref{fig:blueshift_low} show that the exciton peak blueshifts when the charge density is increased in electron-doped MLs, whereas the peak position is essentially unaffected when the charge density is increased in  hole-doped samples. Another intriguing feature not captured by the dynamical BSE is the optical sideband that emerges in electron-doped W-based MLs but is conspicuously absent in hole-doped systems and/or Mo-based MLs  (Figs.~\ref{fig:Gating}, \ref{fig:blueshift_low} and \ref{fig:blueshift_high}).  

Using the previous theoretical framework, the only way to fit the observed blueshift in electron-doped ML-TMDs and its absence in hole-doped MLs is to evaluate the BGR by the use of density-dependent integration cutoff energies that are further different for electron- and hole-doped conditions. However, this approach is clearly non-physical and it requires many density-dependent fitting parameters to balance the BGR and screening so that the calculation matches the experiment. The facts that the electron and hole effective masses are similar in ML-TMDs, and neither the conduction nor valence band is degenerate are strong indications that one should expect a similar roles of BGR and screening in electron- and hole-doped conditions. Therefore, the observed blueshift in electron-doped ML-TMDs and its absence in hole-doped MLs should not be attributed to dynamical-screening  effects from the long-wavelength limit of the Coulomb potential.

 
\section{The coupling between intervalley plasmons and excitons}\label{sec:exciton_intervalley}

We present an analytical model in this section that quantifies the coupling between low-energy exciton states and intervalley plasmons in ML-TMDs. This model calculates many-body interactions in the exciton spectrum without the need to invoke a computationally-intensive dynamical BSE model \cite{VanTuan_PRX17}. Most importantly, the interaction between the exciton and the shortwave plasmons is shown to explain on equal footing both the blueshift of $X^0$ in electron-doped MLs and the emergence of the optical sideband in electron-doped W-based  MLs. Furthermore, the theory captures the observations that the blueshift of $X^0$ is stronger in ML-MoSe$_2$, that it is absent in hole-doped ML-TMDs, and that the optical sideband neither emerges in hole-doped MLs nor in electron-doped Mo-based MLs.


Similar to the previous analysis, the behavior of $X^0$ is studied from the relation between absorption of a photon with energy $\hbar \omega$ and the pair Green's function of the direct exciton. From the numerical solutions we got so far for the pair Green's function when applying the quasistatic approximation (Sec.~\ref{Sec:QuasiStatic}), Shindo  approximation (Sec.~\ref{Sec:Shindo}), or the fully dynamical BSE model (Sec.~\ref{Sec:ExcitonsDyn}), we notice that the pair Green's function of a bound exciton can be written as
\begin{equation}\label{Eq:G0}
G_{d}^{0}(\bm{q},E=\hbar \omega - E_\mathrm{g,d})= \left[ \hbar \omega - E_\mathrm{g,d} - E_{d,\bm{q}} + i \Gamma(E) \right]^{-1},
\end{equation}
after integrating out the Fermion degrees of freedom and performing analytical continuation. $E_\mathrm{g,d}$ is the optical band-gap energy between the valence and conduction bands from which the direct exciton arises,  $\Gamma(E)$ is the broadening function given in Eq.~(\ref{Eq:broadening}), and $\bm{q}$ is the exciton's center-of-mass wavevector. The limit $q \rightarrow 0$ applies for direct excitons in the light cone, and $E_{d,\bm{q}}$ is the energy of the direct exciton,
\begin{equation}\label{Eq:D_ENG}
E_{d,\bm{q}}= E_{d} + \frac{\hbar^2q^2}{2M_d}\,.
\end{equation}
$E_{d}$ is the direct-exciton energy level below the continuum (i.e., $|E_{d}|$ is the binding energy).  The direct-exciton mass is  $M_d = m_{ct} +m_{vt}$ for W-based MLs and $M_d = m_{cb} +m_{vt}$ for Mo-based MLs, where $m_{ct}$($m_{cb}$) denotes the electron effective mass in the top (bottom) valley of the conduction band, and $m_{vt}$ is the hole effective mass in the top valley of the valence band. Conversely, the mass of the indirect exciton is $M_i = m_{cb} +m_{vt}$ for W-based MLs or $M_i = m_{ct} +m_{vt}$ for Mo-based MLs, as shown in Fig.~\ref{fig:Scheme}. The pole of $G_d(\mathbf{q}=0,E=\hbar \omega - E_\mathrm{g,d})$ is at $\hbar \omega = E_\mathrm{g,d} + E_{d} $, which is largely unaffected when the charge density in the ML increases because of the offset between BGR and screening. 

\begin{figure}[htb!]
\centering
\includegraphics*[width=\textwidth]{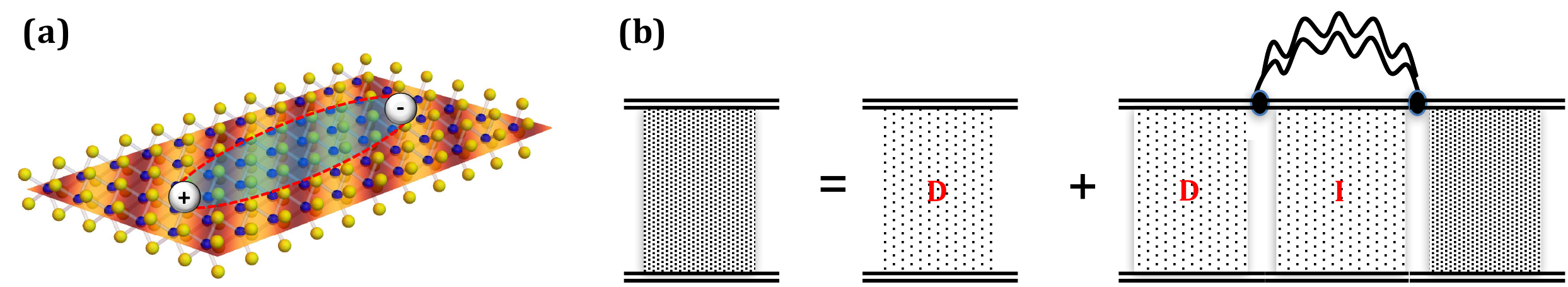}
\caption{(a) Illustration of a neutral exciton and a shortwave charge fluctuation in real space. (b) Feynmann diagram representation for the energy renormalization of the direct exciton due to its interaction with these shortwave charge fluctuations. It involves virtual emission and absorption of intervalley plasmons (wiggly lines), through which the exciton becomes indirect at the intermediate state. The diagram is an infinite sum of the above processes: D + DID + DIDID + ... where D and I denote direct and indirect exciton states, respectively.
}\label{fig:ReDiagram}
\end{figure}

The blueshift in the absorption spectrum can be modeled  from renormalization of the exciton's energy,
\begin{equation}\label{Eq:G_norm}
G_{d}(\bm{q},E)= \frac{ G^0_{d}(\bm{q},E)}{1 - G^0_{d}(\bm{q},E)\Sigma_\mathrm{s}(\bm{q},E) }.
\end{equation}
where $\Sigma_\mathrm{s}$ is the self-energy correction from virtual transitions between direct and indirect excitons mediated by shortwave (intervalley) plasmons \cite{VanTuan_PRX17}. Using the finite-temperature Green's function formalism \cite{Mahan_book}, the self-energy of direct excitons due to shortwave plasmons follows from
\begin{eqnarray}\label{Eq:SE_Omega}
\Sigma_{s}(\bm{q},\Omega)= - k_\mathrm{B}T \sum_{\bar{\bm{q}},\Omega'} \left| \mathcal{M}_{\bar{\bm{q}}}\right|^2  D(\Omega - \Omega',\bar{\bm{q}}) G_{i}(\bar{\bm{q}}+ \bm{q},\Omega'). \,\,\,\,\,\,\,
\end{eqnarray}
$\Omega$ and $\Omega'$ denote even (boson) imaginary Matsubara energies that will eventually be analytically continued into the real-energy axis ($\Omega \rightarrow E + i\Gamma_{\Sigma}$).  The sum over $\bar{\bm{q}}$ is restricted to the range of free-plasmon propagation. $D(\Omega,\bar{\bm{q}})$ is the intervalley-plasmon propagator
\begin{eqnarray}\label{Eq:D}
D(\Omega,\bar{\bm{q}})=\frac{2\hbar \omega_{s,\bar{\bm{q}}}}{\Omega^2-\hbar^2 \omega^2_{s,\bar{\bm{q}}}}\,\,,
\end{eqnarray}
expressed in terms of the collective intervalley plasma frequency, $\omega_{s,\bar{\bm{q}}}$, provided by Eq.~(\ref{eq:wqs}). $G_{i}(\bar{\bm{q}},\Omega)$ is the unperturbed indirect-exciton Green's function (prior to renormalization by intervalley plasmons)
\begin{eqnarray}\label{Eq:Gi}
G_{i}(\bar{\bm{q}},\Omega)  = [\Omega-E_{i,\bar{\bm{q}}}]^{-1}.
\end{eqnarray}
$E_{i,\bar{\bm{q}}}= E_{i} + \hbar^2\bar{q}^2/2M_i$ is defined similarly to $E_{d,\bm{q}}$ in Eq.~(\ref{Eq:D_ENG}), but with indirect exciton parameters. Finally, $\mathcal{M}_{\bar{\bm{q}}}$ is the exciton-plasmon interaction matrix element \cite{VanTuan_arXiv18},
\begin{eqnarray}\label{Eq:Mq}
|\mathcal{M}_{\bar{\bm{q}}}|^2=   \frac{ \pi  \alpha_0 \hbar^3 }{m_{cb} A} \cdot \frac{   r_s(\bar{\bm{q}}) }{  \omega_{s,\bar{\bm{q}}}} ,\,\,\,\,\,\,\,\,\,\,
\end{eqnarray}
where $A$ is the sample area and the other parameters were defined in Eqs.~(\ref{eq:alpha0}), (\ref{eq:rs}), and (\ref{eq:wqs}).  The exciton interaction with shortwave plasmons is unique because the Coulomb interaction associated with intervalley plasmons does not allow for one charge in the exciton to screen the interaction of the opposite charge with the plasmon: $K_0a_X \gtrsim 10$ where $a_X$ is the exciton Bohr radius. When the coupling is between direct and indirect excitons (Fig.~\ref{fig:Scheme}), the electron component of the exciton is scattered between valleys while the hole is a spectator. Figure~\ref{fig:ReDiagram} shows this physical picture. The opposite scenario is relevant when the coupling is between type-A and type B excitons, governed by the large spin-splitting energy of the valence band ($|\Delta_v| \gg |\Delta_c|$). In the latter case, however, the pole of the self-energy $\Sigma_s$ is far-apart from the ground-state energy of the direct exciton.  

The hole-doped case is similar where conduction-band parameters are replaced by valence-band ones, $\Delta_{c} \rightarrow \Delta_{v}$, and effective masses are considered with reversed roles of the bottom and top valleys: $m_{cb} \rightarrow m_{vt}$ and $m_{ct} \rightarrow m_{vb}$. Following the discussion after Eq.~(\ref{eq:Fq}), the interaction with intervalley plasmons is larger in electron-doped samples, coming from larger contributions and constructive interference of umklapp processes when local-field effects are governed by the orbital $d_{z^2}$ compared with smaller contributions and destructive interference when they are governed by $d_{(x \pm iy)^2}$. This behavior is manifested in the value of $\eta$ in $|\mathcal{M}_{\bar{\bm{q}}}|^2 \propto \alpha_0^2 \propto 1/\eta^2$, where analytical calculations by using hydrogen-like $5d$ or $4d$ orbitals yield $\eta \simeq 0.2$ for electron doping and $\eta \simeq 0.45$ for hole doping \cite{VanTuan_arXiv19}.

The self-energy computation is greatly simplified by using the approximated form of $G_{i}(\bar{\bm{q}},\Omega)$ instead of calculating its values from an intensive dynamical BSE model \cite{VanTuan_PRX17}. Considering direct excitons in the light cone [$\bm{q} \rightarrow 0$], we transform the summation over $\Omega'$ in Eq.~(\ref{Eq:SE_Omega}) into contour integration by using the identity in Eq.~(\ref{eq:identity}), and get that 
\begin{eqnarray}\label{Eq:SE_supp}
\!\!\!\!\!\!\!\!\!\!\! \Sigma_\mathrm{s}(\Omega) = -  \sum\limits_{\bar{\bm{q}}}\left|\mathcal{M}_{\bar{\bm{q}}}\right|^2 && \left[\frac{g(E_{i}(\bar{\bm{q}}))-g(\hbar \omega_{s,\bar{\bm{q}}})}{\Omega+\hbar \omega_{s,\bar{\bm{q}}}- E_{i}(\bar{\bm{q}}) } - \frac{g(E_{i}(\bar{\bm{q}}))-g(-\hbar \omega_{s,\bar{\bm{q}}})}{\Omega-\hbar \omega_{s,\bar{\bm{q}}}-E_{i}(\bar{\bm{q}})}\right] . \,\,\,\,\,\,\,\,\,
\end{eqnarray} 
Assuming low temperatures and recalling that $E_{i}(\bar{\bm{q}})$ is negative, the Bose-Einstein distributions follow $g(E_{i}(\bar{\bm{q}})) \rightarrow -1 $, $g(\hbar \omega_{s,\bar{\bm{q}}}) \rightarrow 0$, and $g(-\hbar \omega_{s,\bar{\bm{q}}}) \rightarrow -1$. The resulting self-energy reduces to
\begin{eqnarray}\label{Eq:SE}
\Sigma_\mathrm{s}(\Omega)=  \sum\limits_{\bar{\bm{q}}} \frac{\left|\mathcal{M}_{\bar{\bm{q}}}\right|^2}{\Omega+\hbar \omega_{s,\bar{\bm{q}}}- E_{i,\bar{\bm{q}}}}.
\end{eqnarray} 

\begin{figure}[t]
\centering
\includegraphics*[width=12cm]{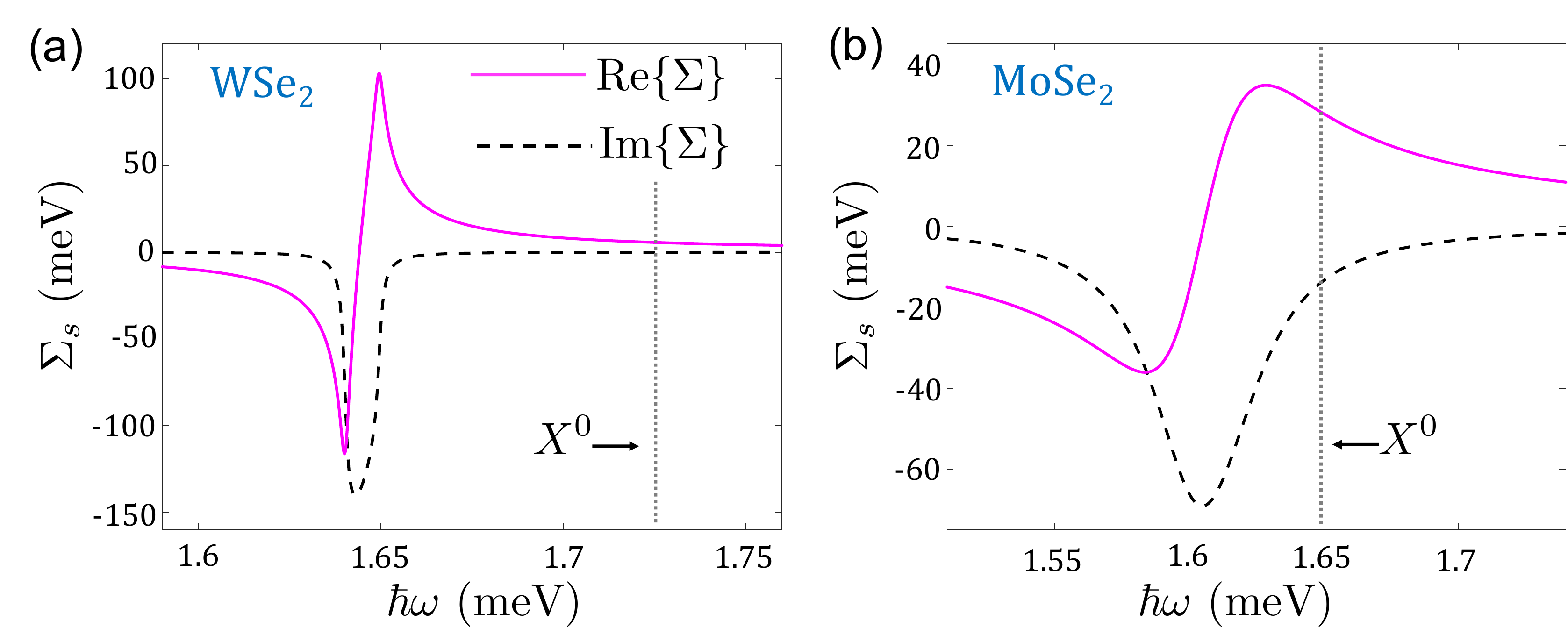}
\caption{Self-energy in electron-doped ML-WSe$_2$ and ML-MoSe$_2$ at $n=5\times10^{12}$ cm$^{-2}$. The energy difference between the direct exciton ($X^0$) and the self-energy pole is larger in WSe$_2$. Taken from Ref.~\cite{VanTuan_arXiv18}.}\label{fig:sigma}
\end{figure}

The final piece of the analysis deals with the broadening we employ to evaluate the self-energy, $\Sigma_\mathrm{s}(\Omega \rightarrow E + i\Gamma_{\Sigma})$. This broadening is dealt differently from the one employed to evaluate the pair Green's function [substituting Eq.~(\ref{Eq:broadening}) in Eq.~(\ref{Eq:G0})]. In the case of ML-MoSe$_2$, a large value for $\Gamma_{\Sigma}$  is needed because of the energy proximity of direct and indirect excitons. In more detail, the indirect exciton in ML-MoSe$_2$ is heavier than the direct one because $m_{ct}\approx 0.58m_0$ whereas  $m_{cb} \approx 0.5m_0$ \cite{Kormanyos_2DMater15}. The resulting larger binding energy of the indirect exciton is offset by a larger band-gap energy, and consequently,  $E_{\mathrm{g},i} + E_i$ is close to $E_{\mathrm{g},d} + E_d$. Further support for this spectral overlap can be found from the absence of a spectrally resolved dark exciton in ML-MoSe$_2$ \cite{Wang_PRL17}. Accordingly, a large broadening is needed to mitigate the nearby singularities in the renormalized Green's function when $E_{\mathrm{g},i} + E_i$ and $E_{\mathrm{g},d} + E_d$ are nearly degenerate. This problem does not arise in ML-WSe$_2$, where $m_{cb}\approx 0.4m_0$ and $m_{ct}\approx 0.29m_0$ \cite{Kormanyos_2DMater15}, and as a result, $E_{\mathrm{g},i} + E_i$ is well below $E_{\mathrm{g},d} + E_d$. Indeed, experiments find that the dark-exciton energy is 40~meV below the bright one in ML-WSe$_2$ \cite{Wang_PRL17,Robert_PRB17,Zhou_NatNano17,Zhang_NatNano17,Barbone_NatComm18}.  Figure~\ref{fig:sigma} shows the calculated self energies in ML-WSe$_2$ and ML-MoSe$_2$ embedded in hBN with $\Gamma_{\Sigma}=1$ and 20~meV, respectively. The band-gap at charge-neutrality conditions is chosen such that the direct-exciton peak emerges at 1.725~eV in ML-WSe$_2$ and 1.65~eV in ML-MoSe$_2$. These values are just reference energy levels in the simulations.

\subsection{The exciton blueshift and emergence of the optical sideband}

We evaluate the absorption profile of neutral excitons in ML-WSe$_2$ and ML-MoSe$_2$ embedded in hBN, using the experimental results in Figs.~\ref{fig:Gating}, \ref{fig:blueshift_low} and \ref{fig:blueshift_high} as a benchmark. The absorption is evaluated from the imaginary part of Eq.~(\ref{Eq:G_norm}), and it requires knowledge of the band-gap energies, broadening parameters, binding energies in charge neutrality conditions, and spin-splitting energies. Each of these parameters yields different values in electron and hole doped conditions.  We first discuss these parameters and then present results.

\subsubsection{Band-gap energies:} $E_{\mathrm{g},d}$ and $E_{\mathrm{g},i}$ include the effect of BGR following the parameters we have used in Sec.~\ref{Sec:BGR} and Fig.~\ref{fig:ch}. The BGR is needed to define the position of the redshifting continuum from which we evaluate the energy-dependent broadening function when we substitute Eq.~(\ref{Eq:broadening}) in Eq.~(\ref{Eq:G0}) and then Eq.~(\ref{Eq:G0}) in Eq.~(\ref{Eq:G_norm}).

\subsubsection{Broadening parameters:} The only free parameters pertain to broadening. The values of $\Gamma_{\Sigma}$ are 1~meV in ML-WSe$_2$ and 20~meV in ML-MoSe$_2$ and they are used in $\Sigma_\mathrm{s}(\bm{q},E=\hbar\omega - E_{g,d} + i\Gamma_{\Sigma})$. These values are similar to the ones used to calculate the exciton's self-energy in Fig.~\ref{fig:sigma}. As mentioned above, a large value is needed in ML-MoSe$_2$ due to the energy degeneracy of its direct and indirect excitons.  

The broadening employed in $G^0_{d}(\bm{q},=\hbar\omega - E_{g,d} + i\Gamma(\hbar\omega))$ follows Eq.~(\ref{Eq:broadening}), where $\Gamma_{1} = 3$~meV, $\Gamma_{2} = 30$~meV, and $\Gamma_{3} = 10$~meV. These values are similar to the ones used in the calculations of the dynamical BSE (Figs.~\ref{fig:qsAbsorption}-\ref{fig:DynAbsorptBefore}). The energy-dependence of $\Gamma(\hbar\omega)$ leads to a strong suppression of photon absorption close to the band-gap energy compared with photons whose energies are far below the continuum. As a result, an artificial redshift of up to 5~meV is introduced in $G^0_{d}(\bm{q},=\hbar\omega - E_{g,d} + i\Gamma(\hbar\omega))$ when the density increases from 0 to 5$\times$10$^{12}$~cm$^{-2}$. To compensate for this small energy shift, a density-dependent energy is added to $|E_d|$  in order to keep the peak position constant in the absorption spectrum  when the charge density is increased and $\Sigma_\mathrm{s}=0$. This correction has no bearing on the many-body effects, and it is not needed if one employs an energy-independent broadening function instead of the one in Eq.~(\ref{Eq:broadening}). 

\subsubsection{Binding energies of direct and indirect excitons in charge-neutrality conditions:}  $|E_d|$ and $|E_i|$ at zero charge density are calculated by the stochastic variational method \cite{VanTuan_arXiv18,Zhang_NanoLett15}. Following Ref.~\cite{VanTuan_arXiv18},  the binding energies in charge neutrality conditions are $|E_d| = 178$~meV and $|E_i| = 195$~meV in ML-WSe$_2$ embedded in hBN. The respective values in embedded ML-MoSe$_2$  are $|E_d| = 203$~meV and $|E_i| = 211$~meV. The value of $|E_d|$ in WSe$_2$ close to charge neutrality conditions is also available experimentally \cite{Stier_PRL18}, and it matches the calculated value.   

\subsubsection{Spin-splitting energies:} The density dependence of the spin-splitting energy is calculated from Eq.~(\ref{Eq:Delta_c}). The density dependence stems  from the exchange contributions in the long-wavelength and shortwave limits. The spin-splitting energies in charge-neutrality conditions are governed by the spin-orbit interaction ($\Delta_{c,0}$ in the conduction band and $\Delta_{v,0}$ in the valence band). Their values are $|\Delta_{c,0}|=23$~meV  and $|\Delta_{v,0}|=427$~meV in ML-WSe$_2$ or $|\Delta_{c,0}|=8$~meV $|\Delta_{v,0}|=185$~meV in ML-MoSe$_2$.  The spin-splitting energies in the valence band are relevant in hole-doped MLs, and they are taken from DFT-based values \cite{Kormanyos_2DMater15}. These values match very well the empirical energy difference between type-A and type-B excitons (optical transition from the top and bottom valleys of the valence band). The spin-splitting energies in the conduction band are relevant in electron-doped conditions. They are extracted by assuming that dark excitons have the same binding energies as the indirect ones (because the electron effective masses are the same in both cases: Dark excitons are formed when the electron and hole reside in the same valley but their spin configuration forbids optical transitions for out-of-plane propagating photons). Using the empirical value for the dark excitons: 40~meV below the neutral direct-bright exciton in WSe$_2$ \cite{Robert_PRB17,Zhang_NatNano17,Zhou_NatNano17}, and about the same energy as that of the neutral direct-bright exciton in MoSe$_2$ \cite{Wang_PRL17}, we have extracted the spin-orbit contribution to the spin-splitting energy from $|\Delta_{c,0}| = 40 - |E_i - E_d| = 23$~meV in ML-WSe$_2$ and $|\Delta_{c,0}| = |E_i - E_d| = 8$~meV in ML-MoSe$_2$.  
 
 \begin{figure*}[ht]
\centering
\includegraphics*[width=15cm]{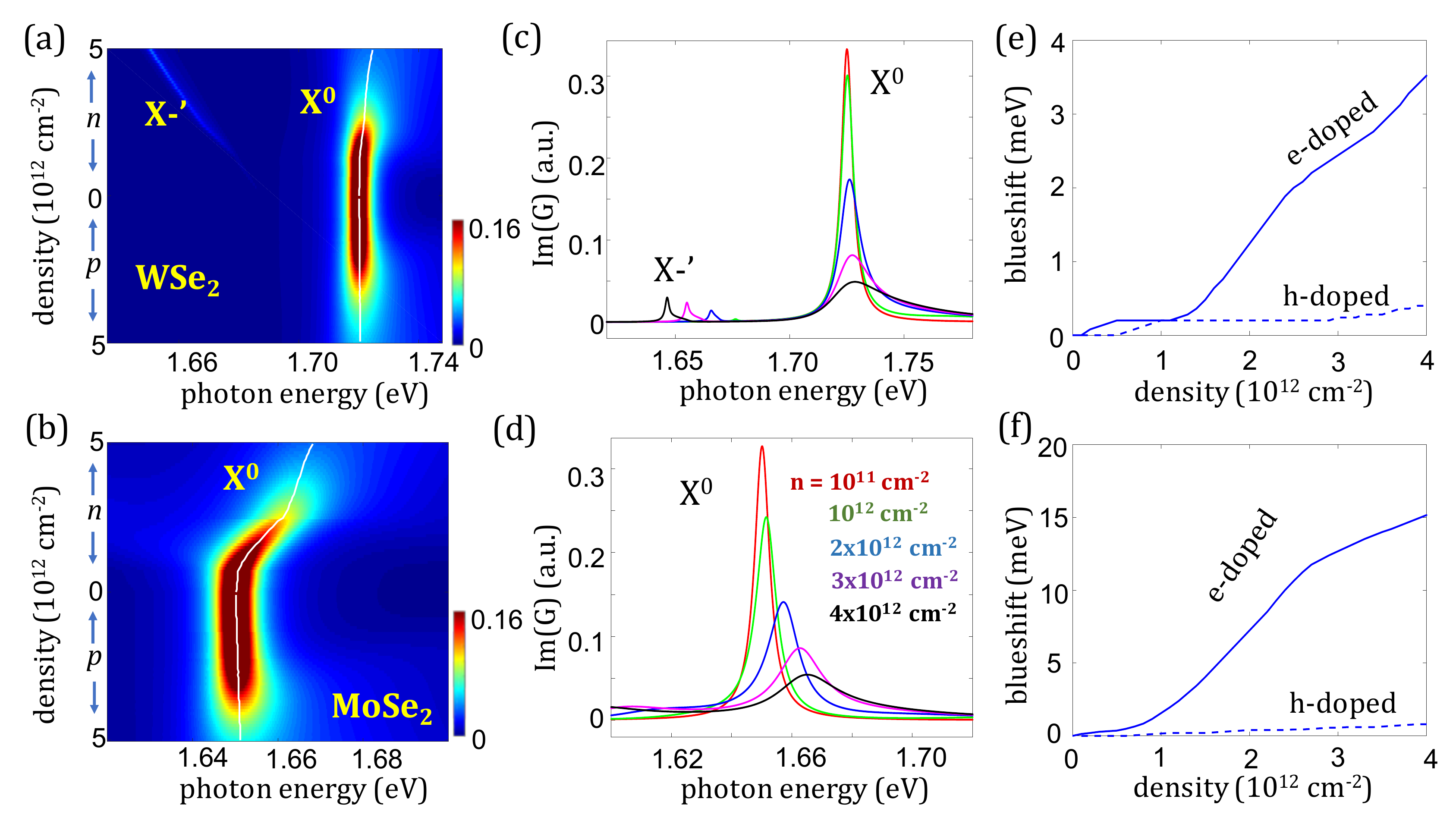}
\caption{Calculated absorption spectrum of the neutral exciton, $X^0$,  for ML-WSe$_2$ (top) and ML-MoSe$_2$ (bottom) embedded in hBN. (a,b) The absorption as a function of charge density and photon energy. The white lines trace the peak position. In addition, the exciton-plasmon interaction corresponds to $X-$' in the low-energy side of the spectrum in electron-doped ML-WSe$_2$. (c,d) Cross sections from panels (a) and~(b) for different electron densities. (e,f) The blueshift dependence of $X^0$ on charge density, where solid (dashed) lines denote electron (hole) doping. Taken from Ref.~\cite{VanTuan_arXiv18}. }\label{fig:Abs}
\end{figure*}
 
\subsection{Putting it all together} 

Figure~\ref{fig:Abs} shows the calculated absorption profile of neutral excitons in ML-WSe$_2$ and ML-MoSe$_2$ embedded in hBN. Comparing these results with the experimental ones in Figs.~\ref{fig:Gating}, \ref{fig:blueshift_low} and \ref{fig:blueshift_high}, we find compelling evidence for the coupling between excitons and intervalley plasmons. For example, the theory confirms that the blueshift of $X^0$ is observed only in electron-doped TMDs [Figs.~\ref{fig:Abs}(e) and (f)], and that it is larger in ML-MoSe$_2$ than in ML-WSe$_2$ [Figs.~\ref{fig:Abs}(c) and (d)]. The latter stems from the proximity between energies of direct and indirect excitons in ML-MoSe$_2$. The blueshift is weaker in hole-doped TMDs because of a smaller local-field effect and a mismatch between the plasmon energy when it is governed by $\Delta_{v,0}$ and the ten-fold smaller energy difference of direct and indirect excitons, governed  by $\Delta_{c,0}$. In addition, the analytical model captures the observed emergence and redshift of the optical sideband in electron-doped ML-WSe$_2$ ($X-$'), and its absence in ML-MoSe$_2$ or in hole-doped conditions. The spectral position of this many-body feature is about one plasmon energy below the indirect exciton, which in ML-WSe$_2$ lies at a lower energy than the direct exciton. A clear advantage of this theoretical model is its exceptional efficiency: All of the density-dependent many-body effects in Fig.~\ref{fig:Abs} are computed within seconds on a simple computer. 

\section{Conclusions and outlook}\label{Sec:Con}

We have provided an overview of dynamical screening in monolayer TMDs. After establishing the form of the non-local dielectric function, we have analyzed the random-phase approximation forms of the dynamically-screened potential in the long-wavelength and shortwave limits. We have discussed the resulting two types of charge excitations governed by intravalley and intervalley plasmons, finding that local-field effects are important in the shortwave limit, especially in electron-doped conditions due to the orbital composition of conduction-band states. Using the single-plasmon pole approximation,  the relatively cumbersome random-phase approximation form of the dynamically-screened dielectric function was then replaced with compact expressions in the long-wavelength and shortwave limits. The self-energies of electrons and holes and their charge-density dependence were  evaluated, and the ensuing band-gap renormalization was calculated along with the density-dependent changes of the spin-spitting energies in the conduction and valence bands. 

After establishing single-particle properties of electrons and holes,  we have shown how one can compute neutral exciton states in electrostatically-doped  ML-TMDs from the Bethe-Salpeter Equation under the screened-ladder approximation. Here, the binding energy between electrons and holes is governed by the long-wavelength part of the dynamically-screened Coulomb potential. Already in the quasistatic approximation, where only the statically screened potential is used in the Bethe-Salpeter Equation, two important and competing effects can be discerned: Increased screening results in a reduction of the exciton binding energies (and hence a blueshift), whereas the band-gap renormalization due to correlation and exchange effects gives rise to a reduction of the band gap (and hence a redshift). Taking into account dynamical effects via the Shindo approximation or through a fully-dynamical Bethe-Salpeter Equation do not qualitatively alter these results. Furthermore,  the coupling between excitons and long-wavelength plasmons is qualitatively similar in ML-TMDs and conventional semiconductor quantum wells, where the only change is that the binding energies and involved densities are much larger in ML-TMDs. 

Going beyond these approximations, we have calculated the self-energy of neutral-excitons due to their interaction with intervalley plasmons. The interaction couples between direct and indirect excitons, and can be viewed as a quasiparticle arising from a neutral exciton and collective intervalley charge excitation.  This quasiparticle reveals the unique role of intervalley plasmons in the absorption spectra of ML-TMDs. The interaction with intervalley plasmons can explain the empirically found blueshift of the neutral exciton peak when the charge density is increased in electron- but not in hole-doped conditions. In addition, the confluence of the intervalley plasmon energy and energy difference between direct and indirect excitons elucidate the emergence of an optical sideband in electron-doped ML-WS$_2$ and ML-WSe$_2$, and its conspicuous absence in hole-doped systems and/or ML-MoS$_2$ and ML-MoSe$_2$. 

\subsection{Outlook}
The study of many-body effects in ML-TMDs is in its infancy. We list several open questions in this relatively nascent field.  
\begin{enumerate}
\item \textit{Beyond the screened-ladder approximation}

The experimental results in Figs.~\ref{fig:Gating}, \ref{fig:blueshift_low} and \ref{fig:blueshift_high} clearly show that the exciton peak disappears without any signs of excitonic enhancement due to the 2D nature of the monolayer  \cite{SchmittRink1986:PRB}. Furthermore, Fig.~\ref{fig:blueshift_low}   shows that the measured decay cannot be attributed solely to the rise of trions (i.e., transfer of the oscillator strength from neutral to charge excitons), since this rise is not commensurate with the much stronger decay in the absorption of neutral excitons. The simulated results only partially capture this behavior. While the exciton absorption peak decays when the charge density increases [e.g., Figs.~\ref{fig:qsAbsorption}(a), \ref{fig:qsBGRcomp}(a) and \ref{fig:DynAbsorptBefore}], the decay is mostly governed by the phenomenological broadening function we have employed [Eq.~(\ref{Eq:broadening})]. That is, the exciton peak decays because the broadening parameter in the analytical continuation step is enhanced when the red-shifting continuum approaches  the exciton energy level [see Figs.~\ref{fig:qsAbsorption}-\ref{fig:DynAbsorptBefore}]. The contribution to the decay from the oscillator strength of the optical transition at elevated charge densities is relatively small, as shown in Fig.~\ref{fig:qsBGR}(b), governed by the fact that an attractive potential in a 2D system always generates a bound state. Thus, the dynamical BSE model directly captures the BGR and reduction in the exciton binding energy when the charge density increases. But the decay of the exciton peak is only indirectly captured through the assumption of enhanced scattering rates when the exciton energy level gets closer to the continuum. 

The inability of the dynamical BSE model to directly capture the exciton decay is rooted in the use of the screened-ladder approximation.  This physical picture is that of an exciton in a Fermi sea of electrons (or holes), where the Fermi sea affects the exciton through the screened potential and the self-energies of its electron and hole components. Using the Feynman diagram in Fig.~\ref{fig:BSE} as a guide, the screened interaction between the electron and hole is encoded in the vertical wiggly double-lines, and the BGR is encoded in the double-line horizontal propagators of the exciton's electron and hole components. When the exciton radius is comparable to the average distance between electrons in the Fermi sea, this diagram does not capture the viable possibility that the hole equally interacts with all of the electrons in its vicinity rather than with a single one.  The problem becomes acute at elevated densities in which the reduced binding energy of the exciton due to screening increases its radius ($E_b \propto a_X^{-2}$), while the average distance between electrons in the Fermi sea decreases ($r_s$). In ML-TMDs, for example, $r_s$ and $a_X$ are comparable when the charge density is a few 10$^{12}$~cm$^{-2}$. 

In summary, the BSE under the screened-ladder approximation has its limitation regardless of whether we employ a quasistatic or a dynamically-screened potential. Further investigations that branch out of the screened-ladder approximation are needed to correctly model the decay of the exciton peak. A theoretical framework in which the hole interacts with the surrounding electron Fermi sea can be used to study both the decay of the exciton and the observed behavior of trions. As mentioned in Sec.~\ref{sec:exciton_behavior}, such a theory can clarify if trions in ML-TMDs evolve into a muted Fermi-edge singularity at elevated densities \cite{Hawrylak_PRB91}.
\item \textit{Detection of intervalley plasmons}

To the best of our knowledge, a direct detection of intervalley plasmons in ML-TMDs has not been demonstrated yet (i.e., not through the exciton optical transitions). In addition to reflection electron
energy loss spectroscopy, resonant Raman or THz spectroscopies are possible experiments to detect these plasmons. 
In order to conserve momentum in Raman or THz experiments, the shortwave nature of intervalley plasmons requires the coupling of the incident photon to two counter-propagating shortwave plasmons.  Electrostatic doping can be used to tell apart the signature of these plasmons from that of optical phonons in the far-infrared spectrum. The gate voltage tunes the charge density and hence the plasmon energy and its amplitude (wider damping-free propagation range). The energies of optical phonons, on the other hand, are much less affected by the gate voltage. 
\item \textit{Many-body and localization effects in the photoluminescence spectrum}

When comparing absorption/reflectance experiments and photoluminescence experiments, such as in Fig.~\ref{fig:Gating}, one observes a notable difference in the relative amplitude of the signal. The largest oscillator strength in absorption/reflectance measurements of ML-TMDs belongs to the neutral exciton $X^0$ at low charge densities, consistent with the calculated absorption spectrum in Fig.~\ref{fig:Abs}(a). On the other hand, the strongest emission in the photoluminescence comes from the optical sideband at elevated electron densities \cite{Jones2013:NN}, as seen from Fig.~\ref{fig:Gating}(a).  A possible explanation is that the plasmon-assisted emission involves the ground state (the indirect exciton in W-based MLs), whereas emission from direct excitons involves the excited state. Furthermore, the plasmon momentum counteracts that of the indirect exciton during the emission process, so that unlike the direct-exciton optical transition, the initial state of indirect excitons is not restricted to a minuscule light-cone (kinetic energy of few $\mu$eV). Another explanation can be that the quantum efficiency of light emission is improved at elevated electron densities because of screening of charged nonradiative recombination centers. Further investigations are needed to pinpoint the microscopic origin for the strong emission from the optical sideband. 

An additional open question deals with the interplay between localization effects and many-body interactions in ML-TMDs.  At low charge densities, electrons or holes in the ML can readily become localized in regions close to charged defects in the substrate or the encapsulated materials. These localization centers can affect the emission spectrum \cite{Godde_PRB16}, and often lead to confusion between emission from delocalized few-body complexes such as trions and phonon-assisted emission from excitons or biexcitons next to localization centers \cite{VanTuan_arXiv18b}.  Many-body effects such as the ones studied in this work become relevant when the charge density in the ML is large enough to screen out the charged defects and to delocalize the electrons (or holes) in the ML. Further experiments are needed to check if changes in the optical spectrum are accompanied with metal-to-insulator transition of the ML (when the charge density decreases). Such experiments should combine both transport and photoluminescence measurements in electrostatically gated structures.

\item \textit{Dynamical screening due to phonons}

The dynamical screening effects discussed in this review come from electron-electron interactions. In the case of vanishing charge densities, the exciton states are calculated through the bare non-local  Coulomb potential (Sec.~\ref{Sec:BareCoulomb}). The common practice in this case is to assume that the exciton binding energy is the largest energy scale in the problem, and therefore, to choose the high-frequency limit of the  dielectric constants in the layers below and above the ML. In other words, the assumption is that the electric-field lines generated by the relative motion of the electron and hole change rapidly, and thus, they  do not induce lattice vibrations in hBN or SiO$_2$. However, this assumption is questionable for the excited exciton states whose binding energies are actually comparable or even smaller than the energies of optical phonons in hBN or SiO$_2$ (up to 160 meV in hBN and more than 100 meV in  SiO$_2$). One then faces a difficulty whether to choose the static or high-frequency limits of these dielectric constants in a calculation that can capture both the ground and excited exciton states (e.g., $\epsilon_{\omega=\infty} = 2.1$ or $\epsilon_{\omega=0} =3.9$ in SiO$_2$). A similar problem arises when choosing a value for the dielectric screening parameter in the monolayer (e.g., the value of $r_0$ in the Rytova-Keldysh potential \cite{Chernikov2014:PRL,Meckbach_PRB18}). 

The dynamical screening models presented in Secs.~\ref{Sec:Shindo} and \ref{Sec:ExcitonsDyn} can be extended to incorporate phonon-induced dynamical effects in the non-local dielectric function. We believe that such analysis is better suited to study the nonhydrogenic Rydberg series of exciton states in ML-TMDs because of its ability to correctly model the ionic contribution to the dielectric screening parameters.  Experimentally, one can look for or engineer materials with similar dielectric constants to those found in hBN (or SiO$_2$), and examine whether the energies of excited exciton states differ from the ones seen in ML encapsulated in hBN (or supported on SiO$_2$).

A related open question deals with the polaron effect in ML-TMDs. The Fr\"{o}hlich interaction between excitons and optical phonons in ML-TMDs is expected to be weak because of the charge neutrality
and small size of the exciton as well as the similar effective masses of electrons and holes. Combined together, the interactions of the electron and hole with the phonon-induced macroscopic electric  field cancel out  \cite{VanTuan_arXiv18b}. One can then assume that the exciton does not distort the polar ML-TMD crystal in its vicinity, and accordingly, the effective masses of the electron and hole are not subjected to polaron effects. This scenario changes for electrons, holes, trions and other charged complexes because their nonzero charge can distort the crystal in their vicinity, and the resulting phonon cloud increases their effective masses. 
Further experiments and theoretical models are needed to quantify how the polaron effect in ML-TMDs affects the mobility of electrons and holes, the binding energy of trions, and the band-gap renormalization in doped MLs. These studies can also resolve discrepancies between experimentally measured effective masses and DFT-based calculations of these masses at the edges of the conduction and valence bands. 

\item \textit{Proximity effects}

We expect dynamical screening to be important in the emerging area of magnetic proximity effects in ML-TMDs \cite{Zutic_MT18}.  Until recently \cite{Scharf2017:PRL}, magnetic proximity effects in all materials were theoretically studied by neglecting the role of Coulomb interactions, even without considering the simpler quasistatic approximation. While early experiments have focused on the out-of plane magnetization demonstrating the removal of the valley degeneracy in transition-metal dichalcogenides \cite{Zhao2017:NN,Zhong2017:SA}, it would be also interesting to explore other magnetization orientations in a substrate which could strongly alter the optical selection rules. For example, in-plane magnetization could yield additional spectral features and transform dark into bright excitons \cite{Scharf2017:PRL}. Recent experiments on magnetic proximity effects in WS$_2$/EuS show a large valley splitting of $\sim 20$~meV at 1~T \cite{Petrou_PC18}, larger splitting than what is available with a static applied magnetic field even at specialized facilities. Since two-dimensional ferromagnetism is available in monolayer van der Waals materials \cite{Huang_N17,Gong_N17} and with the gate-controlled critical temperature \cite{Deng_N18}, there are many unexplored opportunities for probing the role of dynamical screening in van der Waals heterostructures. Magnetic proximity offers another way to control and study many-body interactions in the time-reversed valleys of ML-TMDs \cite{Scharf2017:PRL}. For example, by competing with the influence of the intrinsic spin-orbit coupling, it would change the energy of shortwave plasmons \cite{VanTuan_PRX17}. 

Given many possibilities to fabricate transition-metal dichalcogenides van der Waals heterostructures with superconducting regions, both with conventional \cite{Yoshida_APL16} and unconventional pairing symmetry \cite{Li_SR17}, one can envision other directions where the studies of dynamical screening will play a crucial role. The gate-dependent properties have been separately demonstrated for both normal and superconducting regions of such heterostructures \cite{VanTuan_arXiv18,Yoshida_APL16} . Virtual exchange of shortwave plasmons may give rise to pairing mechanisms in superconductors whose Fermi surfaces comprise distinct pockets in the Brillouin zone. By employing dynamical screening, similar as we have considered here, it is then possible to model plasmon-mediated Cooper pairing \cite{Pines_CJP56,Radhakrishnan_PL65,Pashitskii_SPJETP69,Ruvalds_PRB87} which was argued to facilitate superconductivity in cuprates \cite{Krasnov_NatComm12} and other transition-metal systems through plasmon-phonon hybridization \cite{Ganguly_PRL72,Pashitskii_SPJETP93}, as well as in semiconductors and electron-hole liquids \cite{Takada_JPSJ78,Ruhman_PRB16,Vignale_PRB85,Blank_PSSB95}. 

\item

\textit{Devices}

Dynamical screening and intervalley plasmons may find use in optical devices in which the gate voltage controls the wavelength of the emitted light or the absorption amplitude at a given wavelength.  Achieving wavelength tunability this way has a great cost advantage compared with the use of external optical cavities. Furthermore, these devices can be integrated with silicon circuits for optics-on-chip applications without having to worry about lattice constant matching due to the van der Waals nature of ML-TMDs \cite{Geim2013:N}. These expectations are well supported by a variety of applications that rely on unique optical response of ML-TMDs, from tunable excitonic light-emitting diodes \cite{Ross_NN14} and ultrasensitive photodetectors \cite{LopezSanchez2013:NN}, to single photon emitters \cite{He_NN15, Chakraborty_NN15} important for quantum information processing. Owing to their spin-valley coupling \cite{Xiao2012:PRL}, ML-TMDs were proposed for spin lasers \cite{Lee_APL14} in which the spin imbalance and the conservation of angular momentum between the carriers and photons could enhance performance of conventional (spin-unpolarized) lasers \cite{FariaJunior_PRB15}. Significant improvements in dynamical operation and power consumption using III-V semiconductor-based spin lasers \cite{Lindemann_P18}, as compared to the state-of-the art conventional counterparts, suggest also important opportunities for spin lasers in ML-TMDs \cite{Zutic_MT18} where lasing has already been demonstrated \cite{Wu_15N, Ye_15NP}. In addition to the common approach of using optical or electrical spin injection to generate spin-polarized carriers in spin laser, magnetic proximity effects in ML-TMDs could provide a viable alternative \cite{Zutic_MT18}.

\end{enumerate}

\section*{Acknowledgements}
This work was supported by the U.S. DOE, Office of Science BES, under Awards No. DE-SC0014349 (Rochester) and DE-SC0004890 (Buffalo), the German Science Foundation (DFG) via Grants No. SCHA 1899/2-1 and SFB 1170 ``ToCoTronics'', and by the ENB Graduate School on Topological Insulators. The computational work in Rochester was also supported by the National Science Foundation (Grant No. DMR-1503601). The computational work in Buffalo was supported by the UB Center for Computational Research and the Unity Through Knowledge Fund, Contract No. 22/15.

\end{document}